\documentclass[a4paper,11pt]{article}
\pdfoutput=1 
\usepackage{jcappub} 
\usepackage[T1]{fontenc}
\usepackage{graphicx}
\usepackage{amssymb}
\usepackage{amsmath}
\usepackage{color}
\usepackage{cancel}
\usepackage{hyperref}
\hypersetup{urlcolor=blue, colorlinks=true,citecolor=blue}
\usepackage[utf8]{inputenc}
\usepackage{url}
\usepackage[normalem]{ulem}
\usepackage{cancel}

\newcommand{\be}{\begin{equation}}
\newcommand{\ee}{\end{equation}}
\newcommand{\bea}{\begin{eqnarray}}
\newcommand{\eea}{\end{eqnarray}}
\newcommand{\bfig}{\begin{figure}}
\newcommand{\efig}{\end{figure}}
\newcommand{\nn}{\nonumber}

\newcommand{\ylm}[1]{Y_{\ell m}(\hat #1)}
\newcommand{\ylmc}[1]{Y^*_{\ell m}(\hat #1)}

\def\dt{{\rm d}^3}

\def\l{\left(}
\def\r{\right)}
\def\lr{\left[}
\def\rr{\right]}
\def\alm{a_{\ell m}}

\def\O{\Omega}
\def\don{{\rm d}\O_{\hat n}}
\def\vk{\vec{k}}

\def\z{\l z\r}
\def\di{{\rm d}}

\def\H{{\mathcal H}}

\def\odm{{\Omega_{\rm cdm  0}}}
\def\ob{{\Omega_{\rm b 0}}}

\title{Anti-symmetric clustering signals in the observed power spectrum}

\author[1,2]{Jos\'e Fonseca}
\author[1,2,3]{Chris Clarkson}
\affiliation[1]{School of Physics \& Astronomy, Queen Mary University of London, London E1 4NS, UK}
\affiliation[2]{Department of Physics \& Astronomy, University of the Western Cape, Cape Town 7535, South Africa}
\affiliation[3]{Department of Mathematics \& Applied Mathematics, University of Cape Town, Cape Town 7701, South Africa}

\emailAdd{josecarlos.s.fonseca@gmail.com}

\abstract{

In this paper, we study how to directly measure the effect of peculiar velocities in the observed angular power spectra. We do this  by constructing a new anti-symmetric estimator of Large Scale Structure using different dark matter tracers. We show that the Doppler term is the major component of our estimator and we show that we can measure it with a signal-to-noise ratio up to $\sim 50$ using a futuristic SKAO HI galaxy survey. We demonstrate the utility of this estimator by using it to provide constraints on the Euler equation.
}

\begin{document}
\maketitle
\flushbottom

\section{Introduction}\label{sec:into}

Observing the Large-Scale Structure (LSS) of the Universe to further constrain the standard cosmological model is one of the driving forces of future spectroscopic galaxy surveys such {as} Euclid \cite{Laureijs:2011gra}, the Dark Energy {Spectroscopic} Instrument (DESI) \cite{Levi:2013gra} and the Square Kilometre Array Observatory (SKAO) \cite{Bacon:2018dui}, for example. An extensive body of work has gone into transforming the observed clustering of tracers of Dark Matter into measurements of cosmological parameters. But the observed clustering is a projection into the past light-cone of the tracer's density fluctuations which is not uniquely given by the underlying density contrast \cite{Yoo:2012se,Challinor:2011bk,Bonvin:2011bg}. These so-called ``General Relativistic (GR) effects'' leave imprints in the observed linear power spectrum of any LSS tracer that are often small \cite[see, e.g.,][]{Alonso:2015uua} but can be sizeable in certain conditions (see \cite{Montanari:2015rga} for a good example on how magnification lensing becomes dominant at large redshift separations). 

In practice, in galaxy surveys, one measures angular positions and estimates redshifts which then we translate into physical distances and positions in the cosmic grid. But any perturbation in the photon's energy between the galaxy and the observer will alter the measured redshift, and perceived volume \cite[see][for a review]{Bonvin:2014owa}. The most well-known of such effects is the Redshift Space Distortions (RSD) \cite{Kaiser:1987qv}. Although not considered a GR effects \emph{per se}, {the line-of-sight gradient of} peculiar velocities perturb the assumed volume and change the observed clustering. RSD is the largest such correction but others may also be relevant and detectable, such as lensing magnification \cite{Jelic-Cizmek:2020pkh} and {effects which are linearly dependent on the velocity, commonly referred to as the Doppler term }\cite{Bonvin:2013ogt}. Peculiar velocities should alter the observed redshift solely from the Doppler effect, but also change their apparent size \cite{Bacon:2014uja,Bonvin:2016dze,Andrianomena:2018aad}. Smaller corrections come from potential terms, both local and line-of-sight integrated. Such relativistic effects only become important on the largest scales and for most purposes can be safely neglected in standard cosmological parameter estimation from single tracer clustering \cite{Lorenz:2017iez}. However, for combinations of multiple tracers, this is no longer true \cite{Viljoen:2021ocx}.

Irrespective of their importance on improving constraints on cosmological parameters, the detection of such effects are exquisite probes and tests of General relativity. As an example, the authors of \cite{Alam:2017izi} attempted to identify the Shapiro time delay. Others, e.g. \cite{Metcalf:2020tgt}, have attempted to recover the lensing potential using the Lyman-$\alpha$ forest. More importantly, the authors of \cite{McDonald_2009,Gaztanaga:2015jrs,Giusarma:2017xmh} have studied the observational asymmetries in galaxy cross-correlations induced by the Doppler term, which will be the focus of our paper. 

Peculiar velocities can affect the observed clustering of galaxies in several ways not only changing their redshift and inducing a magnification but also being important in the gauge change between the synchronous gauge and the Newtonian gauge where the bias is defined \cite{Challinor:2011bk}. More importantly, is how they affect or could be measured in the traditional summary statistics used in cosmology: be it the 2-point function $\xi$ (or its multipoles $\xi_\ell$); the 3D power spectrum $P(k)$ (or its multipoles $P_\ell(k)$); or the angular power spectrum $C_\ell$. The effects of peculiar velocities have been extensively studied using the two-point function, both in auto-correlations \cite{Bonvin:2014owa,Raccanelli:2016avd} and in cross-correlations between different tracers \cite{Bonvin:2013ogt,Gaztanaga:2015jrs,Hall:2016bmm,Giusarma:2017xmh}. The authors of \cite{Abramo:2017xnp} study the effect of the Doppler effect in the multipoles of the 3D power spectrum in single and the multiple-tracer of galaxy surveys, while \cite{Hall:2016bmm} also includes HI intensity mapping (IM). Several authors have also looked at the detectability of the Doppler effect in LSS using the angular power spectra in single tracer \cite{Alonso:2015uua} and multi-tracer\cite{Alonso:2015sfa,Fonseca:2015laa,Fonseca:2018hsu,Viljoen:2021ypp}. But in all these cases the most promising way of capturing the Doppler effect is using cross-correlations between two different galaxy samples. Such ``dipolar'' or ``asymmetric'' structure arises from a broken line-of-sight symmetry between different tracers of dark matter. This effect had originally been identified and well studied by \citep{McDonald_2009}, as it induces a non-zero imaginary part in the cross-correlation power spectrum.

We will keep the same spirit in this paper and build upon these previous works. The question is, can an equivalent estimator of this line-of-sight asymmetry  be built using the observed angular power spectrum? The angular power spectrum naturally incorporates all the GR effects without any complicated modelling, including wide-angle effects. These need to be taken into account in future wide sky area surveys.  Therefore, we looked for ways of capturing this asymmetry using $C_\ell$. In the literature, there are already available anti-symmetric clustering estimators, such as the one proposed in \cite{Jalilvand:2019bhk} to reconstruct the lensing potential using magnification from cross-correlations between photometric surveys and HI intensity mapping. We repurpose this estimator for spectroscopic surveys, but now using thin bins to increase the sensitivity to peculiar velocities. It is well established that surveys with large redshift uncertainties such as photometric surveys, bin the data into large redshift bins which average out the RSD, or any effect of peculiar velocities. For this reason, we need to use a very fine tomographic bin which increases the sheer number of redshift bins, reducing the detectability of any individual anti-symmetric estimator. Despite this, as we will see, we can use this to our advantage and stack anti-symmetric estimators to increase the individual signal to noise. Using the angular power spectra with such narrow bins adds extra practical and theoretical complications \cite{Jalilvand:2019brk,Taylor:2021bhg,Matthewson:2021rmb} which we do not consider here. 

Detecting the Doppler term is not only a mild curiosity, it is indeed a very important test of the consistency of General Relativity. The shape of the Doppler contribution depends on how peculiar velocities couple with the gravitational potential, which in GR is governed by the Euler equation. While the traditional tests of GR focus on the Poisson equation and on  anisotropic stress \cite[see, e.g.][]{Baker:2011jy}, the Euler equation is poorly tested, especially because of the lack of observables of peculiar velocities on linear scales. But as pointed out in \cite{Bonvin:2018ckp} probing the Doppler contribution in LSS provides an exquisite window onto this. Here we will not review the extensive modified gravity literature and advise the reader to refer to \cite{Bonvin:2018ckp} for theories that alter the Euler Equation. We follow their steps and apply our anti-symmetric estimator to the same problem. Furthermore, as pointed out by \cite{Umeh:2020cag}, a velocity bias between tracers or between baryons and the Dark Matter mimics an Euler equation modification. 

Thus, in this paper, we propose a clustering anti-symmetric estimator which can be used to both detect the Doppler term in the observed power spectrum and provide further tests to Einstein's theory of General Relativity. Throughout this paper we will assume a Planck-like fiducial cosmology \cite{Aghanim:2018eyx} $A_s=2.142\times 10^{-9}$, $n_s=0.967$, $\odm=0.26$, $\ob=0.05$, $w=-1$, $H_0=67.74\,$ km/s/Mpc. The paper is organised as follows. In \S \ref{sec:ObsClustAnti} we review the observed power spectrum and identify where and when anti-symmetries arise. We then construct an estimator of these anti-symmetries, show how they behave in theory{, and study when and how the Doppler term can be the dominant contributor to such estimator. Subsequently in \S \ref{sec:zetaexemples} we show  examples of the behaviour of the anti-symmetric estimator for different choices of the biases of a galaxy sample. In \S \ref{sec:detectsignal} we compute the covariance of the estimator and} present calculations of the signal-to-noise ratio for different surveys{, as well prospects for detecting the Doppler term with the estimator. We study} a case example with a more futuristic galaxy survey, an upgrade to the SKAO, {where we look at how one reconstructs such anti-symmetric estimator and how detectability can be improved.} 
In \S \ref{sec:testEuler} we look into how our anti-symmetric estimator can be used to test modifications to the cosmological Euler equation. We conclude in \S \ref{sec:conclusion} and present in two appendices all the astrophysical details assumed \S\ref{app:survdets} and the changes we needed to implement in CAMB\_sources \cite{Challinor:2011bk} \S\ref{app:eulermods}.

\section{Observational clustering anti-symmetries} \label{sec:ObsClustAnti}

\subsection{Observed angular power spectra}

Let us assume that the galaxy type A has an average proper number density of sources $\bar n^A$. Let $\delta^A_{{\rm O}}(z,\hat n)=[n^A(z,\hat n)-\bar n^A(z)]/\bar n^A(z)$ be the observed number density contrast of a galaxy type, at a comoving distance $\chi(z)$ and direction $\hat n$. We decompose the density contrast on the sky using spherical harmonics as
\be
\delta^A_{{\rm O}}(z,\hat n)= \sum^{\infty}_{\ell = 0} \sum^{\ell}_{m = -\ell} \alm^A(z) \ylm{n}\,.
\ee
The only dependence on the distance is encoded in the $\alm$, which are the decomposition of the number density contrast into a spherical harmonic basis, i.e., 
\be \label{eq:alm_delta}
\alm^A\z=\int \don\ \delta^A_{{\rm O}}(z,\hat n) \ylmc{n}\,.
\ee
The $\alm$ will keep the same statistical properties of $\delta$, i.e., 
\bea
\langle \alm \rangle&=& 0\,,\\
\langle \alm a^*_{\ell'm'} \rangle&=&C_\ell\ \delta_{\ell\ell'}\delta_{mm'}\,.
\eea
The $C_\ell$ is the angular power spectrum and characterises the angular statistical distribution of the $\alm$ and is the main carrier of the cosmological information. The full expression for the linear density perturbation in real space has been computed in \citep{Challinor:2011bk,Bonvin:2011bg} where the relevant terms are
\bea \label{eq:delta_real}
\delta^A_{{\rm O}}(z,\hat n)&=&b^A\delta_{{\rm N}}(z,\hat n)-\frac1{{\cal H}(z)}\frac{\partial}{\partial\chi}(\vec{v}(z){\cdot}\hat{n})-A^{{A}}_{{\rm D}}(z)\vec v(z){\cdot}\hat n\nn\\
&&-{A^{A}_{{\rm L}}(z)}\int^{\chi(z)}_0\di\tilde\chi\frac{\chi(z)-\tilde \chi}{2\chi(z)\tilde \chi}\Delta_{\hat n}\l\Phi(\tilde \chi,\hat n)+\Psi(\tilde \chi,\hat n)\r+\mathrm{\ldots} 
\eea
Here $\delta_{{\rm N}}$ is the density contrast of dark matter, $v$ is the velocity field and $\Phi,\Psi$ are the metric potentials of the perturbed Friedman-Robertson-Walker metric
\be \label{eq:frwpertmetric}
ds^2=a^2\left[-(1+2\Psi)d^2\eta+(1-2\Phi)\delta_{ij}dx^idx^j\right]\,.
\ee
The first term {of \autoref{eq:delta_real} is} the conventional density contrast term in Newtonian gauge, the second the redshift space distortions (or the first Kaiser term for the \emph{aficionados}), the following is the Doppler term 
and the last is the lensing contribution. 
{For a formal mathematical derivation of \autoref{eq:delta_real} please refer to the review \cite{Bonvin:2014owa} and references therein. Still one can give an heuristic argument for the origin of such corrections. The observed density contrast depends on the observed redshift and observed volume, which may not correspond to the true ones. In fact, Lensing and RSDs come from the perturbations between observed and true volumes. On the other hand, linear velocities affect both the inferred volume (which is equivalent to the second Kaiser term) and the perturbations between true and measured redshifts. It is conventional to bundle all corrections linearly dependent on the peculiar velocity and call it the Doppler term. We follow such convention here. For brevity on the notation, we} also defined
\bea
A^{{A}}_{{\rm D}}(z)&=&\left[\frac{2-5s^A(z)}{{\cal H}(z) \chi(z)}+5s^A(z)-b^A_{{\rm e}}(z)+\frac{ {\cal H}'(z)}{{\cal H}^2(z)}\right]\,,\\
{A^{A}_{{\rm L}}(z)}&{=}&{2-5s^A(z)\,,}
\eea
{as the amplitudes} 
of the Doppler effect {and magnification lensing}. Note that the derivatives are taken with respect to conformal time $\eta$. The other GR terms are safely neglected and are not detectable with future surveys, or combinations of surveys \cite{Viljoen:2021ypp}. The magnification bias $s^A$ is defined as
\be
s^A(z)\equiv\frac{\partial \log_{10}\bar N^A(z,m<m_*)}{\partial m_*}\,,
\ee
where the observed angular number density of sources is
\be
\bar N^A(z,m<m_*)=\bar n^A(z)\ \frac{ \chi^2\z }{H\z}\,,
\ee
and $m_*$ is the flux threshold of the survey. The evolution bias $b^A_{{\rm e}}$ is defined as
\be
b^A_{{\rm e}}\equiv\frac{\partial \ln \bar n^A}{\partial \ln a}\,,
\ee
which measures how the comoving number of sources changes with redshift.

Introducing another tracer of dark matter as galaxy type $B$, the angular power spectrum can be written in terms of the primordial power spectra of the curvature perturbation ${\cal P} (k)$ and transfer functions $\Delta^{\cal W}_\ell$ as \cite{Challinor:2011bk}
\be \label{eq:clgeneral}
C^{AB}_\ell \l z_i,z_j \r=4\pi\!\!\int\!\!\di \ln k\, \Delta^{\cal W_A}_\ell\l z_i,k\r \Delta^{\cal W_B}_\ell\l z_j,k\r\, \mathcal P (k),
\ee
with the primordial power spectrum given by $\mathcal P (k)=A_s\l k/k_0\r^{n_s-1}$. Here we showed the expression of the angular power spectrum in a general form to allow for different redshift bins and different tracers. The transfer function $\Delta^{\cal W_{{A}}}_\ell$ takes into account the fact that any survey will have a redshift distribution of sources {A} $p^{{A}}(z)$ and a window function $W^{{A}}(z,z_i)$. Then 
\be \label{eq:obs_tranf}
\Delta_\ell^{\cal W_{{A}}}( z_i,k)=\int\!\! \di z\,p^{{A}}(z)\ W^{{A}}(z_i,z)\ \Delta^{{A}}_\ell(z,k)\,.
\ee
Note that in most cases, such as galaxy surveys, $\int\!\di z\,p^{{A}}(z)W^{{A}}(z_i,z)=1$ for all $z_i$ to ensure that the observed transfer function is the weighted average of the \emph{theoretical} transfer function. In this paper, we will only consider spectroscopic surveys with narrow bins. In effect, we will take top-hat window functions smoothed at the edges for numerical reasons. 
 The \emph{theoretical} transfer function is effectively constructed from \autoref{eq:delta_real} and \autoref{eq:clgeneral}, and has contributions from the different terms above as
\be \label{eq:deltatot}
\Delta^{{A}}_\ell(k)={b^{A}}~ \Delta^{\delta}_\ell(k)+ \Delta^{{\rm R}}_\ell(k)+ {A^A_{\mathrm{D}}}~\Delta^{{\rm D}}_\ell(k)+ {A^A_{\mathrm{L}}}~\Delta^{{\rm L}}_\ell(k) + \ldots \,,
\ee
{where we omitted the implicit redshift dependence. We have also separated the tracer dependent amplitudes from the tracer independent transfer functions} which are given by
\bea
\Delta^{\delta}_\ell(k)&=& \,\delta^{\rm c}_k\,j_\ell\l k\chi\r\,,\\
\Delta^{{\rm R}}_\ell(k)&=& \frac {k v_k}{\H}\,\frac{\partial^2 j_\ell(k\chi)}{\partial k\chi^2}\,, \\
\Delta^{{\rm D}}_\ell(k)&=& v_k \frac{\partial j_\ell(k\chi)}{\partial k\chi}\,,\\
\Delta^{{\rm L}}_\ell(k)&=& \frac{\ell\l\ell+1\r}2\int_0^{\chi} \di \tilde\chi\frac{(\chi-\tilde\chi)}{\chi\tilde\chi} {\big[\Phi_k(\tilde\chi)+\Psi_k(\tilde\chi)\big]} j_\ell\l k\tilde\chi\r \,,
\eea
where we called ``$\delta$'' for density only, ``R'' for RSD, ``L'' for the lensing contribution from $\kappa$, and ``D'' for Doppler.

\subsection{The anti-symmetric power spectra}

Several authors have taken advantage of line-of-sight asymmetries in two galaxy samples to extract the Doppler term. Most notably \cite{Bonvin:2013ogt} uses the dipolar structure induced in the {two-point} function. Here we will take advantage of the break of symmetry between redshift bins when we consider different tracers of dark matter but using the angular power spectra. While for a single tracer $C^A_\ell(z_i,z_j)=C^A_\ell(z_j,z_i)$, this is no longer true for cross-tracer power spectrum (see \autoref{eq:delta_real}). Let us define 
\be \label{eq:anitsym_est}
\zeta^{AB}_{{\ell,}ij}\equiv C^{AB}_\ell(z_i,z_j)-C^{AB}_\ell(z_j,z_i)=C^{AB}_\ell(z_i,z_j)-C^{BA}_\ell(z_i,z_j)\,.
\ee
It is clearly anti-symmetric as it is zero for $A=B$ or $i=j$, and gains a minus sign under {$A\leftrightarrow B$ or $i\leftrightarrow j$}. This estimator has already been studied to extract the magnification lensing \cite{Jalilvand:2019bhk}. Here we are interested in extracting the Doppler contribution only. To better understand how this estimator can be used to extract the Doppler term let us assume that we have infinitesimal redshift bins (i.e., the redshift integral in \autoref{eq:obs_tranf} can be considered as a delta function integral). This way, all the redshift tracer-dependent ``coefficients'' can be put in evidence. For our heuristic argument, we will abuse the notation and expand the angular power spectrum in \autoref{eq:anitsym_est} as 
\bea
C^{AB}_{{\ell,}ij}&\simeq& b^A(z_i)b^B(z_j)C^{\delta\delta}_\ell(z_i,z_j)+ b^A(z_i) C^{\delta {\mathrm{R}}}_\ell(z_i,z_j)+ b^B(z_j) C^{{\mathrm{R}}\delta }_\ell(z_i,z_j)\nn\\
&&+ C^{{\mathrm{R}} {\mathrm{R}} }_\ell(z_i,z_j) + b^A(z_i){A_{\mathrm{L}}^B(z_j)}C^{\delta {\mathrm{L}}}_\ell(z_i,z_j) +{A_{\mathrm{L}}^B(z_j)} C^{{\mathrm{R}} {\mathrm{L}}}_\ell(z_i,z_j)\nn\\
&&+{ A_{\mathrm{L}}^A(z_i) A_{\mathrm{L}}^B(z_j)}C^{{\mathrm{L}} {\mathrm{L}}}_\ell(z_i,z_j)+ { A_{\mathrm{L}}^A(z_i)}b^B(z_j)C^{{\mathrm{L}} \delta}_\ell(z_i,z_j)\nn\\
&&+{ A_{\mathrm{L}}^A(z_i)}C^{{\mathrm{L}} {\mathrm{R}}}_\ell(z_i,z_j)+A_{{\mathrm{D}}}^A(z_i)b^B(z_j)C^{{\mathrm{D}} \delta}_\ell(z_i,z_j)+A_{{\mathrm{D}}}^A(z_i)C^{{\mathrm{D}} {\mathrm{R}}}_\ell(z_i,z_j)\nn\\
&&+A_{{\mathrm{D}}}^A(z_i){ A_{\mathrm{L}}^B(z_j)}C^{{\mathrm{D}} {\mathrm{L}}}_\ell(z_i,z_j)+A_{{\mathrm{D}}}^A(z_i)A_{{\mathrm{D}}}^B(z_j)C^{{\mathrm{D}} {\mathrm{D}}}_\ell(z_i,z_j)\nn\\
&&+b^A(z_i)A_{{\mathrm{D}}}^B(z_j)C^{\delta {\mathrm{D}}}_\ell(z_i,z_j)+A_{{\mathrm{D}}}^B(z_j)C^{{\mathrm{R}} {\mathrm{D}}}_\ell(z_i,z_j)\nn\\
&&+{ A_{\mathrm{L}}^A(z_i)}A_{{\mathrm{D}}}^B(z_j)C^{{\mathrm{L}} {\mathrm{D}}}_\ell(z_i,z_j)\,.
\eea
Note that here we have called ``D'' to $v_k \partial j_\ell(k\chi)/\partial k\chi$ only and ``L'' to $\kappa_\ell$ only. Despite this slight abuse of terminology it is better to keep a less crowded number of superscripts. We can then write the anti-symmetric estimator as
\bea
\zeta^{AB}_{{\ell,}ij}&\simeq& {\lr b^A(z_i)b^B(z_j)-b^A(z_j)b^B(z_i)\rr} C^{\delta\delta}_\ell(z_i,z_j)\nn\\
&&+{\lr b^A(z_i) -b^B(z_i)\rr} C^{\delta {\mathrm{R}}}_\ell(z_i,z_j)+ {\lr b^B(z_j)-b^A(z_j)\rr} C^{{\mathrm{R}}\delta }_\ell(z_i,z_j)\nn\\
&&+ {\lr b^A(z_i) A_{\mathrm{L}}^B(z_j)-b^B(z_i) A_{\mathrm{L}}^A(z_j)\rr } C^{\delta {\mathrm{L}}}_\ell(z_i,z_j) \nn\\
&&+{\lr  A_{\mathrm{L}}^A(z_i) A_{\mathrm{L}}^B(z_j) -  A_{\mathrm{L}}^B(z_i) A_{\mathrm{L}}^A(z_j) \rr } C^{{\mathrm{L}} {\mathrm{L}}}_\ell(z_i,z_j)\nn\\
&&+ {\lr b^B(z_j) A_{\mathrm{L}}^A(z_i)-b^A(z_j) A_{\mathrm{L}}^B(z_i)\rr } C^{{\mathrm{L}} \delta }_\ell(z_i,z_j) \nn\\
&&+5 {\lr s^B(z_j) - s^A(z_j)\rr} C^{{\mathrm{R}} {\mathrm{L}}}_\ell(z_i,z_j)+5 {\lr s^A(z_i) - s^B(z_i)\rr} C^{{\mathrm{L}} {\mathrm{R}}}_\ell(z_i,z_j)\nn\\
&&+{\lr A_{{\mathrm{D}}}^A(z_i)b^B(z_j)-A_{{\mathrm{D}}}^B(z_i)b^A(z_j)\rr} C^{{\mathrm{D}} \delta}_\ell(z_i,z_j)\nn\\
&&+{\lr b^A(z_i)A_{{\mathrm{D}}}^B(z_j)-b^B(z_i)A_{{\mathrm{D}}}^A(z_j)\rr} C^{\delta {\mathrm{D}}}_\ell(z_i,z_j)\nn\\
&&+{\lr A_{{\mathrm{D}}}^A(z_i)- A_{{\mathrm{D}}}^B(z_i)\rr} C^{{\mathrm{D}} {\mathrm{R}}}_\ell(z_i,z_j)+ {\lr A_{{\mathrm{D}}}^B(z_j)- A_{{\mathrm{D}}}^A(z_j)\rr} C^{{\mathrm{R}}{\mathrm{D}}}_\ell(z_i,z_j) \nn\\
&&+{\lr A_{{\mathrm{D}}}^A(z_i) A_{\mathrm{L}}^B(z_j)-A_{{\mathrm{D}}}^B(z_i) A_{\mathrm{L}}^A(z_j) \rr } C^{{\mathrm{D}} {\mathrm{L}}}_\ell(z_i,z_j)\nn\\
&&+{\lr  A_{\mathrm{L}}^A(z_i) A_{{\mathrm{D}}}^B(z_j) -  A_{\mathrm{L}}^B(z_i) A_{{\mathrm{D}}}^A(z_j) \rr } C^{{\mathrm{L}} {\mathrm{D}}}_\ell(z_i,z_j)\nn\\
&&+ {\lr A_{{\mathrm{D}}}^A(z_i)A_{{\mathrm{D}}}^B(z_j)-A_{{\mathrm{D}}}^A(z_j)A_{{\mathrm{D}}}^B(z_i)\rr} C^{{\mathrm{D}} {\mathrm{D}}}_\ell(z_i,z_j)\,.
\eea
So far we cannot see any advantage is such estimator as only the tracer independent parts have canceled (i.e. the Redshift Space Distortions only). Let us now assume that the biases evolve little in redshift, i.e., if we consider that $z_i$ and $z_j$ are contiguous bins then $b(z_i)\simeq b(z_j)$ (and similarly to the other biases). In addition, in this thin bin next to each other limit all lensing contributions should be negligible. Then {the anti-symmetric} $\zeta$ estimator becomes
\bea \label{eq:zeta_thinnext}
\zeta^{AB}_{{\ell,}ij}{\l z_i\approx z_j\r}&\simeq& \l b^A -b^B\r {\lr C^{\delta {\mathrm{R}}}_\ell(z_i,z_j) - C^{{\mathrm{R}}\delta }_\ell(z_i,z_j)\rr}\nn\\
&&+\l A_{{\mathrm{D}}}^Ab^B-A_{{\mathrm{D}}}^Bb^A\r {\lr C^{{\mathrm{D}} \delta}_\ell(z_i,z_j)- C^{\delta {\mathrm{D}}}_\ell(z_i,z_j)\rr}\nn\\
&&+ \l A_{{\mathrm{D}}}^A- A_{{\mathrm{D}}}^B\r {\lr C^{{\mathrm{D}} {\mathrm{R}}}_\ell(z_i,z_j) - C^{{\mathrm{R}}{\mathrm{D}}}_\ell(z_i,z_j) \rr} \,.
\eea
We can see that the main contributions come from velocity terms, or their correlations with density. Therefore, while in the angular power spectrum the Doppler term is negligible for most purposes, in the anti-symmetric estimator it is a main contributor to the signal. This is only valid for contiguous narrow bins in spectroscopic surveys (or potentially very near bins depending on the surveys). The size of the signal depends on the particular choices of tracers and their clustering, magnification, and evolution bias. For instance, given that the clustering bias is always positive, one can choose (or construct) samples for which their Doppler amplitude has opposite signs. In \autoref{eq:zeta_thinnext} inside the {right-hand} side brackets of each line, we see the ``cosmology'' anti-symmetric part. It is non-zero as expected and is cancelled for single tracers. These cosmological parts depend on the particular redshift chosen and on the window function and, in principle, one can choose selection functions that enhance the signal. 

{It is important to compare what we found in \autoref{eq:zeta_thinnext} with results commonly found in the literature. The most well established result is the Doppler Dipole (see \cite{Bonvin:2014owa} for a review). The mathematical equivalence between the two estimators is not straightforward. On one hand, the dipolar structure in the two-point function is local, i.e., the multipole expansion is done in terms of an angle $\beta$ (see Figure 3 of \cite{Bonvin:2014owa}), which in the plane-parallel approximation is the same as the angle between a given direction and the line-of-sight. On the other hand the summary statistics deal with the data differently, while $C_\ell$ projects the information onto the sphere, the two-point function $\xi$ computes correlations in a 3D space assuming that the observer is sufficiently distant. Irrespective of it, the trends remain similar. The second and third lines of \autoref{eq:zeta_thinnext} involve the same terms as the relativistic dipole of \cite{Bonvin:2013ogt}, i.e., and anti-correlation between the Doppler terms and the conventional density and RSD terms. The first line however does not appear in the relativistic dipole. It is indeed a contamination to the relativistic dipole when one takes into account higher order corrections to the distant-observer approximation (see subsection 3.2.3 of \cite{Bonvin:2014owa}). These come from wide-angle correlations which are naturally included in the angular power spectra.} 

\section{{Examples of the anti-symmetric angular power spectra}} \label{sec:zetaexemples}

To gain some intuition on how the signal of the anti-symmetric clustering estimator works let us go through a handful of examples of combinations of surveys.

\begin{figure}[]
\centering
\vspace*{-0.5cm}
\includegraphics[width=5.cm]{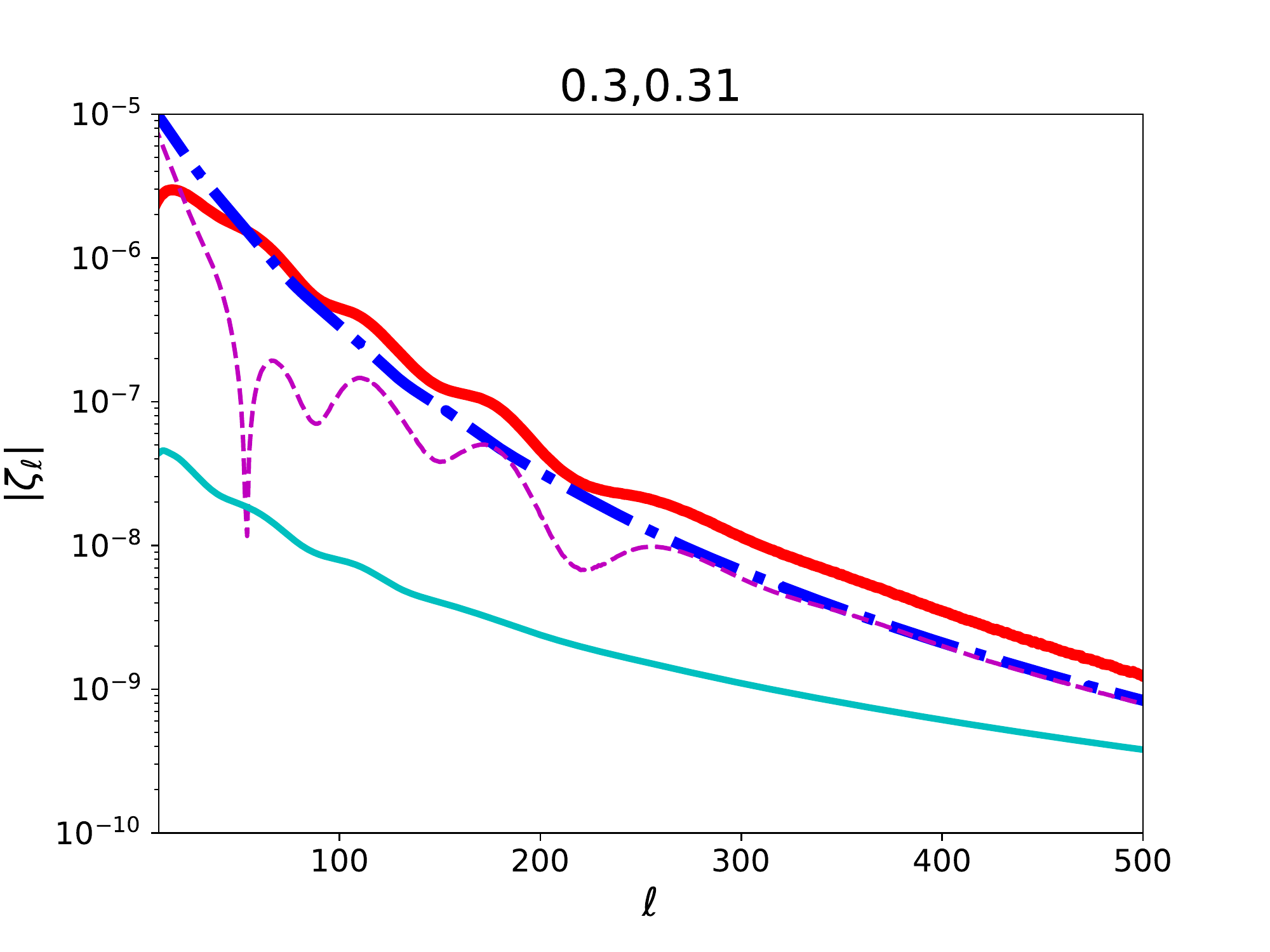}
\includegraphics[width=5.cm]{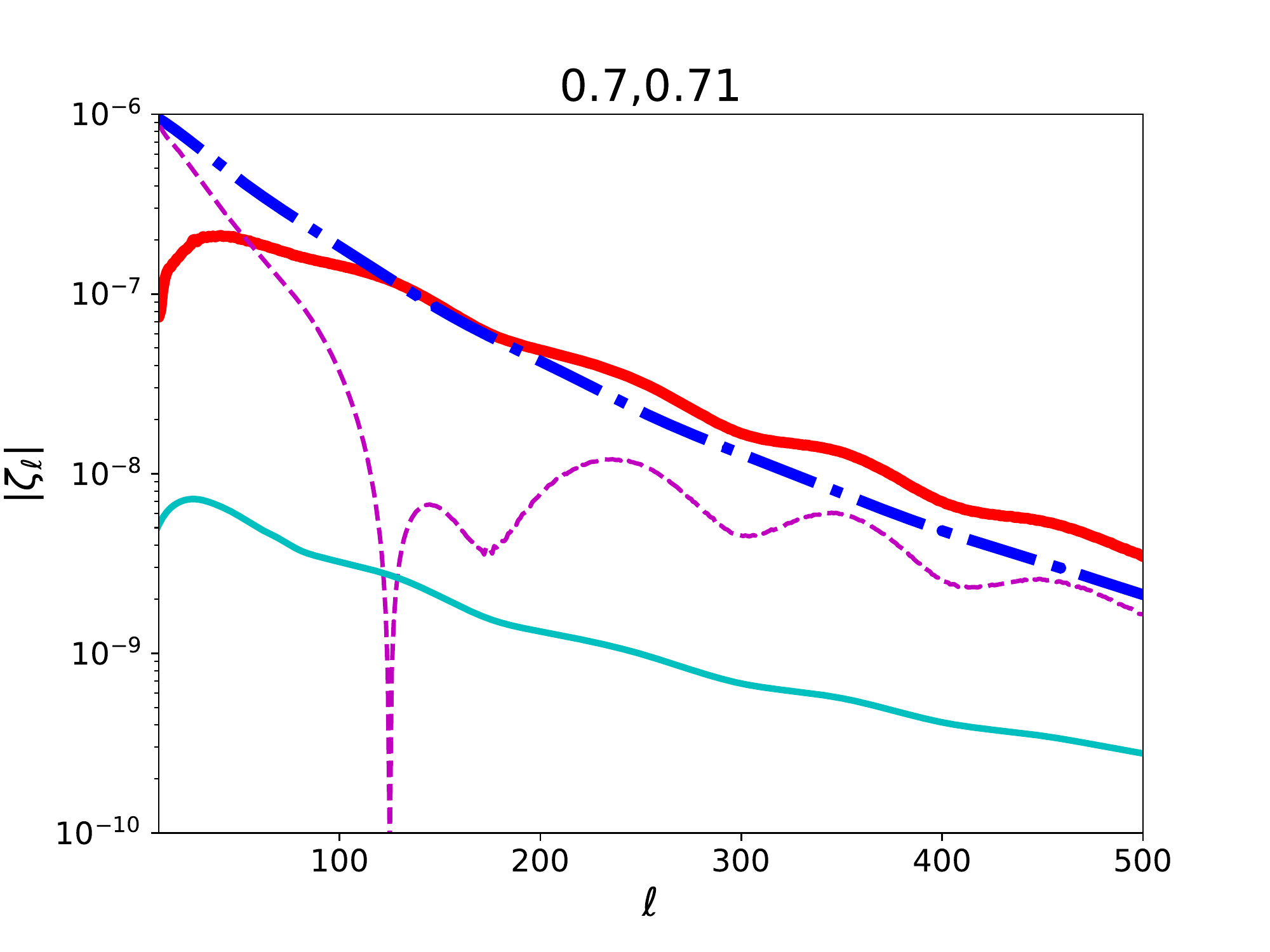}
\includegraphics[width=5.cm]{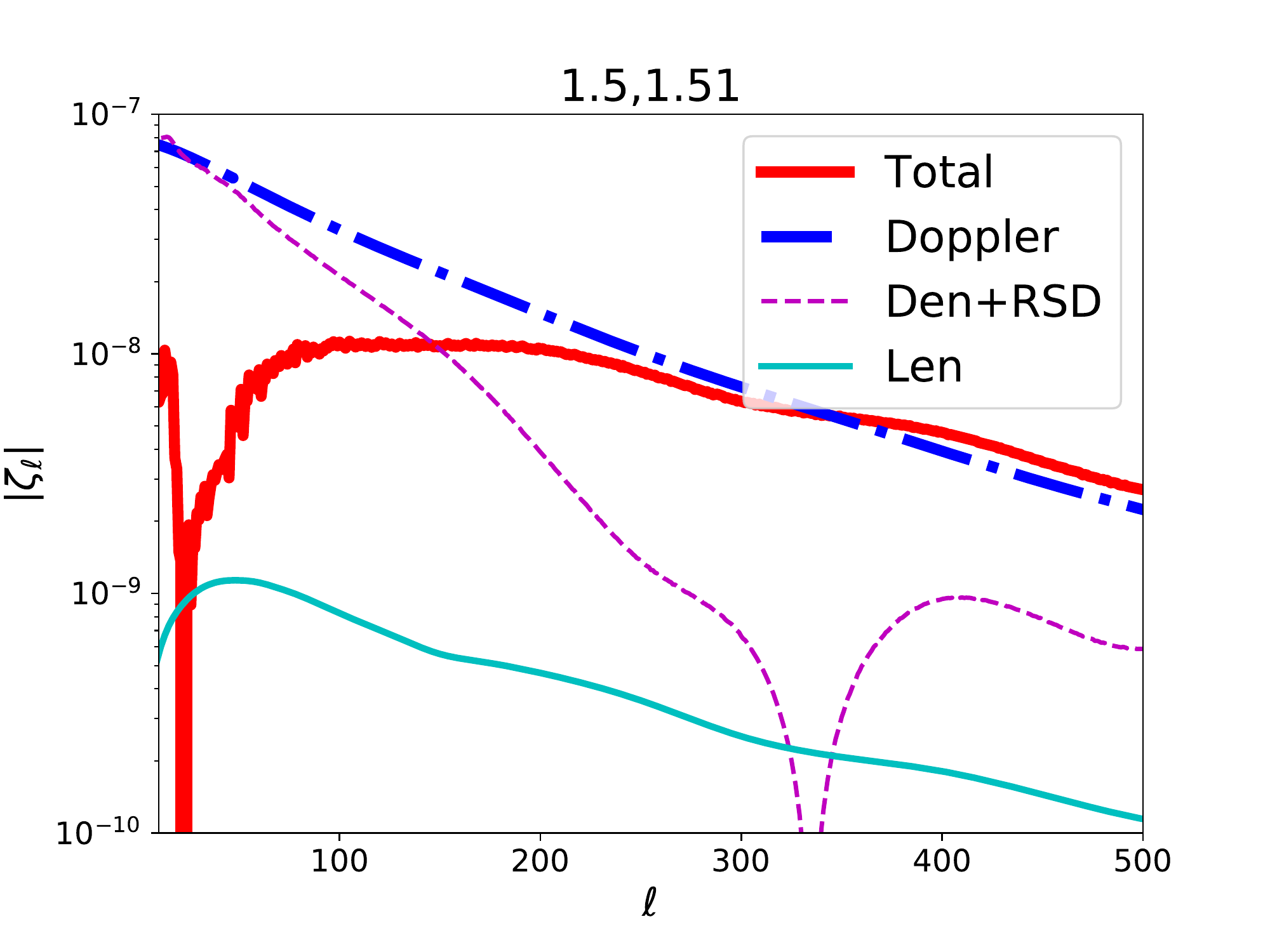}
\includegraphics[width=5.cm]{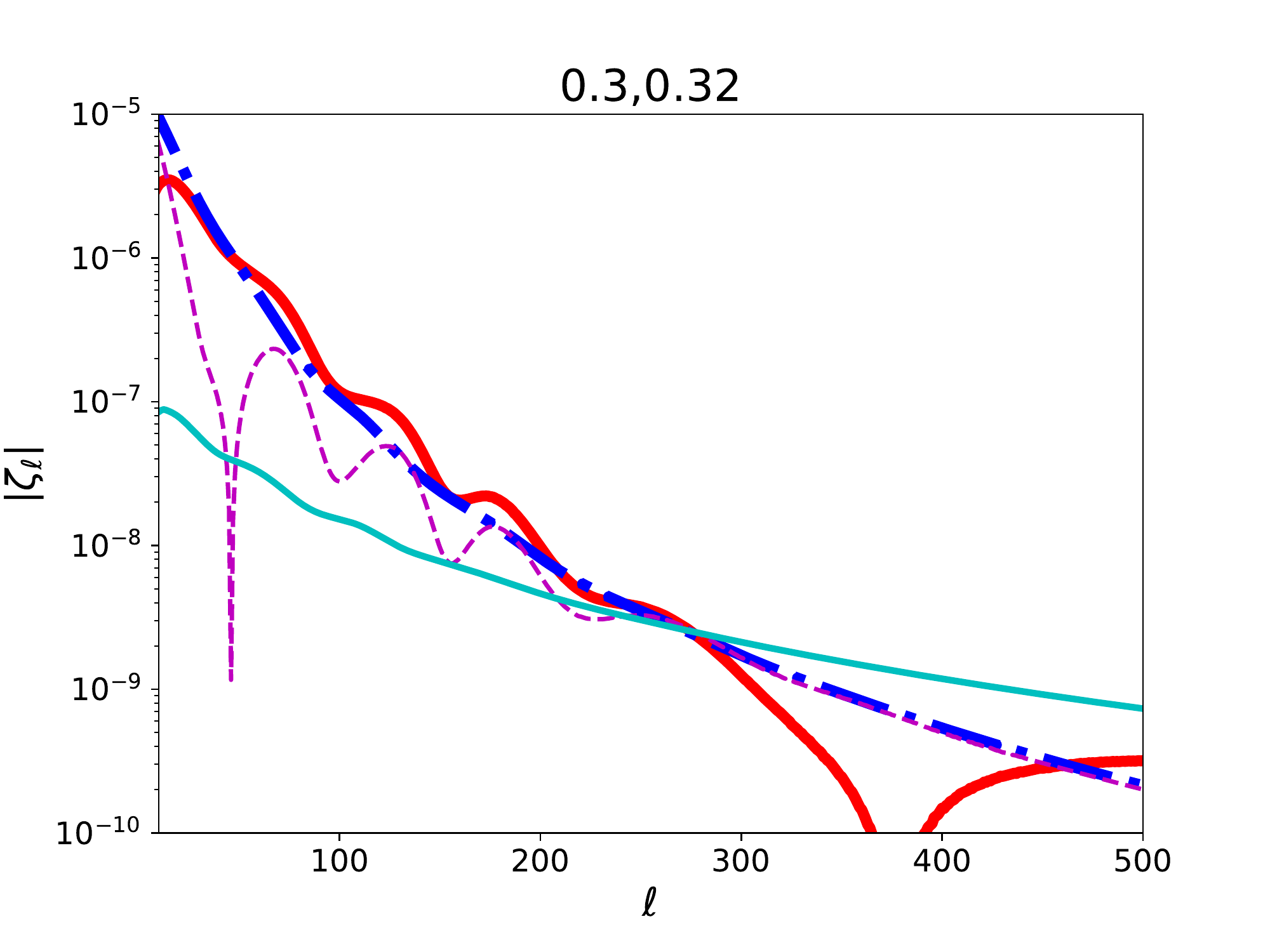}
\includegraphics[width=5.cm]{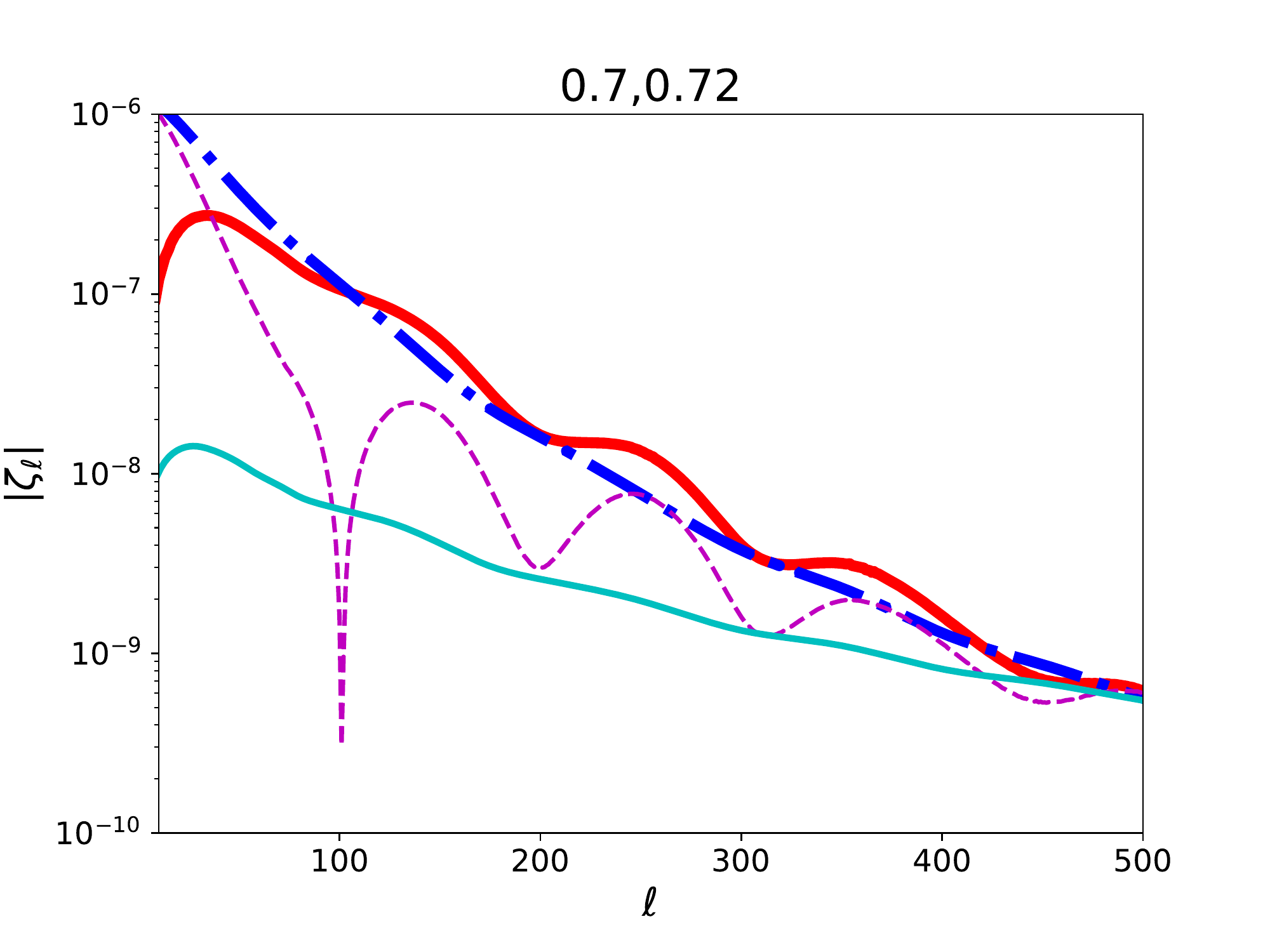}
\includegraphics[width=5.cm]{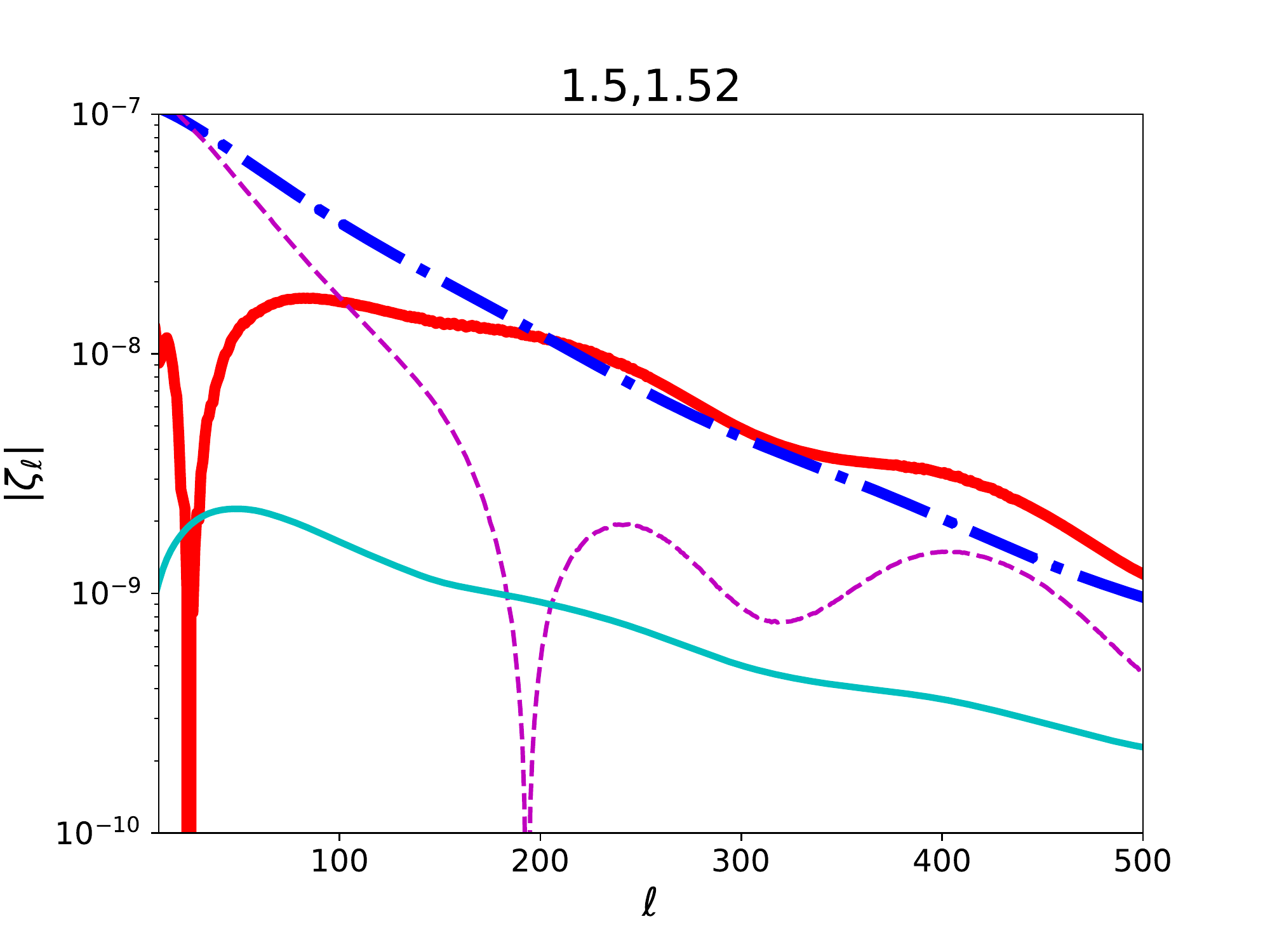}
\includegraphics[width=5.cm]{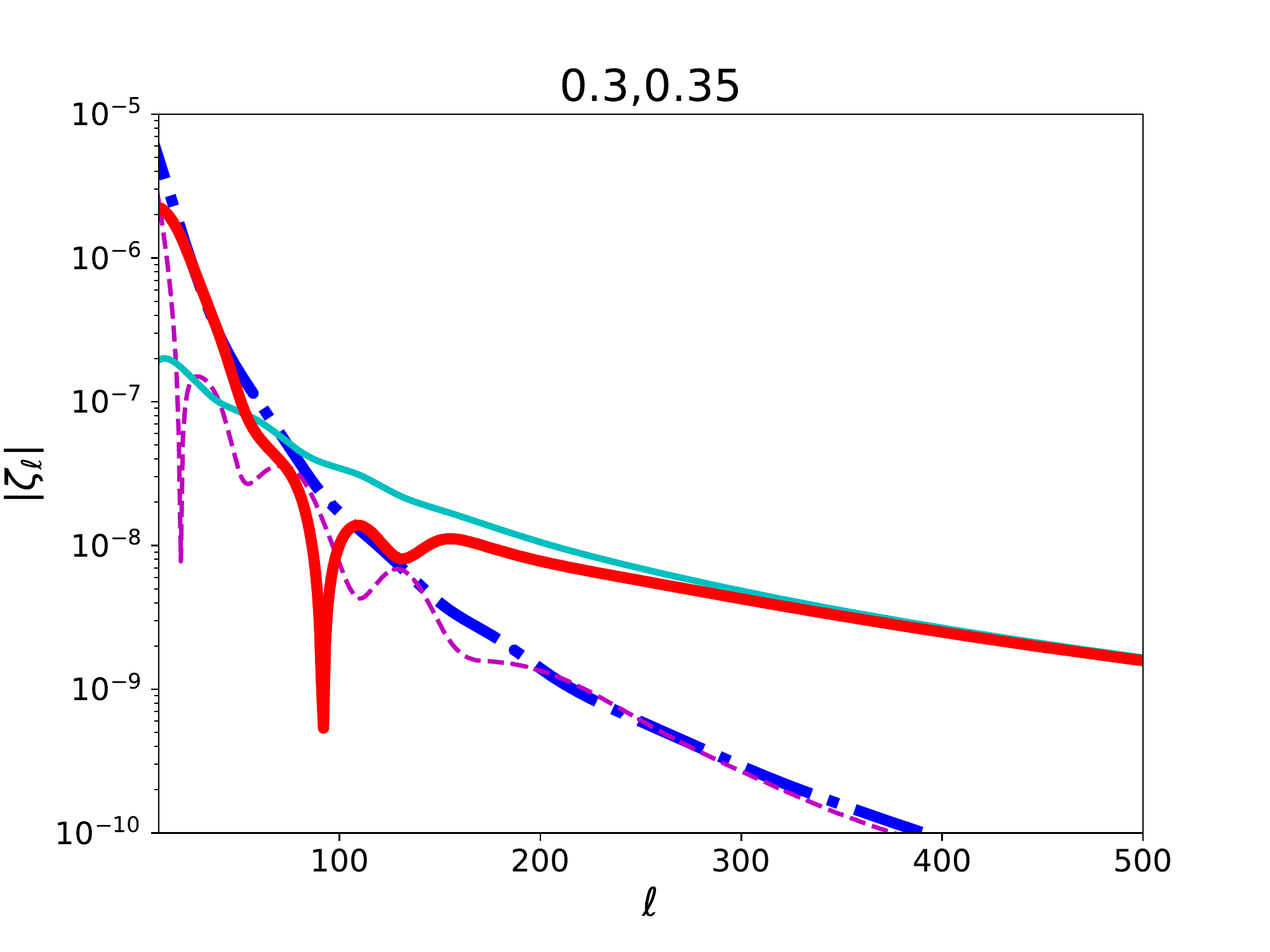}
\includegraphics[width=5.cm]{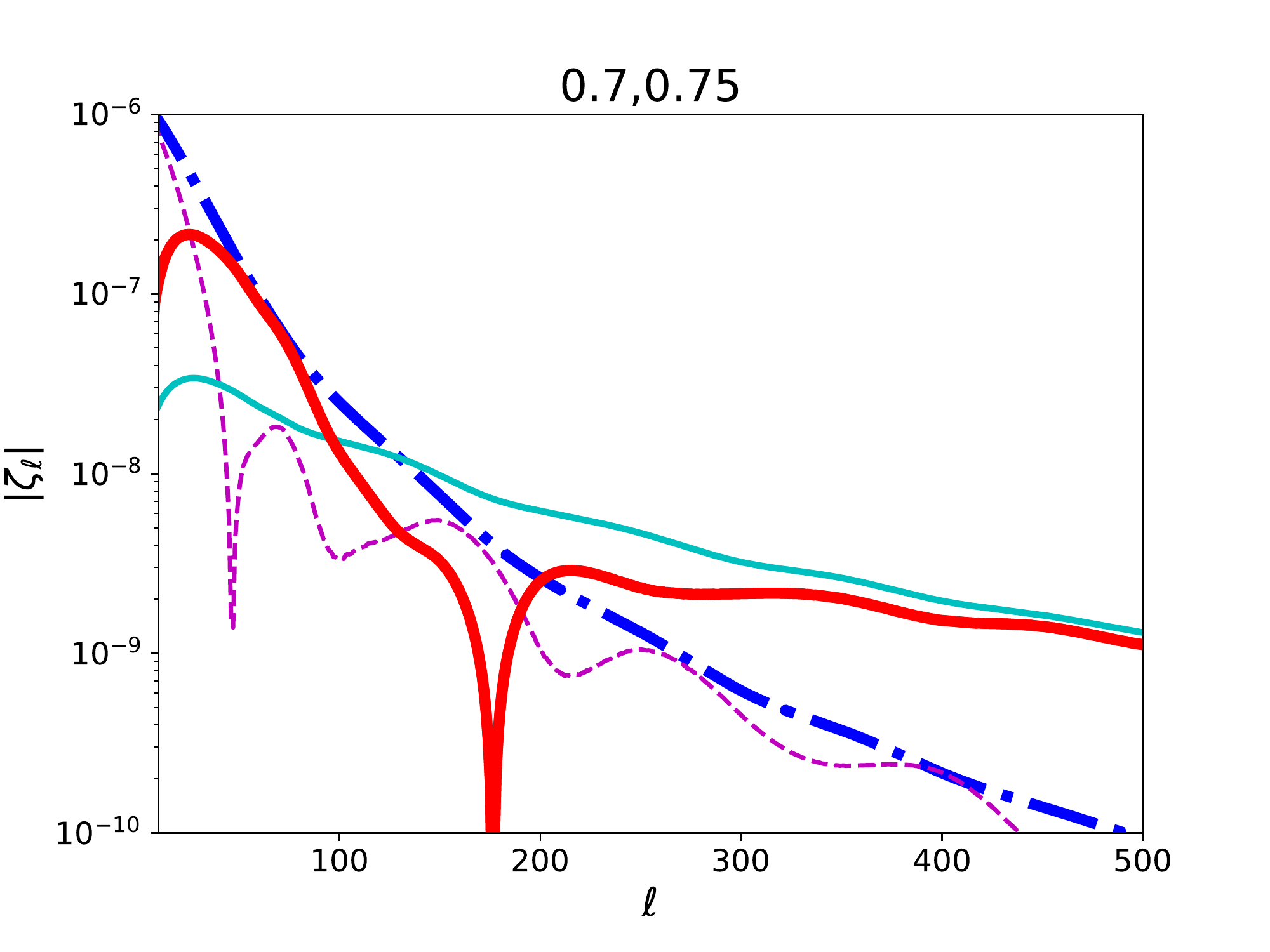}
\includegraphics[width=5.cm]{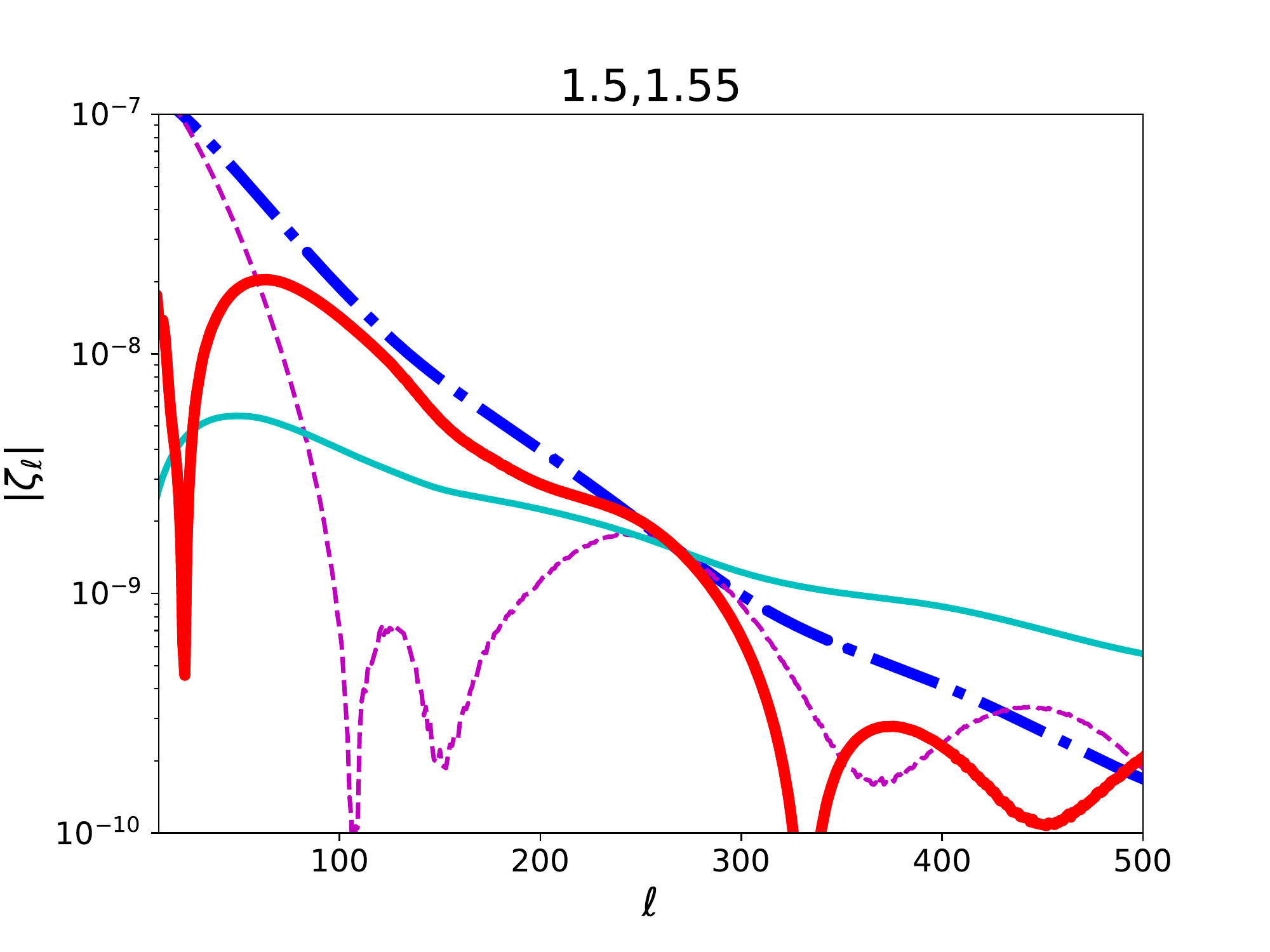}
\includegraphics[width=5.cm]{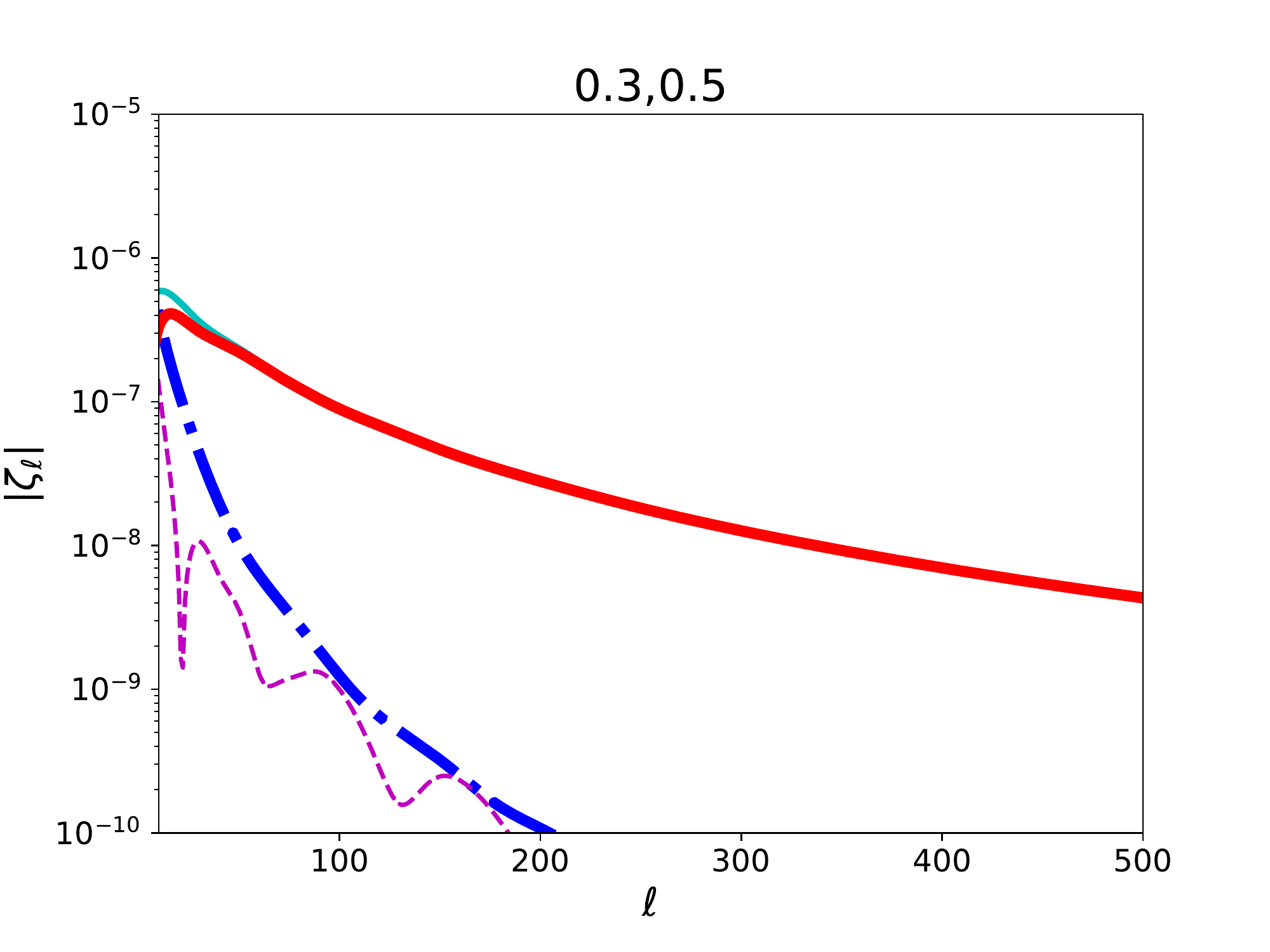}
\includegraphics[width=5.cm]{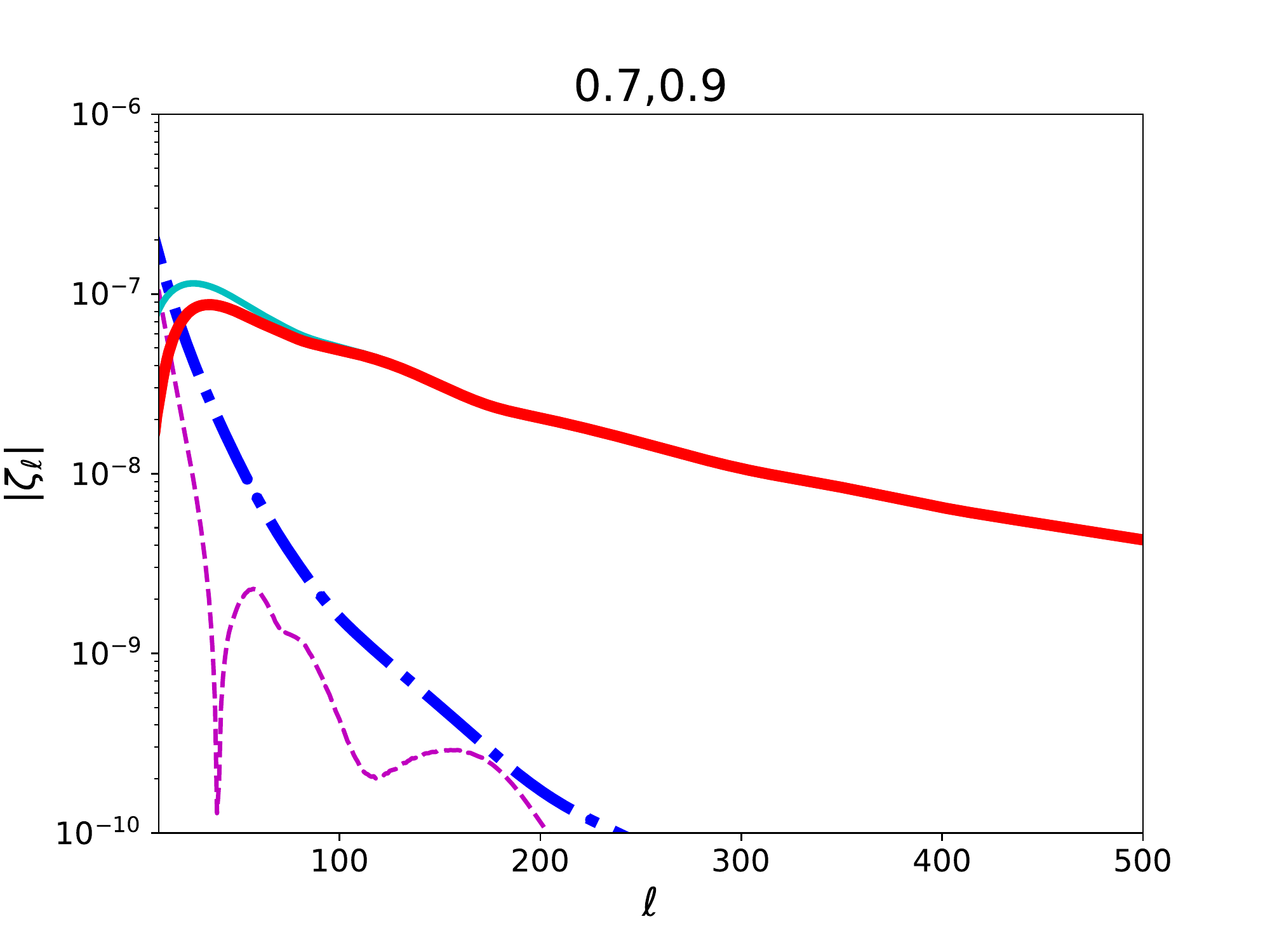}
\includegraphics[width=5.cm]{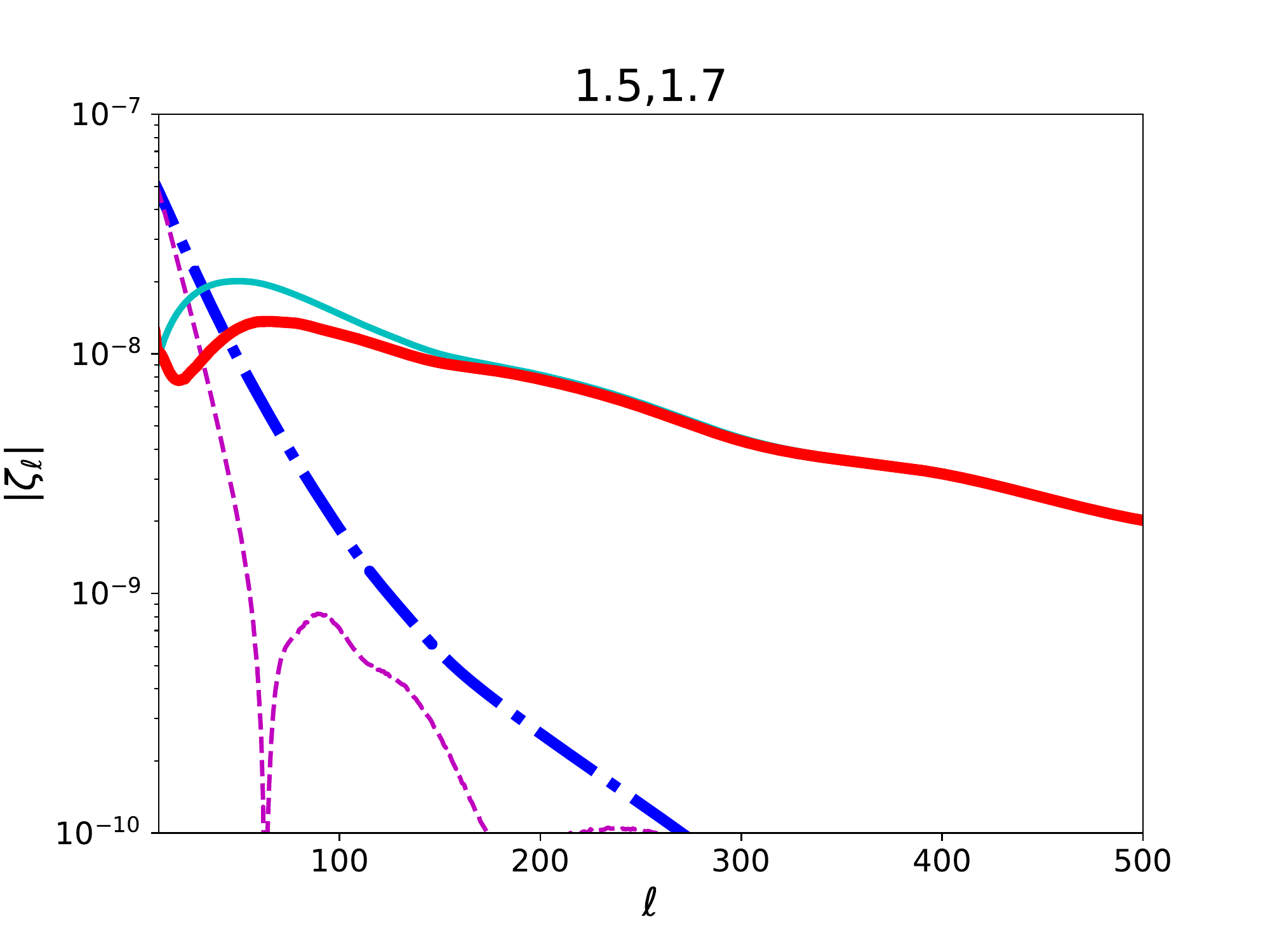}
\caption{Examples of $\zeta$ in solid thick red for tracers with $s=b_{{\mathrm{e}}}=0$ at different redshifts: \emph{Left column}: $z_i=0.3$; \emph{Middle column}: $z_i=0.7$; \emph{Right column}: $z_i=1.5$, for different redshift widths: \emph{Top row}: $\Delta=0.01$; \emph{2nd row}: $\Delta=0.02$; \emph{3rd row}: $\Delta=0.05$; \emph{Bottom row}: $\Delta=0.2$. We also plot individual contributions: density and RSD in thin dashed magenta, Doppler in thick dot-dashed blue and Lensing in solid cyan.}
\label{fig:sbe0}
\end{figure}

For demonstration purposes only, let us choose two tracers with $b_1=1.0\ , b_2=1.5$ (without any redshift evolution) that we fix throughout this subsection. For now let us take the case where $b_{{\mathrm{e}},1}=b_{{\mathrm{e}},2}=s_{1}=s_{2}=0$. In this case, $g(z)$ becomes independent of the tracer and the last line of \autoref{eq:zeta_thinnext} vanishes. Therefore there are only two terms both involving velocity, although only one coming from the Doppler correction. This behaviour is redshift dependent, as for higher redshifts a fixed redshift bin corresponds to a smaller physical size. For the demonstration, we can no longer neglect lensing because bins have thickness. In \autoref{fig:sbe0} we show some examples for different redshifts and bin {widths}. Although we consider a thought example we can see that no case isolates uniquely the Doppler. Despite this, and contrary to the angular power spectrum, the Doppler contribution is one of the leading effects. This is dependent on the thickness of the bin. In \autoref{fig:sbe0} the bin size increases from top to bottom and we can see a nice transition between the relevance of the Doppler term to the Lensing contribution being completely dominant in the large bin regime. This comes of no surprise and has already been the work presented in \cite{Jalilvand:2019bhk}. For a fixed bin width, the higher redshifts become more dominated by the Doppler term, although as expected when the bin width is larger the dominant term becomes lensing. All of this indicates that one can construct a $\zeta$ optimisation to better extract the Doppler term. This will be highly dependent on the galaxy samples and we will leave such study for later.  

\begin{figure}[]
\centering
\vspace*{-0.5cm}
\includegraphics[width=5.cm]{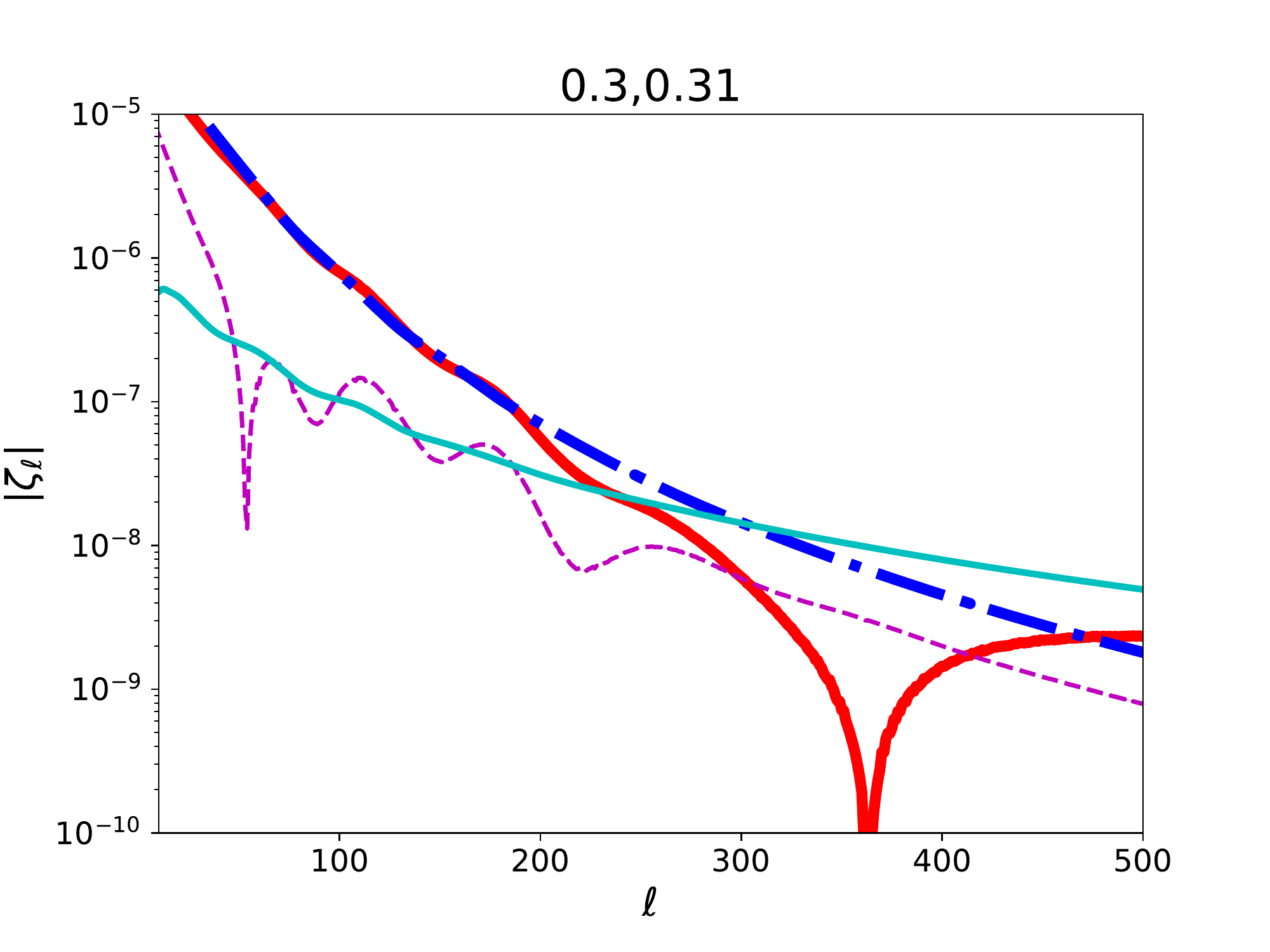}
\includegraphics[width=5.cm]{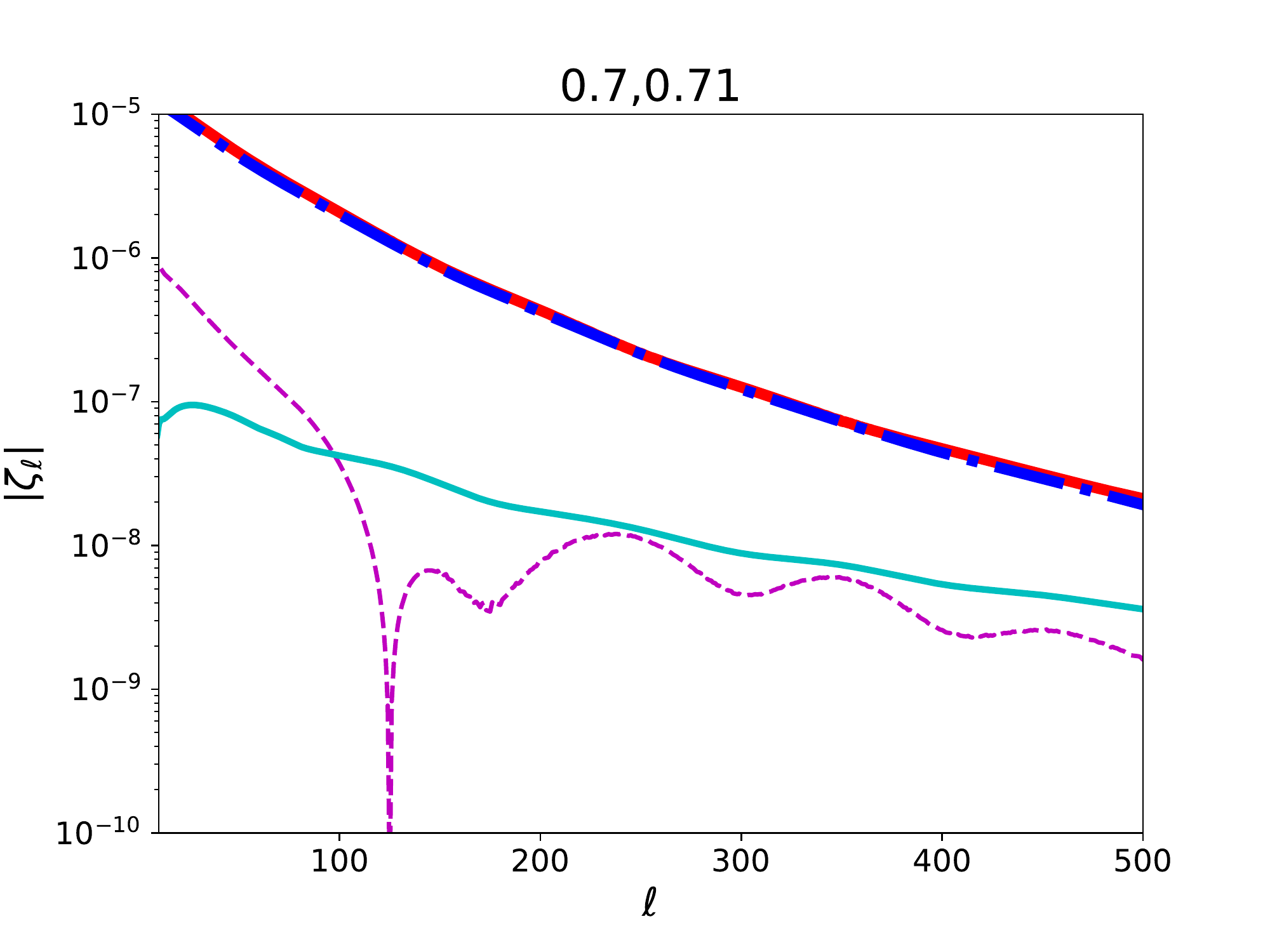}
\includegraphics[width=5.cm]{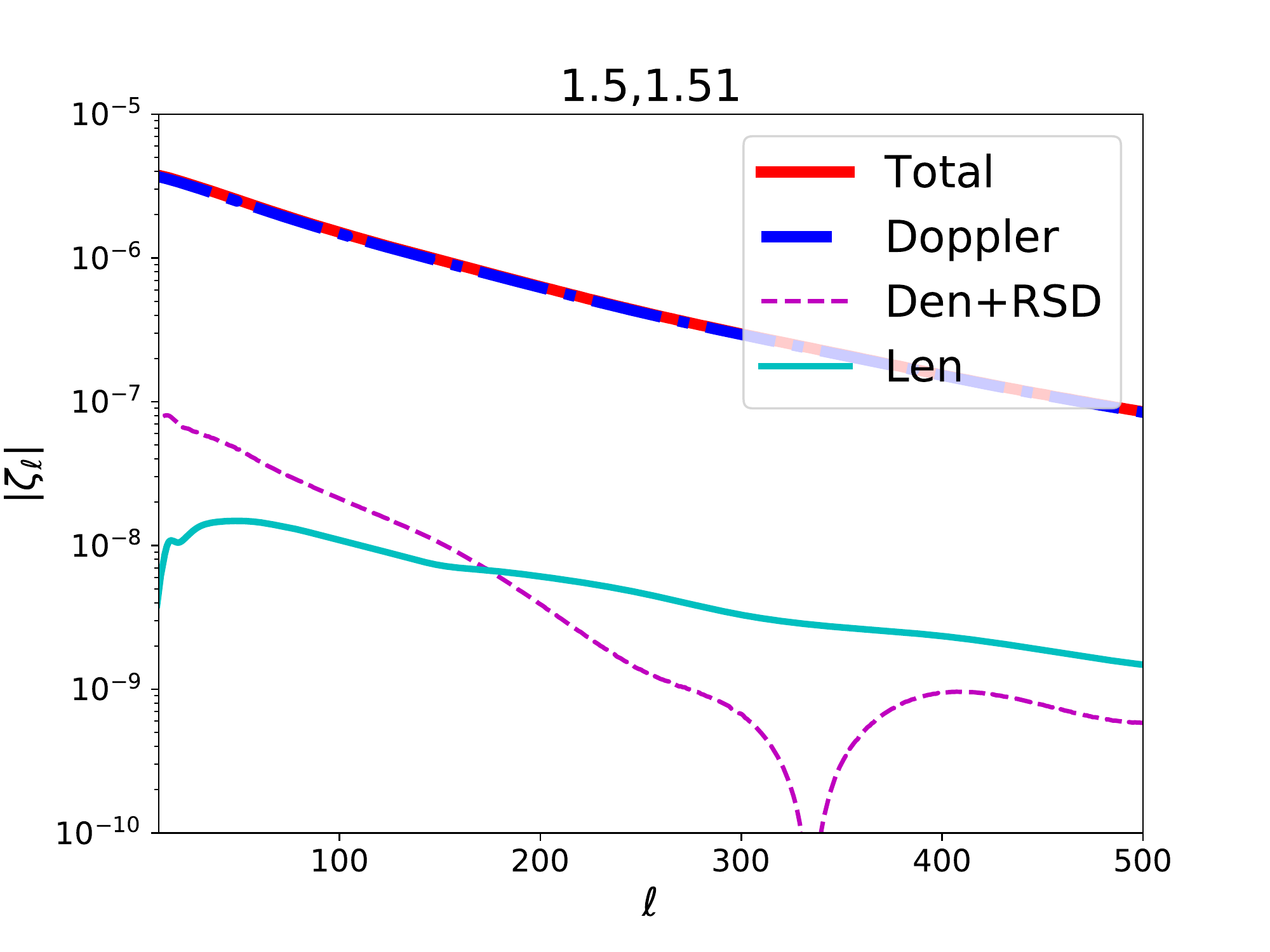}
\includegraphics[width=5.cm]{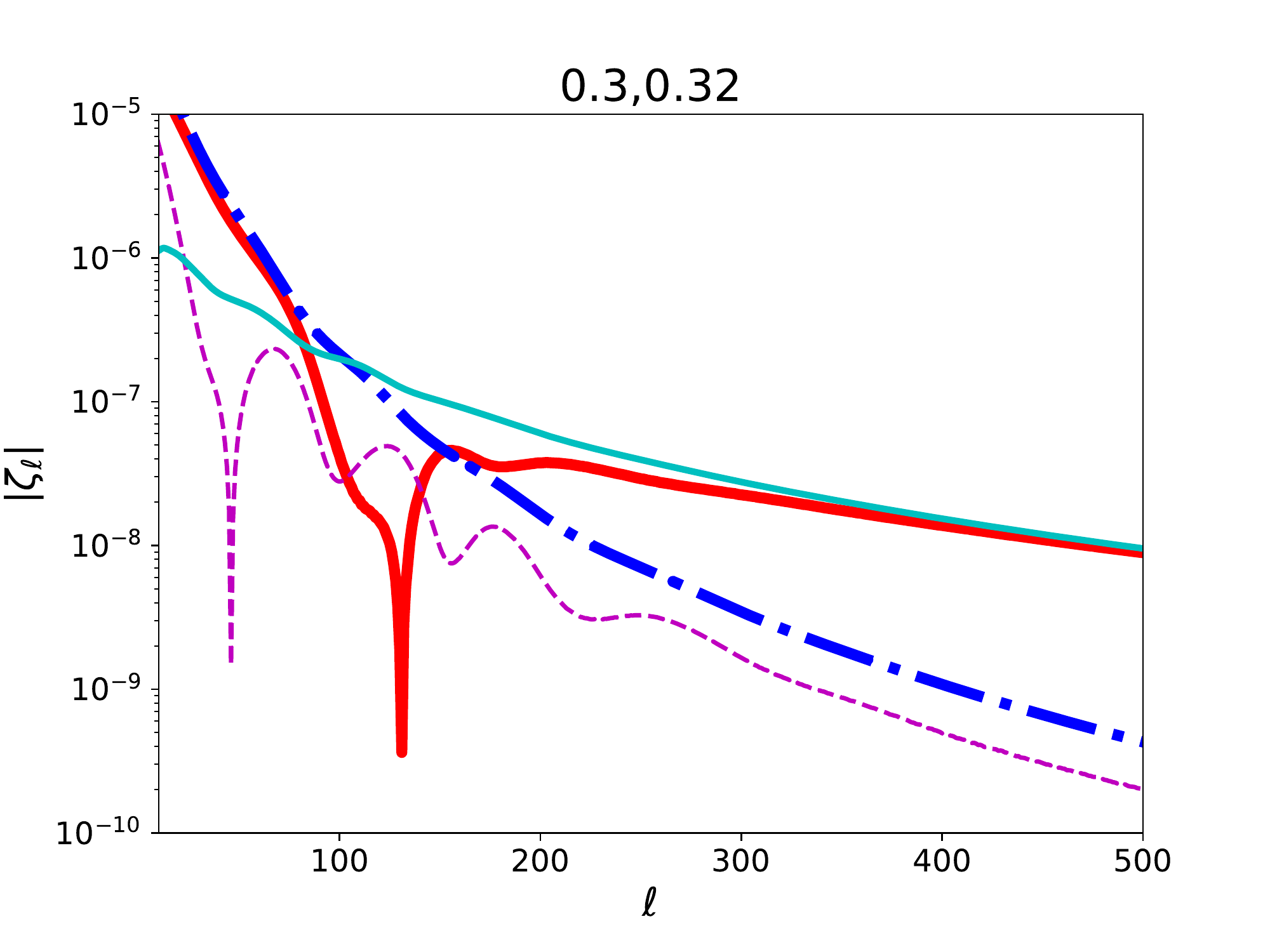}
\includegraphics[width=5.cm]{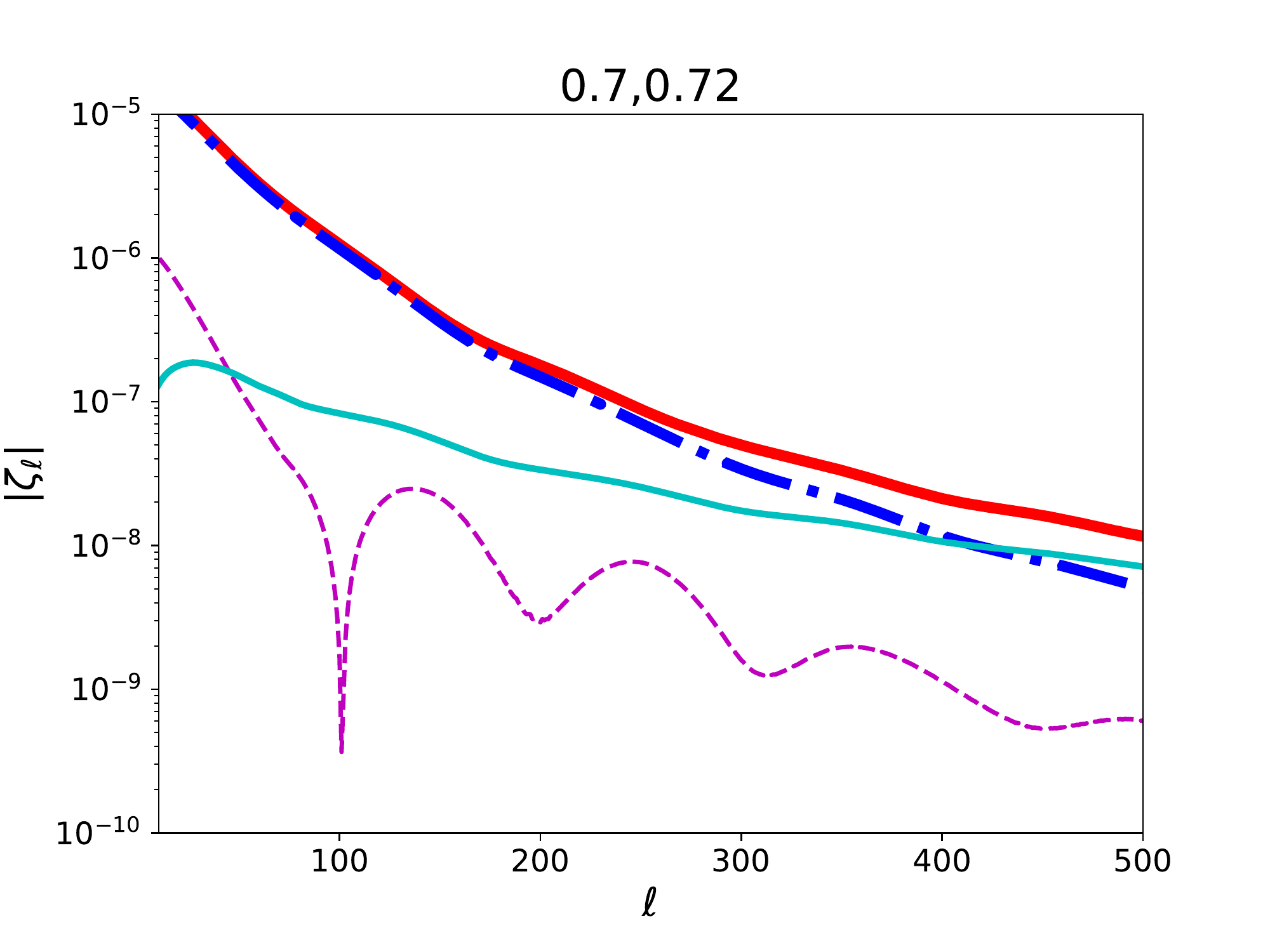}
\includegraphics[width=5.cm]{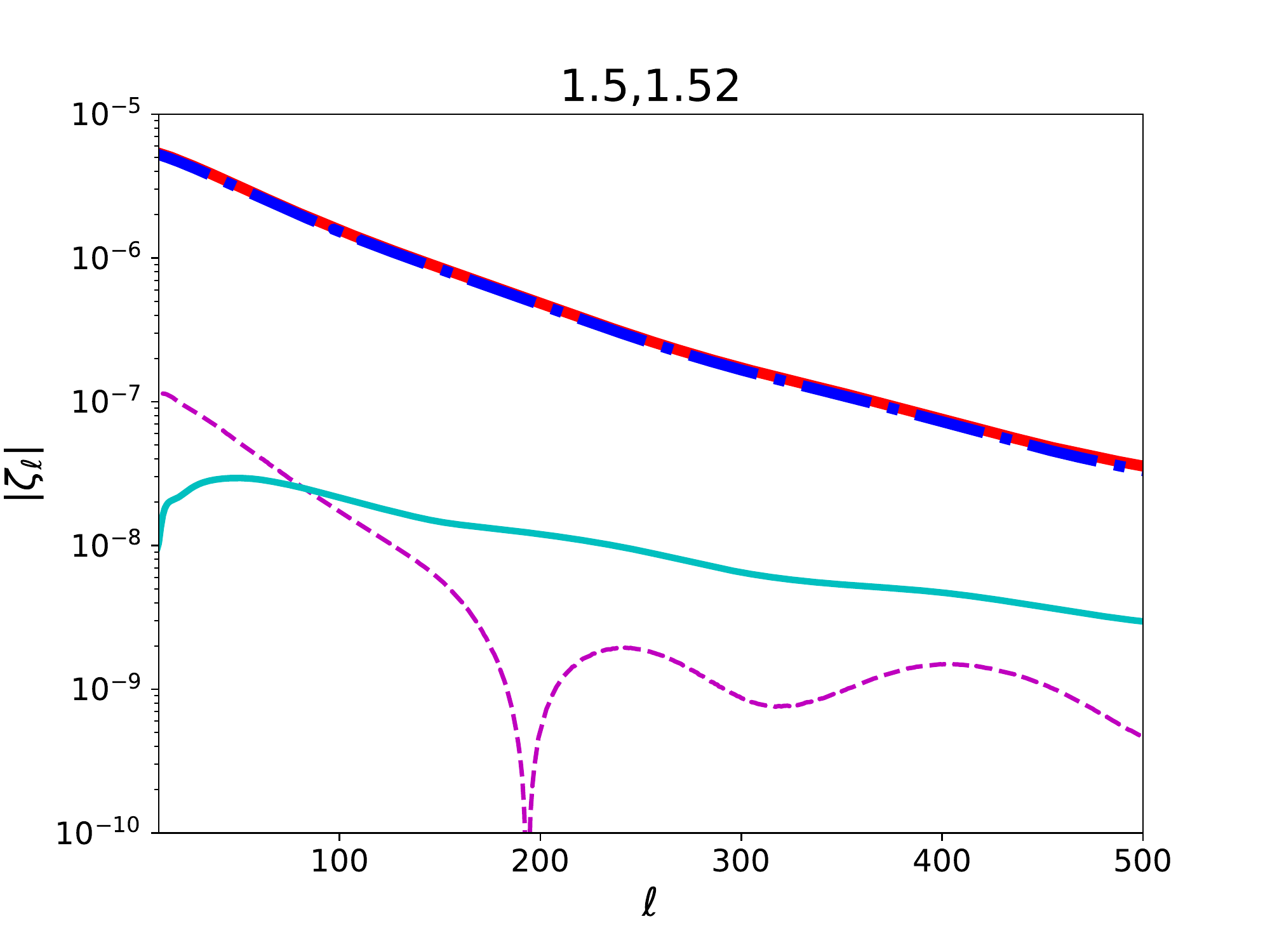}
\includegraphics[width=5.cm]{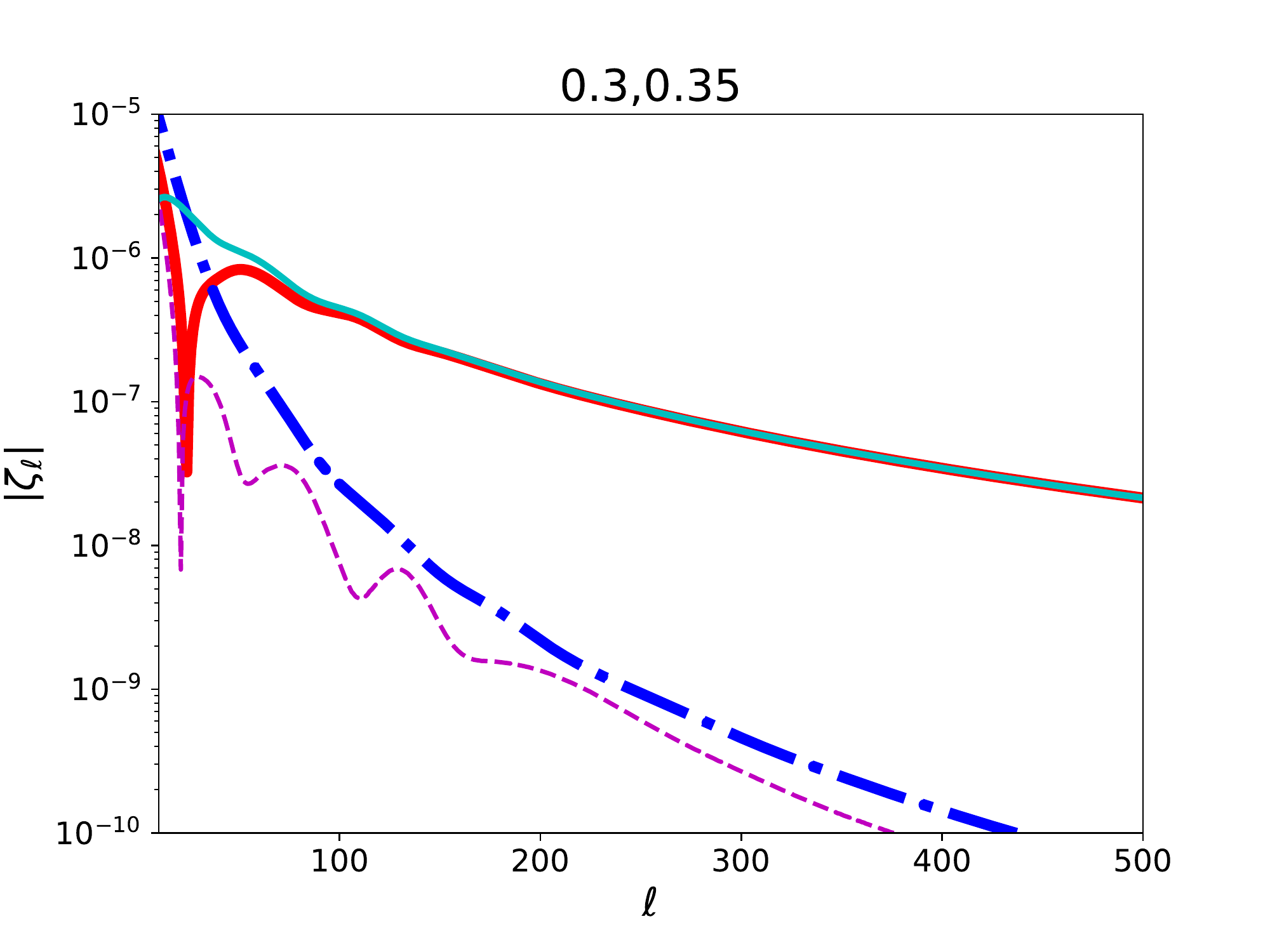}
\includegraphics[width=5.cm]{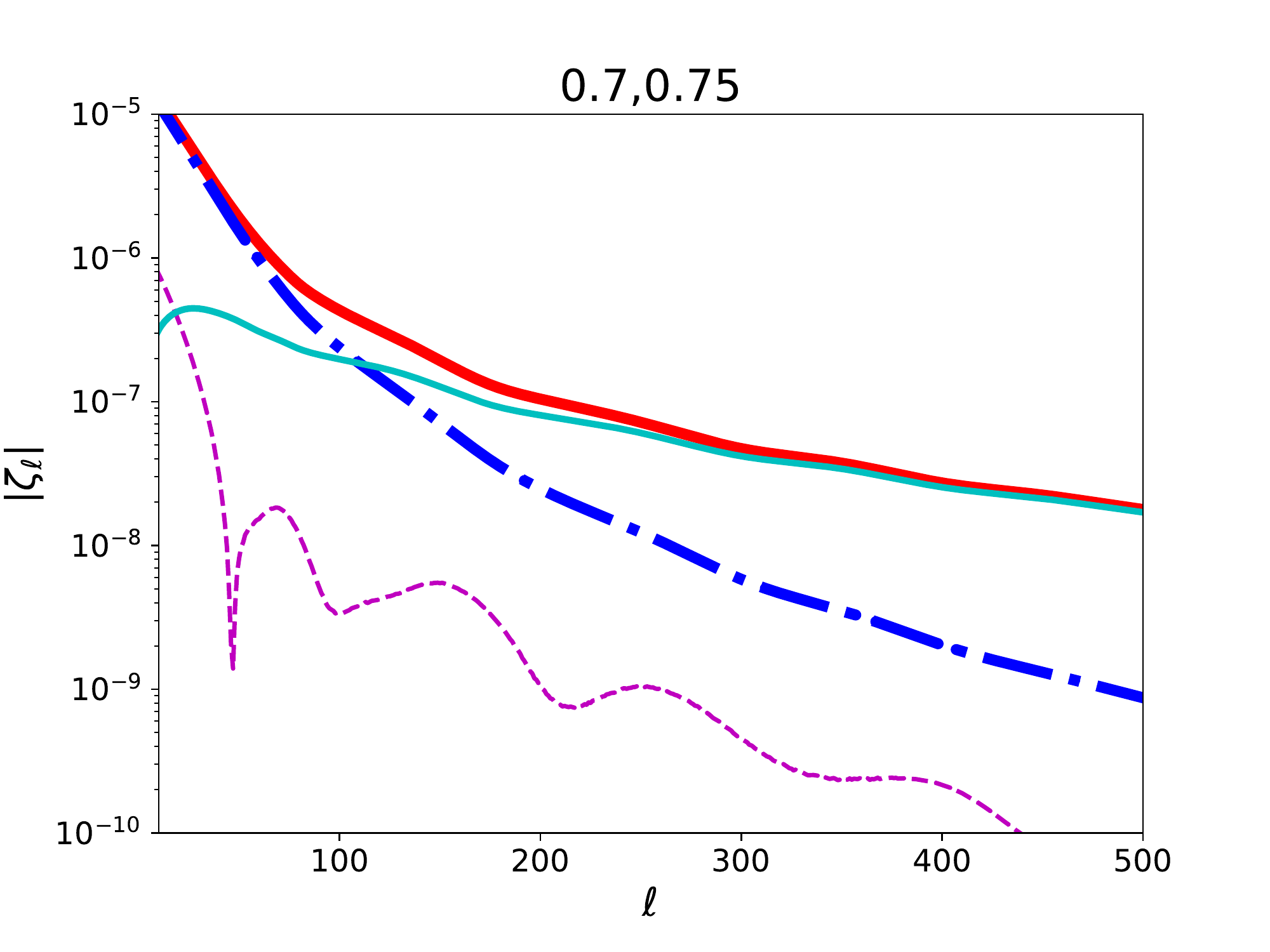}
\includegraphics[width=5.cm]{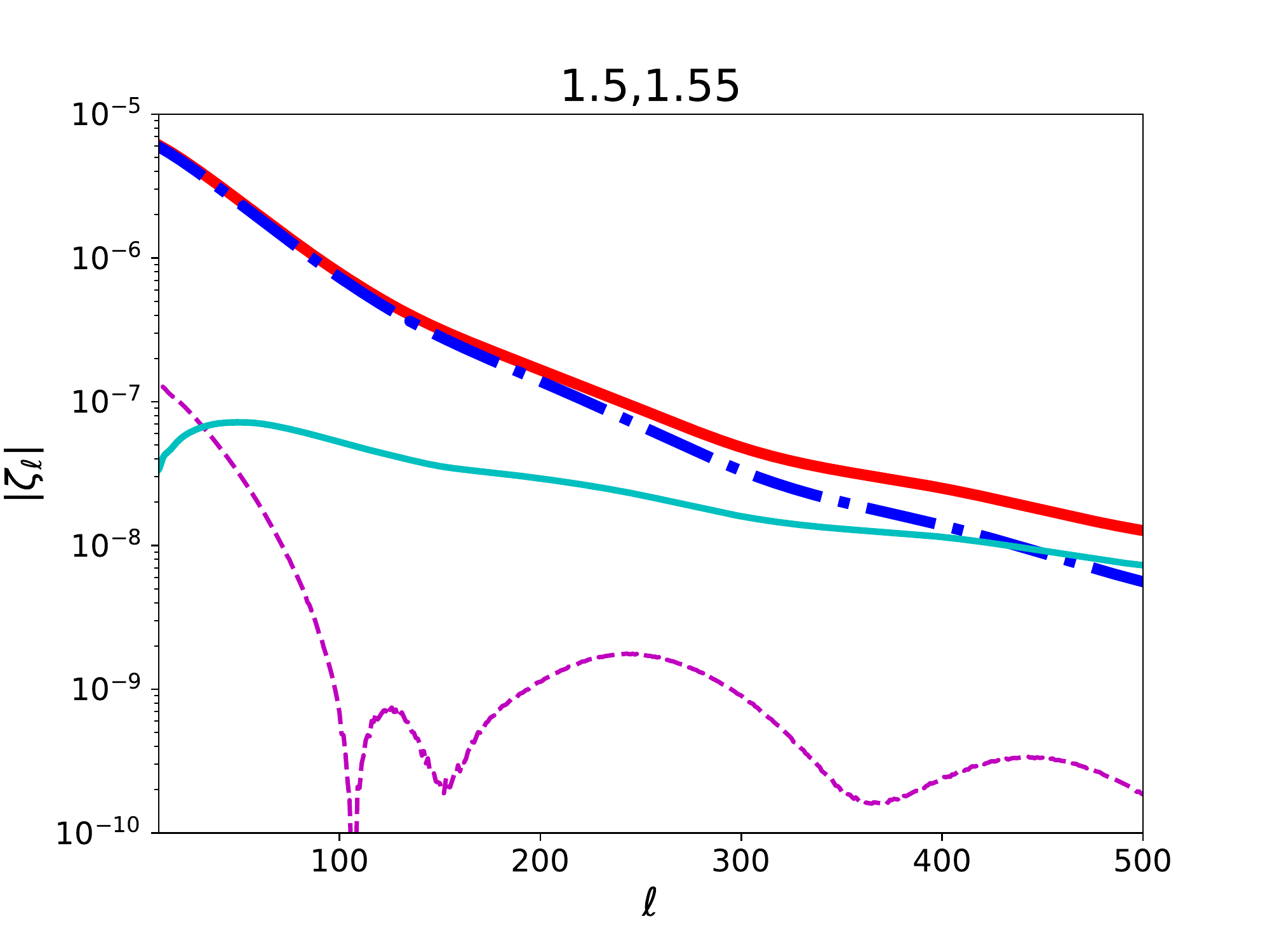}
\includegraphics[width=5.cm]{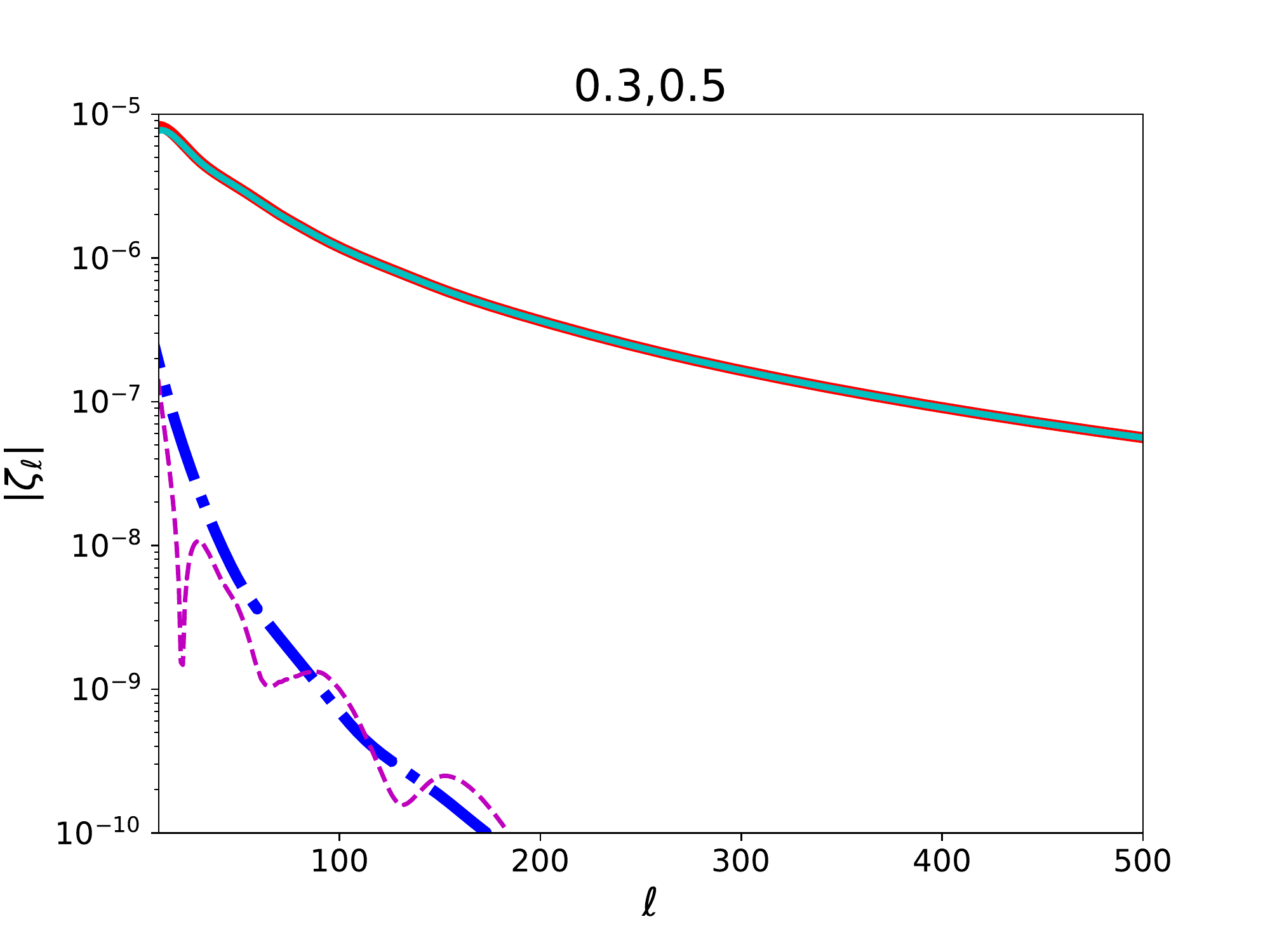}
\includegraphics[width=5.cm]{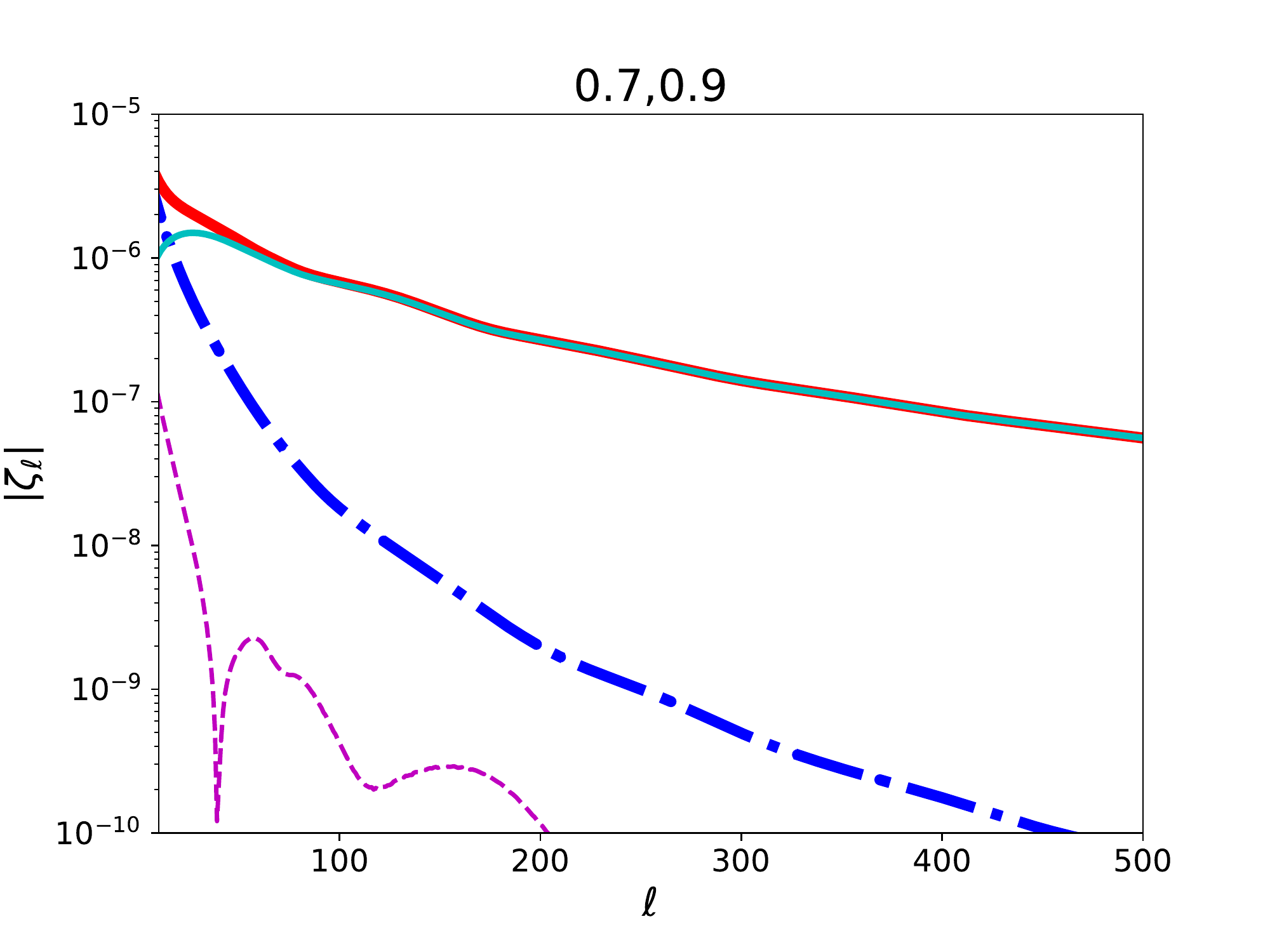}
\includegraphics[width=5.cm]{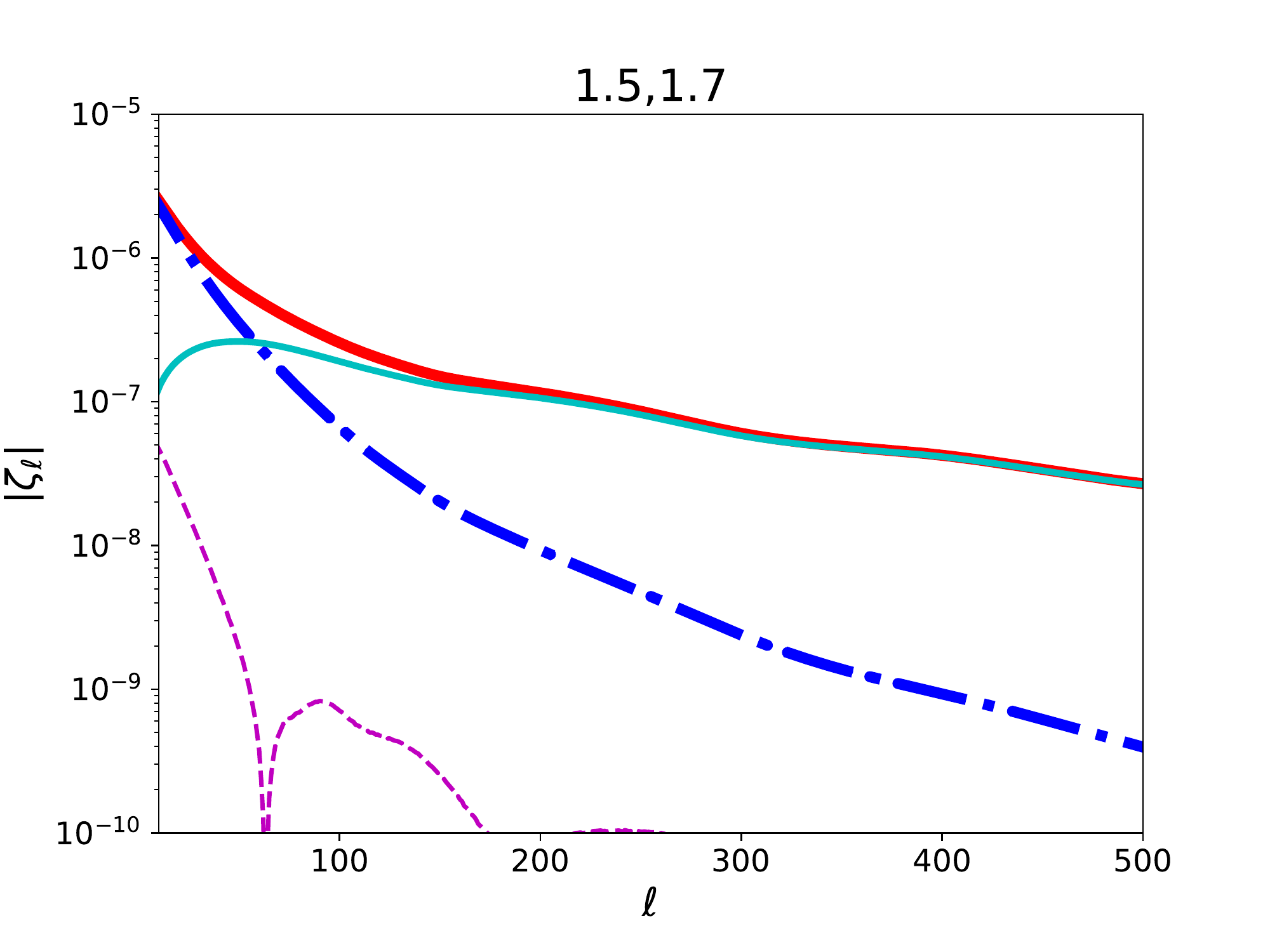}
\caption{Same as \autoref{fig:sbe0} but with evolution and magnification biases which enhance the Doppler term: \emph{Left column}: $z_i=0.3$; \emph{Middle column}: $z_i=0.7$; \emph{Right column}: $z_i=1.5$, for different redshift widths: \emph{Top row}: $\Delta=0.01$; \emph{2nd row}: $\Delta=0.02$; \emph{3rd row}: $\Delta=0.05$; \emph{Bottom row}: $\Delta=0.2$.}
\label{fig:sbemax}
\end{figure}

One can tweak the biases to enhance the significance of the Doppler contributions. Just for the sake of argument, assuming one can construct samples with any sort of biases (which is not necessarily true) let us maximize (or attempt to) the Doppler signal by cooking up magnification biases and evolution biases. Let us make $A_{{\mathrm{D}}}^A$ and $A_{{\mathrm{D}}}^B$ have opposite signs and chose $s^A=-0.3$ (for instance a faint sample may have a negative magnification bias not from the flux threshold but from the flux cut that makes the distinction between faint/bright samples, see Appendix \ref{app:survdets}), $s^B=1.2$, $b_{{\mathrm{e}}}^A=10$, $b_{{\mathrm{e}}}^B=-7$ (some samples can indeed have these kind of numbers see Appendix \ref{app:survdets}). In \autoref{fig:sbemax} we reproduce \autoref{fig:sbe0} but for magnification and evolution biases that enhance the Doppler term. As before, we can see that the regimes in which the Doppler term is the dominant contributor correspond to higher redshifts and thinner bins. In some bin size cases, we can see a transition between Doppler to Lensing as the dominant contributor to $\zeta$. In general, one can claim that Lensing is the leading part of the signal on small scales while Doppler leads on large scales. This kind of trade-off can be used to optimise the bin size for some combinations of galaxy samples/surveys.

\begin{figure}[]
\centering
\vspace*{-0.5cm}
\includegraphics[width=5.cm]{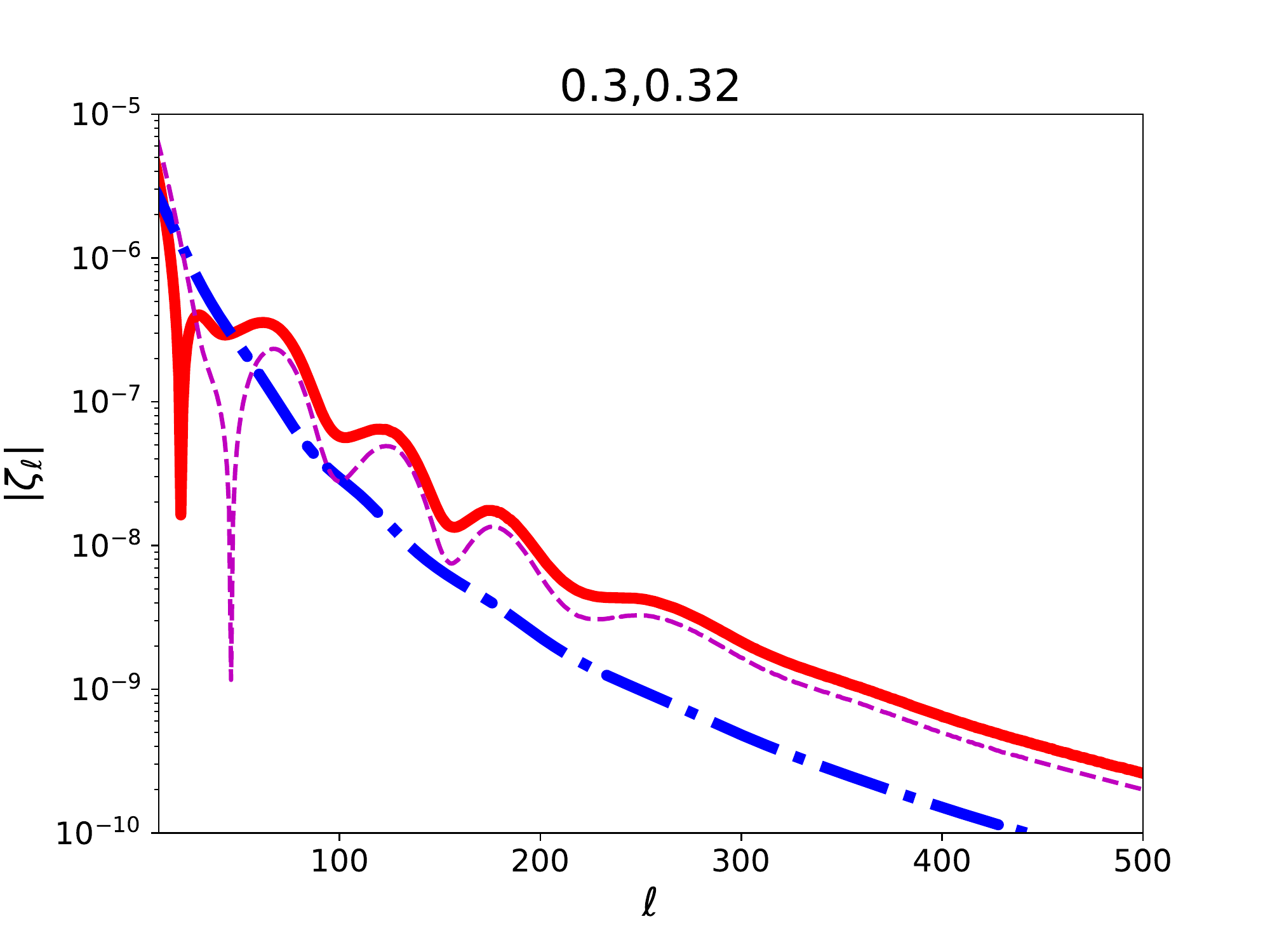}
\includegraphics[width=5.cm]{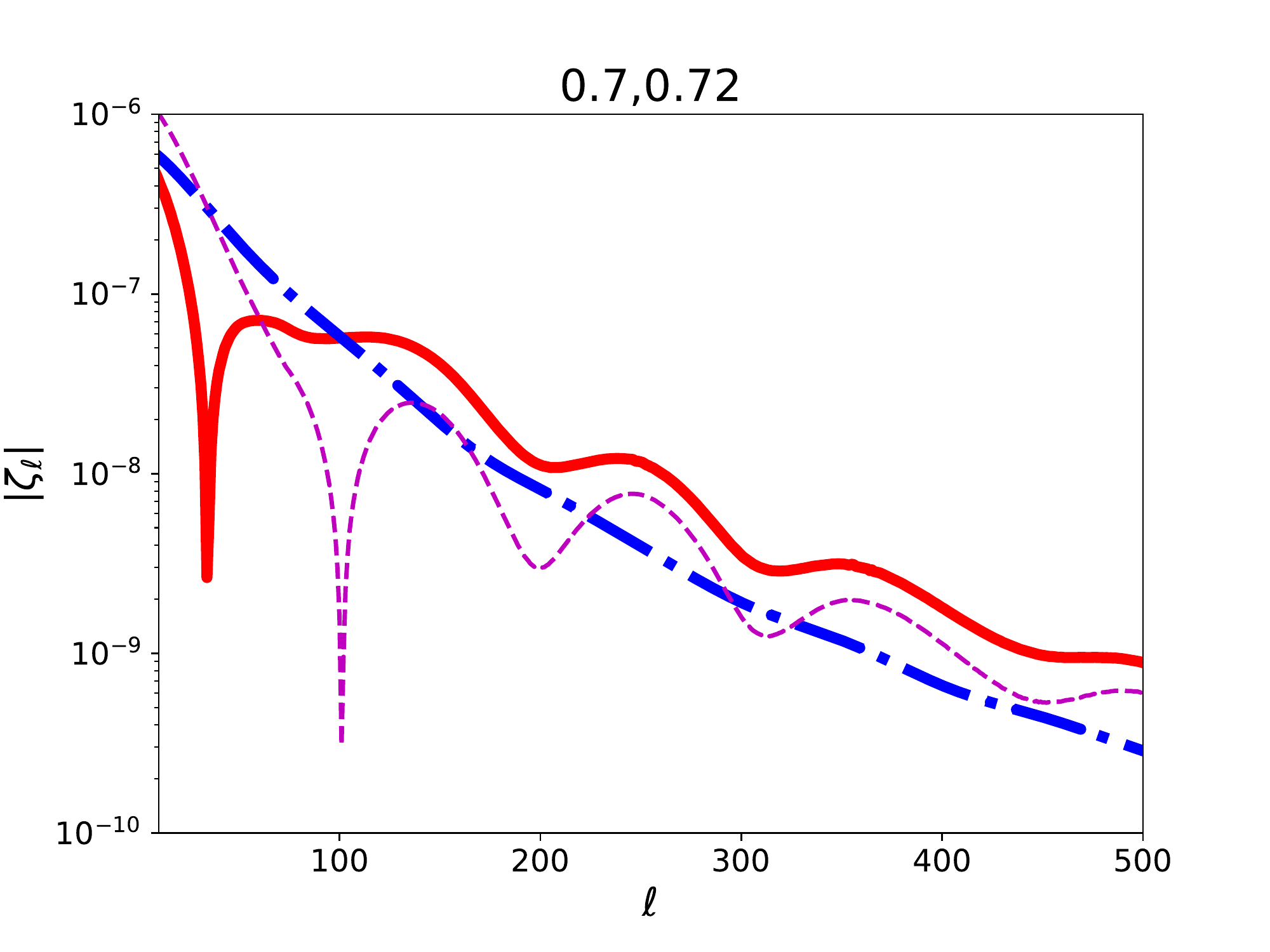}
\includegraphics[width=5.cm]{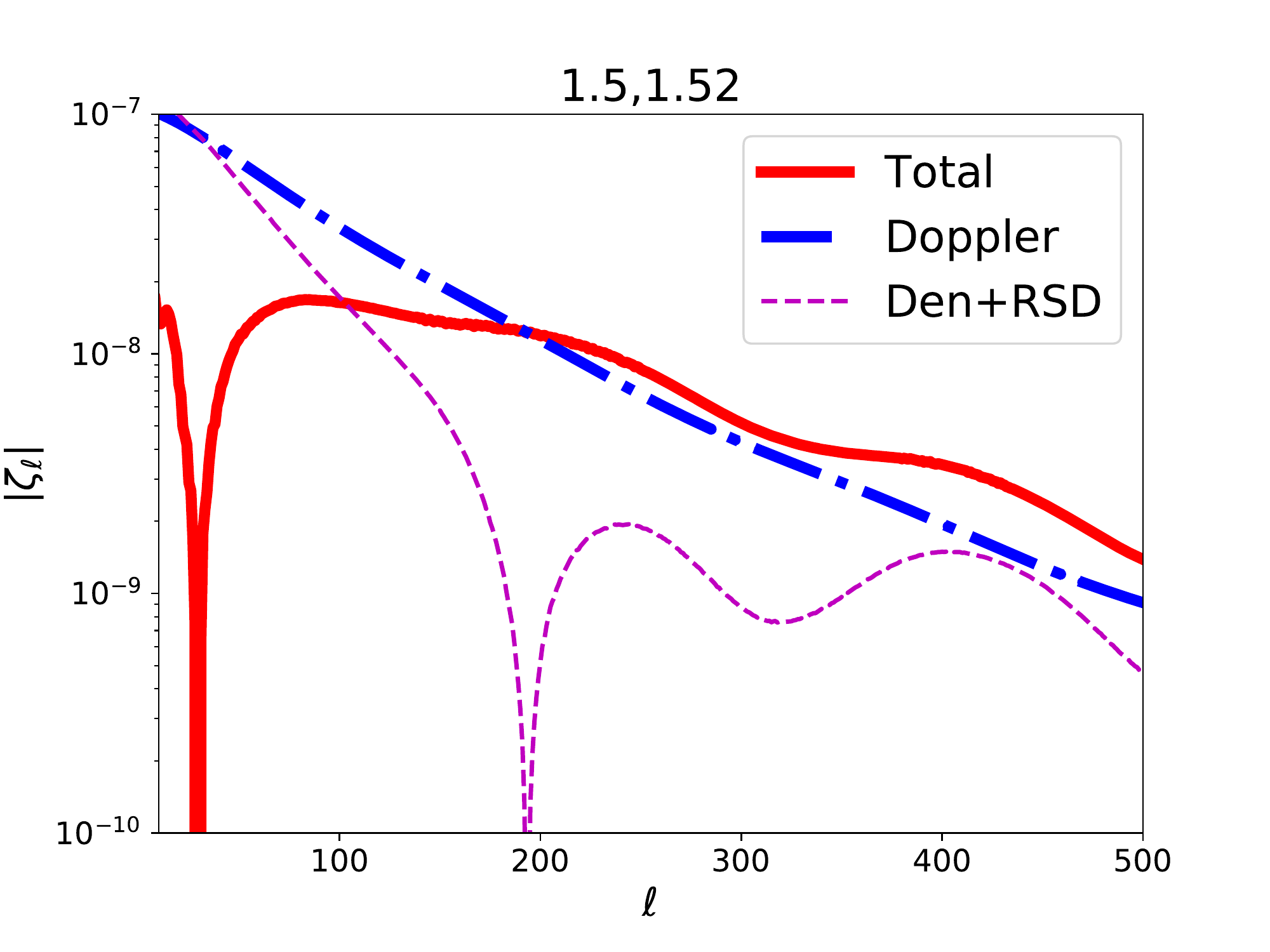}
\includegraphics[width=5.cm]{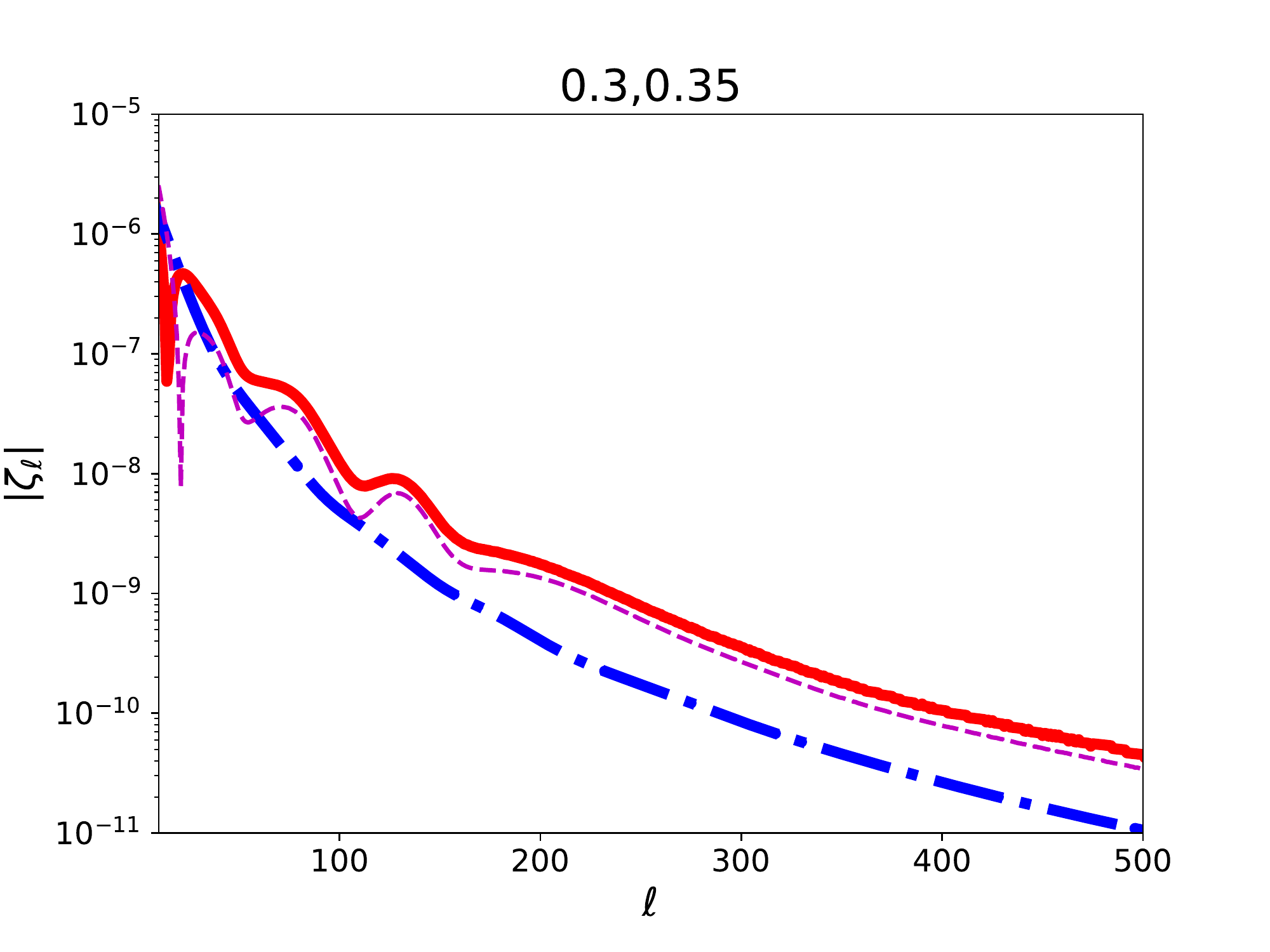}
\includegraphics[width=5.cm]{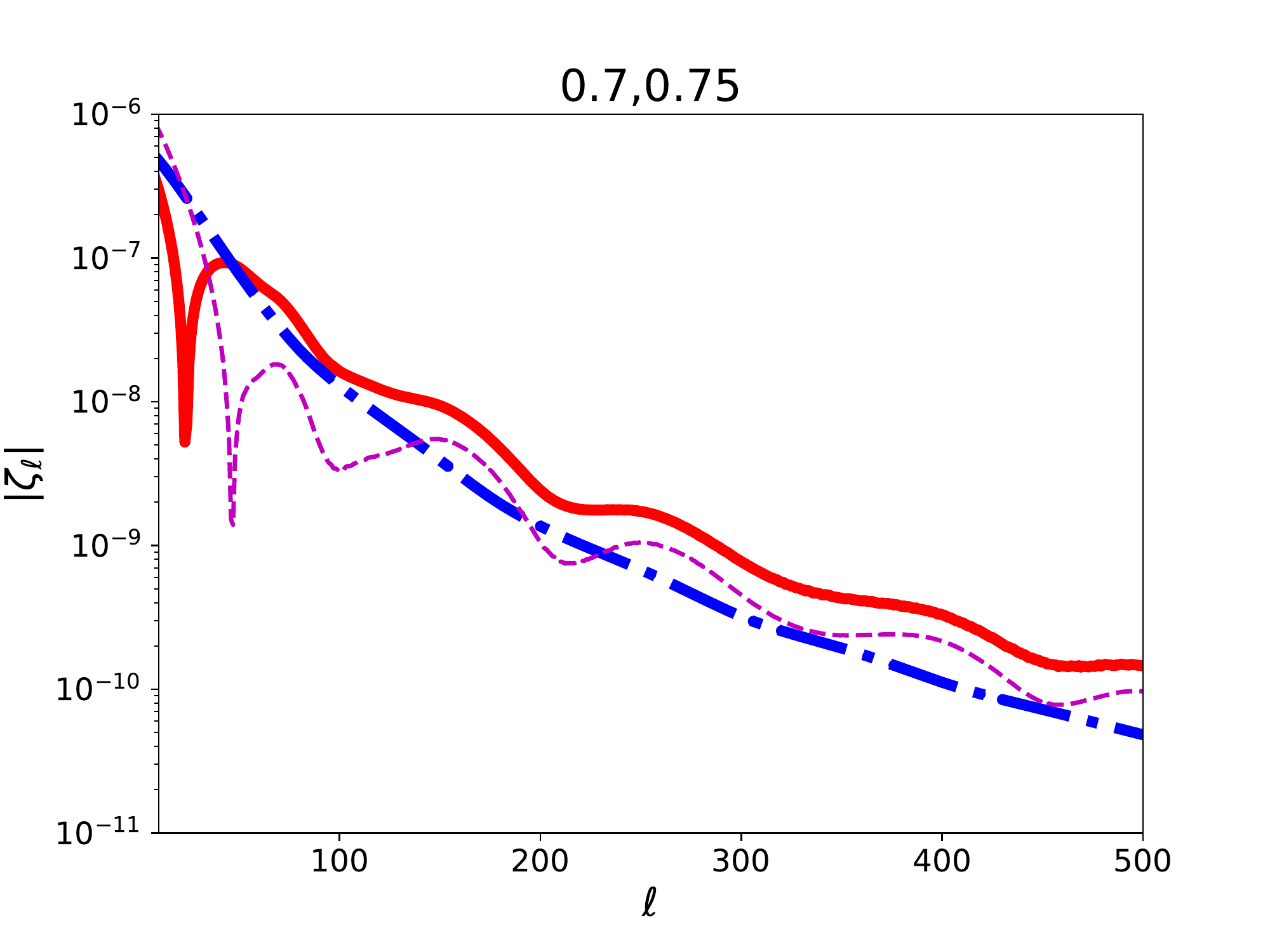}
\includegraphics[width=5.cm]{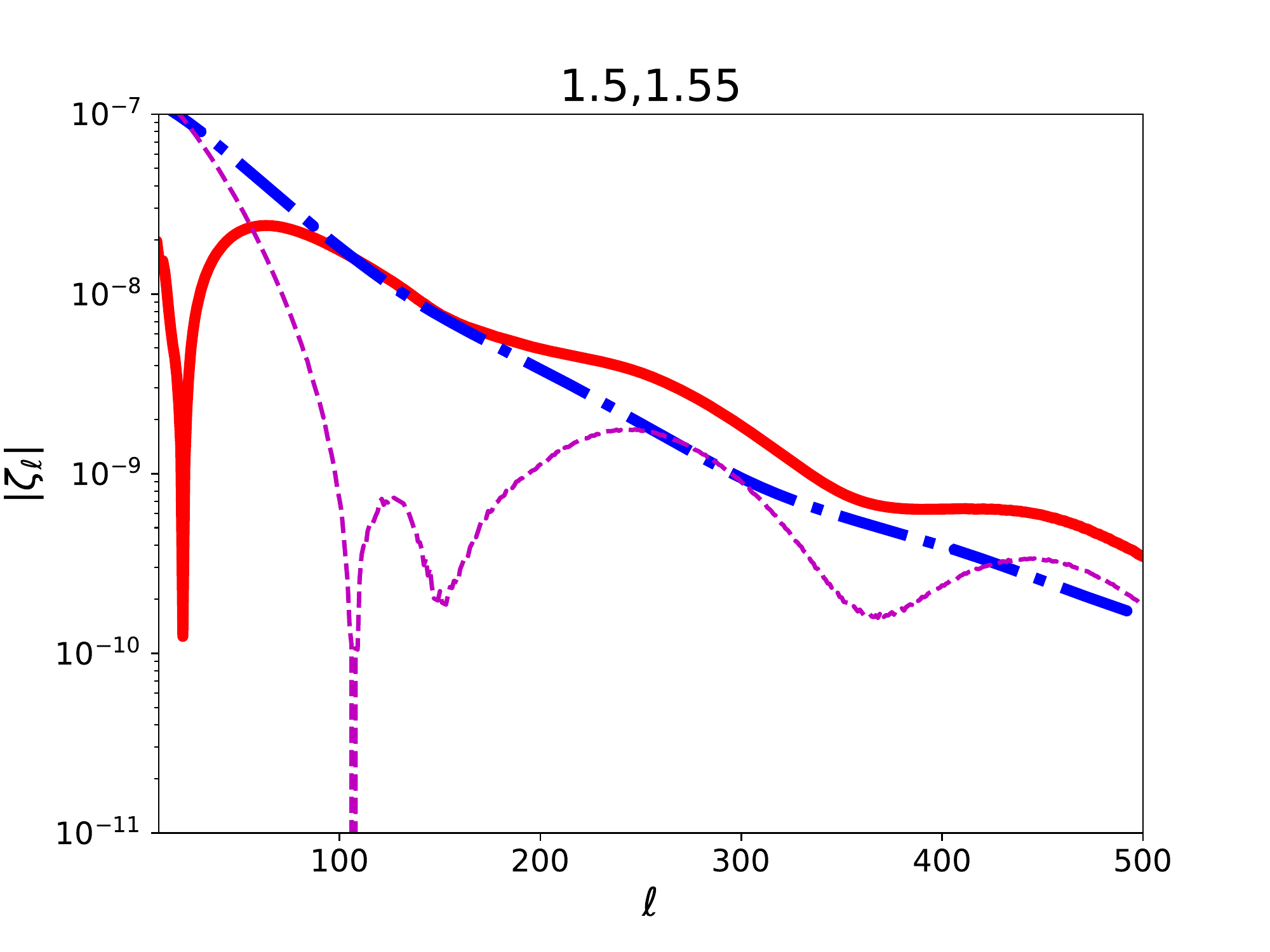}
\caption{Same as \autoref{fig:sbe0} but with evolution and magnification biases which enhance the Doppler term: \emph{Left column}: $z_i=0.3$; \emph{Middle column}: $z_i=0.7$; \emph{Right column}: $z_i=1.5$, for different redshift widths: \emph{Top row}: $\Delta=0.02$; \emph{Bottom row}: $\Delta=0.05$.}
\label{fig:sbeim}
\end{figure}

We see that this anti-symmetric estimator is also sensitive to Lensing on small scales even when we only consider consecutive bins in intermediate size bins. As an example, intensity maps \cite[see, e.g.][]{Fonseca:2016qqw} would not have lensing contribution altogether \cite{Hall:2012wd}. In \autoref{fig:sbeim} we show examples when we fix $s^A=s^B=0.4$ and $b_{{\mathrm{e}}}^A=b_{{\mathrm{e}}}^B=0$. Although at low redshifts the density-RSD anti-correlation is still dominating over the Doppler contribution is non-negligible, while at higher redshifts it becomes the leading contribution. We can also use larger bins to gain signal-to-noise as the relative relevance of the Doppler is not reduced. 

Although the clustering anti-symmetric angular power $\zeta$ does not uniquely recover the Doppler, its contribution is not subdominant as it is with the angular power spectrum only.
It indicates that we can select samples of galaxies such that we enhance the Doppler. For the purpose of this paper, from now on we will select a handful of surveys and use their astrophysical details computed from reasonable assumptions. We present such calculations in Appendix \ref{app:survdets}. 

\section{{Detecting the signal}} \label{sec:detectsignal}

\subsection{{Signal-to-noise}}

It is important to {estimate} how well one can detect the anti-symmetric part of the angular power spectrum for a pair of different galaxy samples. To compute the signal-to-noise ratio one needs the covariance of $\zeta^{AB}_\ell$ which is defined as
\bea
{\mathrm{cov}}(\zeta^{AB}_{ij},\zeta^{AB}_{op})&\equiv&\langle \zeta^{AB}_{ij}\zeta^{AB}_{op}\rangle - \langle \zeta^{AB}_{ij}\rangle\langle \zeta^{AB}_{op}\rangle\,,\\
&=&{\mathrm{cov}}(C^{AB}_{\ell,ij},C^{AB}_{\ell,op})+{\mathrm{cov}}(C^{BA}_{\ell,ij},C^{BA}_{\ell,op})-{\mathrm{cov}}(C^{BA}_{\ell,ij},C^{AB}_{\ell,op})-{\mathrm{cov}}(C^{AB}_{\ell,ij},C^{BA}_{\ell,op})\nn\,,
\eea
where in the second line we used the fact that $\zeta_\ell$ is linear in the angular power spectra $C_\ell$. Assuming a large enough sky coverage ($f_{{\mathrm{sky}}}$) one can show that the covariance of the angular power spectra is given by \cite[see, e.g., the appendix of][]{Fonseca:2020lmi}
\bea
{\mathrm{cov}}(C^{AB}_{\ell,ij},C^{CD}_{\ell,op})=\frac1{(2\ell+1)\Delta\ell f_{{\mathrm{sky}}}}\bigg[&& \l C^{AD}_{\ell,ip}+ \delta_{ip}\delta^{AD}{\cal N}_{\ell,i}^A \r \l C^{BC}_{\ell,jo}+ \delta_{jo}\delta^{BC}{\cal N}_{\ell,j}^B\r \nn\\
&&+\l C^{AC}_{\ell,io}+ \delta_{io}\delta^{AC}{\cal N}_{\ell,i}^A \r \l C^{BD}_{\ell,jp} + \delta_{jp}\delta^{BD}{\cal N}_{\ell,j}^B\r \bigg]
\eea
where ${\cal N}_{\ell}$ is the noise. Note as well that we have assumed that no cross-shot noise exists, i.e., that the samples do not overlap. {Nonetheless,} we expect this term to be small \cite{Viljoen:2020efi}.  
Therefore one can write the covariance as
\bea
{\mathrm{cov}}(\zeta^{AB}_{ij},\zeta^{AB}_{op})=\frac1{(2\ell+1)\Delta\ell f_{{\mathrm{sky}}}}\bigg[&& \l C^{AA}_{\ell,io}+ \delta_{io}{\cal N}_{\ell,i}^A\r\l C^{BB}_{\ell,jp} + \delta_{jp}{\cal N}_{\ell,j}^B\r\nn\\
&&+\l C^{BB}_{\ell,io} + \delta_{io}{\cal N}_{\ell,i}^B\r\l C^{AA}_{\ell,jp}+ \delta_{jp}{\cal N}_{\ell,j}^A\r\nn\\
&& -C^{BA}_{\ell,io} C^{AB}_{\ell,jp} -C^{AB}_{\ell,io} C^{BA}_{\ell,jp}\nn\\
&&+C^{AB}_{\ell,ip} C^{BA}_{\ell,jo}+C^{BA}_{\ell,ip} C^{AB}_{\ell,jo}\nn\\
&&-\l C^{BB}_{\ell,ip}+ \delta_{ip}{\cal N}_{\ell,i}^B\r \l C^{AA}_{\ell,jo}+ \delta_{jo}{\cal N}_{\ell,j}^A\r\nn\\
&& -\l C^{AA}_{\ell,ip}+ \delta_{ip}{\cal N}_{\ell,i}^A \r\l C^{BB}_{\ell,jo} + \delta_{jo}{\cal N}_{\ell,j}^B \r\bigg]\,.
\eea
Then the signal-to-noise ratio (SNR) for each bin is simply given by
\be
{\mathrm{ SNR}}\l\zeta^{AB}_{ij}\r=\sqrt{\sum_{\ell_{{\rm min}}}^{\ell_{\rm max}}\l\frac{(\zeta^{AB}_{\ell,ij})^2}{{\mathrm{cov}}(\zeta^{AB}_{\ell,ij},\zeta^{AB}_{\ell,ij})}\r}\,,
\ee
and the total SNR of the survey (considering only consecutive bins) is just
\be
{\rm SNR}_{\rm tot}=\sqrt{\sum_i {\rm SNR}^2\l\zeta^{AB}_{i,(i+1)}\r}\,.
\ee
Here we focus only on consecutive bins as the signal dies out quickly or becomes dominated by other terms. We also consider each thin bin independently although averaging a bundle of thin bins would not alter the signal-to-noise ratio but increase the detectability of the bundled $\zeta$ (see below). 

In \autoref{tab:snr_comb_surv} we computed the SNR for different combinations of surveys. We consider several surveys although we do not combine them all. We will consider two DESI-like BGS samples, one bright as described in \cite{Aghamousa:2016zmz} and a full sample which includes fainter galaxies \cite{Beutler:2020evf}. We will also consider a Euclid-like H$\alpha$ galaxy sample \cite{Pozzetti:2016cch}. We will also consider several samples from SKAO: from MID we will consider the HI galaxy and the HI IM surveys \cite{Bacon:2018dui}; as well as 3 HI galaxy samples from a futuristic SKAO upgrade which we will call SKAO2 as it has been called traditionally. We present all astrophysical details we considered in Appendix \ref{app:survdets} where we explain how we computed the magnification and evolution biases for each survey or galaxy sample as well as what we assumed or how the bias and distribution of sources were obtained. We only assumed spectroscopic surveys, including HI intensity mapping. For simplicity, we considered a total overlap between surveys and therefore make $f_{{\mathrm{sky}}}$ equal to the survey with the smallest footprint. Also in \autoref{tab:snr_comb_surv} we show two different bin width, $\Delta z=0.02$, and for $\Delta z=0.01$ inside brackets. One can conclude from the table that at lower redshifts it is beneficial to have thinner resolutions while at higher redshifts the opposite happens. It is also interesting to see that in the case of the BGS samples, the Bright sample has higher SNR than the Full sample when combined with SKAO2 HI galaxies using thin redshift bins. In the case of wider bins, this is no longer true. But naively one would expect a comparatively higher SNR as the full sample has a lower shot-noise. This example demonstrates that $\zeta$ is very sensitive to the astrophysical details of the samples. One should also note that despite the high spectral resolution of IM, its low angular resolution degrades the SNR of any combination we consider, making it suboptimal for $\zeta$ at high redshift. In general, forthcoming surveys can provide an SNR $>1$, but one needs to wait for a SKAO2-like HI galaxy survey to obtain enough galaxies to have an SNR above 40. Note that this SNR is of the same order of magnitude as found in \cite{Bonvin:2018ckp}. 

\begin{table}
\caption{\label{tab:snr_comb_surv} Signal-to-noise ratio for different combinations of surveys. All considered a cut at $\ell_{\rm max}=300$ with $\Delta\ell=5$, and $\Delta z=0.02$, and for $\Delta z=0.01$ inside brackets.}
\centering
\begin{tabular}{lcccc}
\\
 & BGS - Full & BGS - Bright & H$\alpha$ & SKAO2 Faint \\
 \hline
SKAO MID HI gal  & 3.5 (4.6) & 2.7 (3.4) & - & -\\
SKAO MID HI IM  & 4.8 (5.4) & 3.4 (4.9) & 6.7 (5.4) & - \\
SKAO2 Full  & 9.2 (11.4) & 8.5 (12.4) & 17.2 (15.5) & - \\
SKAO2 Bright  & - & - & - & 41.9 (53.3) \\
\hline
\end{tabular}
\end{table}

\subsection{{Prospects for detection of Doppler term contributions}}

Additionally one can also check the detectability of the Doppler term itself. Let us include a fudge factor $\epsilon_{{\mathrm{D}}}=1$ in \autoref{eq:deltatot} as done in \cite{Fonseca:2018hsu}, i.e., 
\be
\Delta^{{A}}_\ell(k)={b^A}~ \Delta^{\delta}_\ell(k)+\Delta^{{\mathrm{R}}}_\ell(k)+\epsilon_{{\mathrm{D}}}~{A^A_{\mathrm{D}}}~ \Delta^{{\mathrm{D}}}_\ell(k)+{A^A_{\mathrm{L}}}~ \Delta^{{\mathrm{L}}}_\ell(k) + \ldots \,.
\ee
In principle, this $\epsilon_{{\mathrm{D}}}$ measures the detectability of the Doppler contribution in the transfer function $\Delta_\ell$. 
To constraint the $\epsilon_{{\mathrm{D}}}$ parameter we run a simple Fisher forecast \cite{Fisher:1935b}, where the fisher matrix is given by
\be \label{eq:fisher}
F_{\vartheta_a\vartheta_b}= \sum_\ell \frac{\partial \zeta^{AB}_{ij}}{\partial \vartheta_a} \left[{\mathrm{cov}}(\zeta^{AB}_{ij},\zeta^{AB}_{op})\right]^{-1} \frac{\partial\zeta^{AB}_{op}}{\partial \vartheta_b}\,.
\ee
Here we will assume good constraints on cosmological parameters have been obtained by other means and only compute the conditional error on $\epsilon_{{\mathrm{D}}}$, which is just $1/\sqrt{F_{\epsilon_{{\mathrm{D}}}\epsilon_{{\mathrm{D}}}}}$. We present the results of the conditional error in percentage in \autoref{tab:cond_error_eD}. One should note that if $\zeta$ is uniquely given by the Doppler term then $\epsilon_{{\mathrm{D}}}\simeq 100/{\mathrm{SNR}}$. If this is not the case, it indicates that the density-RSD contribution is large (in some cases even the approximation of slowly evolving biases breaks down). As expected, the SKAO2 HI galaxy survey is the one that would better detect the Doppler term using the anti-symmetric estimator. Similarly, any combination with this future survey is the one that provides better constraints. Here we have not optimised the bin width and solely gave results for two pre-chosen thin bins. We can see from the results in \autoref{tab:cond_error_eD} that again high-z surveys do not require such a fine bin while low-z improve with higher resolution. This is related to the physical scales at which this anti-symmetric estimator peaks. In comparison with a full tomographic analysis \cite{Viljoen:2021ypp}, $\zeta$ has a much worse performance when comparing the combinations MID's HI IM with H$\alpha$ galaxy survey, and full BGS with MID's HI IM. One should note that to construct $\zeta$ we neglect most of the data, which a full tomographic approach takes into account. But, as we will see below, one has an estimator that de-blends the Doppler term from all other contributions. Still, for SKAO2 HI galaxy sample and considering the full covariance between our ``observables'', one gets $\sigma(\epsilon_{{\mathrm{D}}})=2.3\%$ for a $\Delta z=0.02$ binning and $\sigma(\epsilon_{{\mathrm{D}}})=1.8\%$ for a $\Delta z=0.01$. These are substantially smaller than quoted before in the literature. 

\begin{table}
\caption{\label{tab:cond_error_eD} Conditional error on $\epsilon_{{\mathrm{D}}}$ in percentage for different combinations of surveys. All considered a cut at $\ell_{\rm max}=300$ with $\Delta\ell=5$, and $\Delta z=0.02$, and for $\Delta z=0.01$ inside brackets.}
\centering
\begin{tabular}{lcccc}
\\
 \% &BGS - Full & BGS - Bright & H$\alpha$ & SKAO2 Faint \\
 \hline
SKAO MID HI gal  & 37.8 (29.4) & 65.3 (50.8) & - & -\\
SKAO MID HI IM  & 111 (92.3) & 111 (98.7) & 79.0 (93.1) & - \\
SKAO2 Full  & 7.4 (5.8) & 6.4 (4.7) & 6.3 (6.6) & - \\
SKAO2 Bright  & - & - & - & 2.2 (1.7) \\
\hline
\end{tabular}
\end{table}

\subsection{{Reconstructing $\zeta$ using faint and bright SKA HI galaxies}} \label{sec:ska2exemple}

Here we exemplify how the anti-symmetric estimator {$\zeta$} of a sample of faint and bright HI galaxies in SKAO2 reconstructs the Doppler term. Here we only exemplify with SKAO2 HI galaxies as they are the ones with the highest SNR. One should note that another combination of tracers and survey details alters the relevance of the Doppler term in the anti-symmetric estimator. Therefore we cannot extrapolate any general conclusion from this example.

\begin{figure}[]
\centering
\vspace*{-0.5cm}
\includegraphics[width=7.5cm]{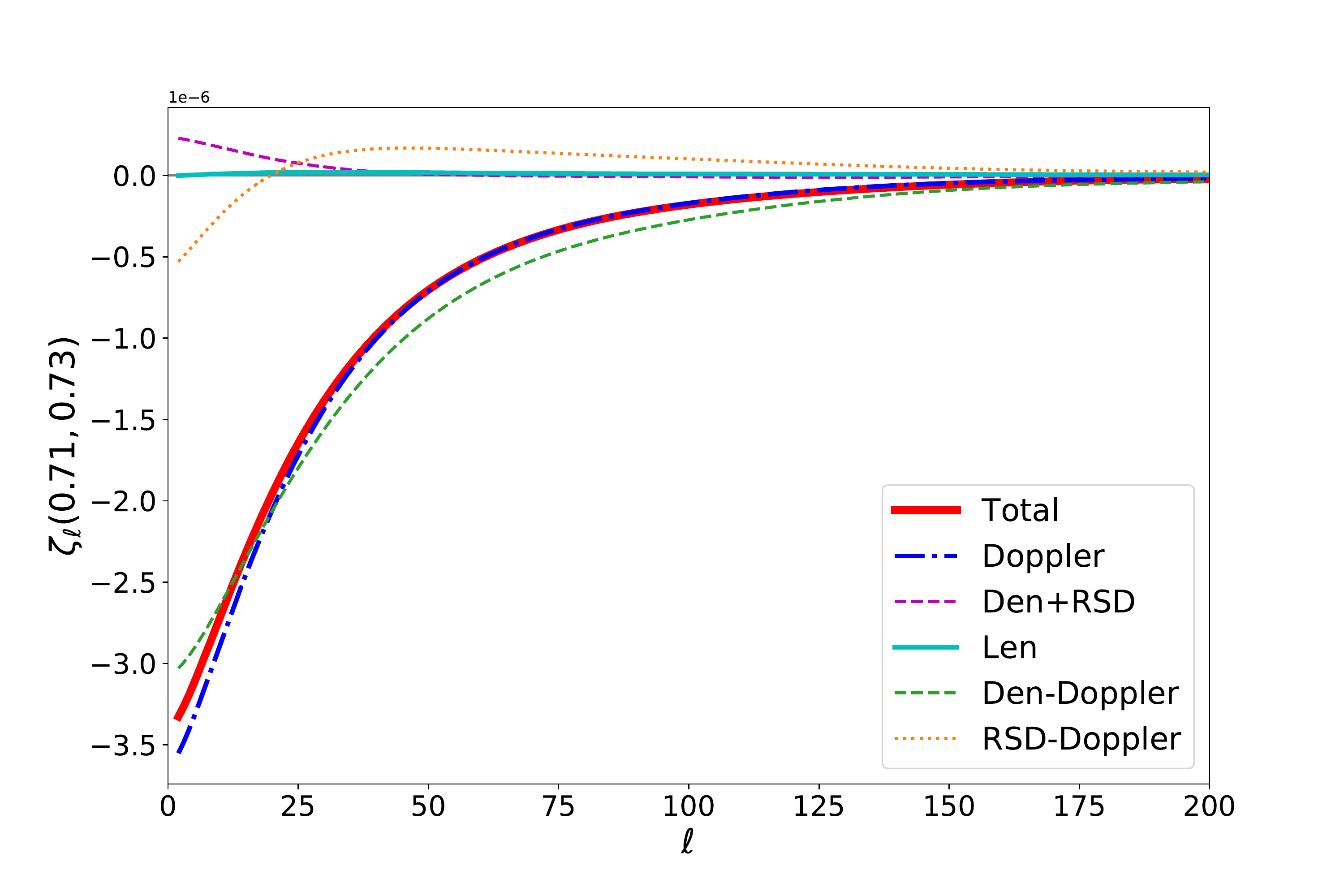}
\includegraphics[width=7.5cm]{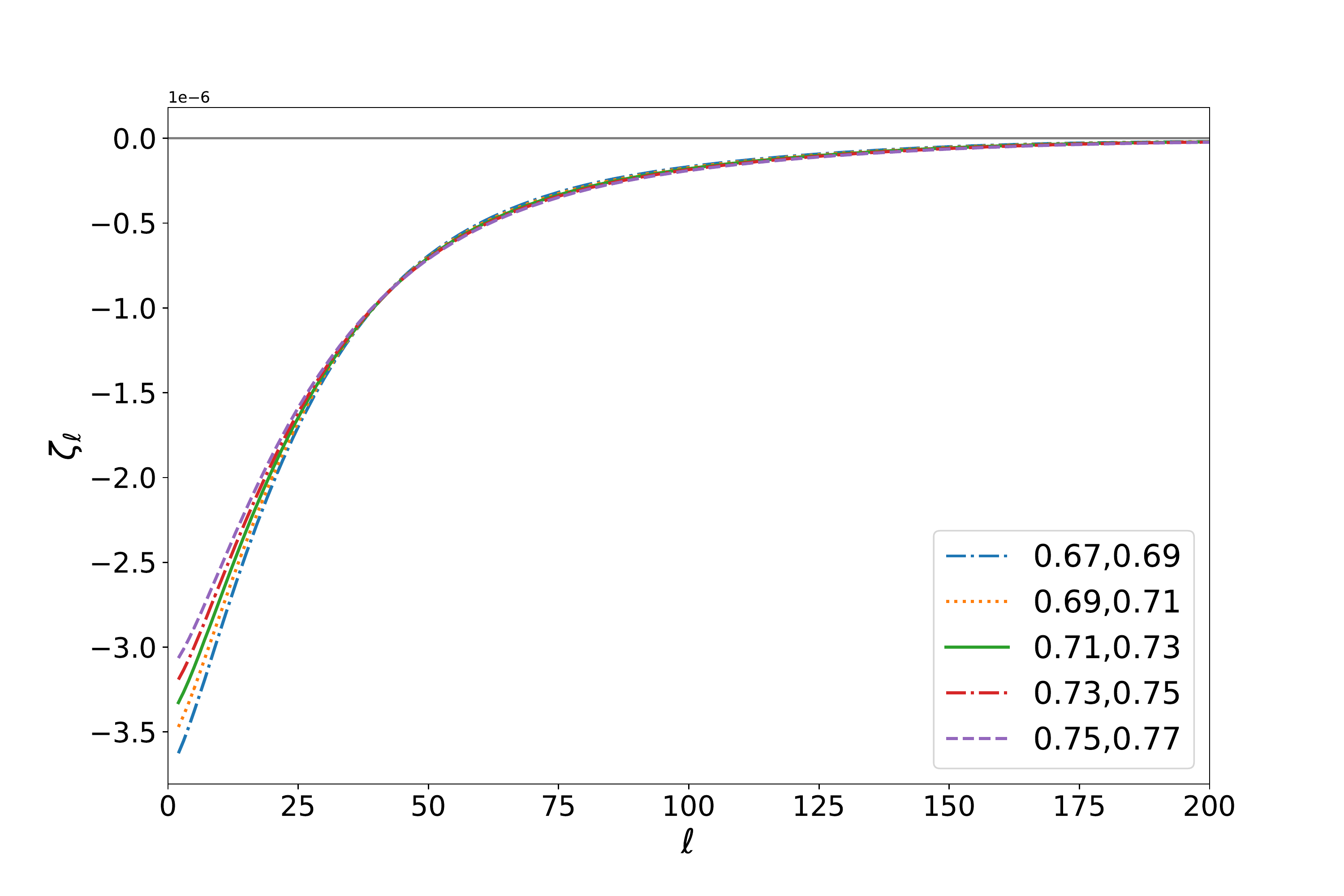}
\caption{\emph{Left}: Contribution of the different terms to the anti-symmetric estimator. ``Den+RSD'' {refers} to the density and Redshift Space Distortions only while the ``Doppler'' includes correlations with density and RSD. Clustering bias of the considered samples. We also plot the anti-symmetry between density and Doppler, and RSD and Doppler. \emph{Right}: Consecutive anti-symmetric estimators around the pair $z_i=0.71$ and $z_j=0.73$.}
\label{fig:zeta_ska2_072}
\end{figure}

On the left panel of \autoref{fig:zeta_ska2_072} we plot in thick solid red an example of how the anti-symmetric estimator looks like for a redshift bin pair of $z_i=0.71$ and $z_j=0.73$, with bins $\Delta z=0.02$. We also plot in dashed magenta the contribution from density and RSD only. In this particular case, they are small and only provide a small correction on large scales. In this setting, lensing is negligible (thick solid cyan line). The Doppler part (blue thick dot-dashed) {closely follows} the total signal. We further plot the anti-symmetric part coming from the density and Doppler in thin dashed green, and the RSD and Doppler in thin dotted orange. In this case, one can see that the anti-symmetric estimator is mainly coming from the anti-symmetry between the Doppler term and the density term of the two galaxy samples. These proportions are not necessarily true in all cases or all redshifts but demonstrate that measuring $\zeta$ is a probe of peculiar velocities.

\begin{figure}[]
\centering
\vspace*{-0.5cm}
\includegraphics[width=7.5cm]{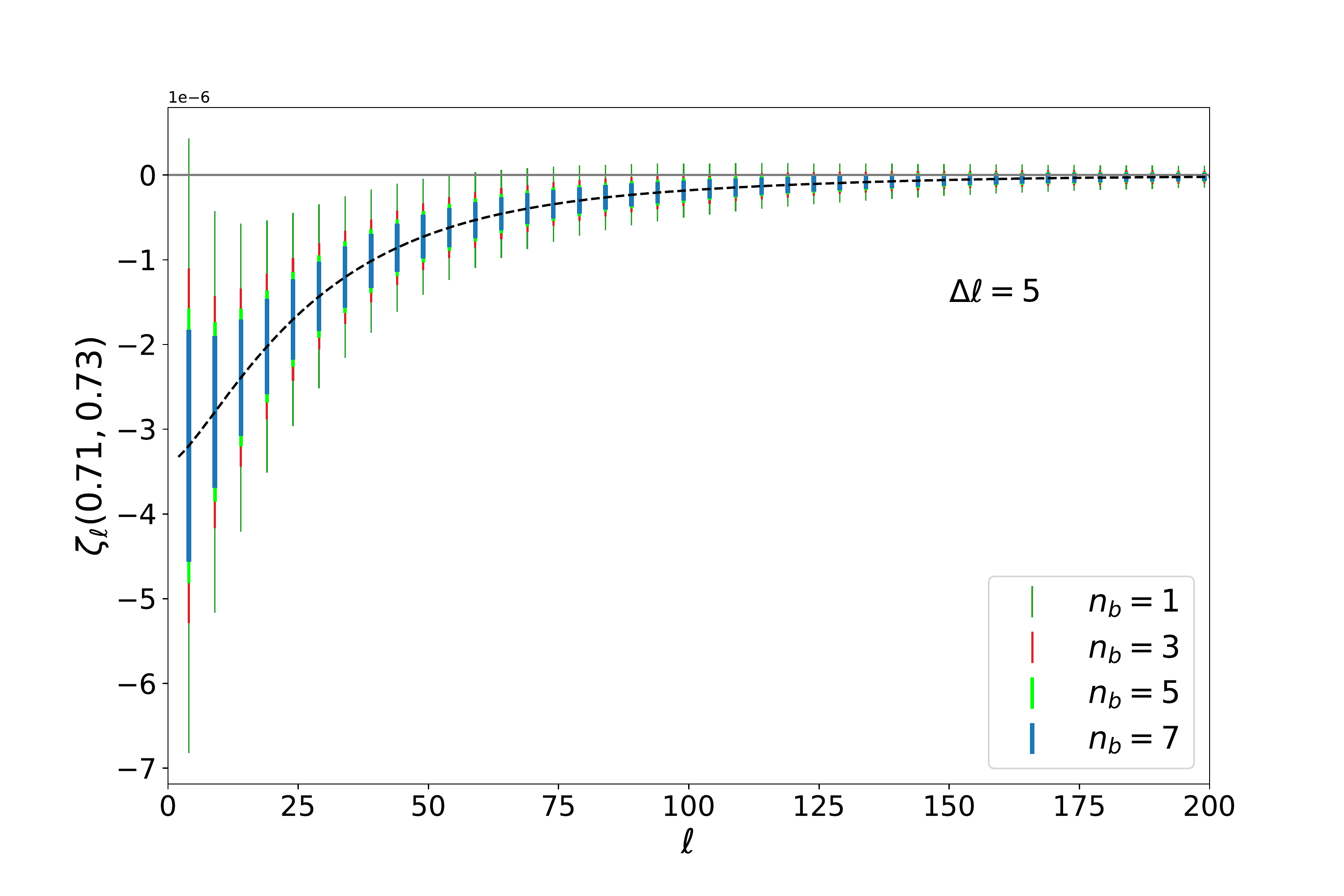}
\includegraphics[width=7.5cm]{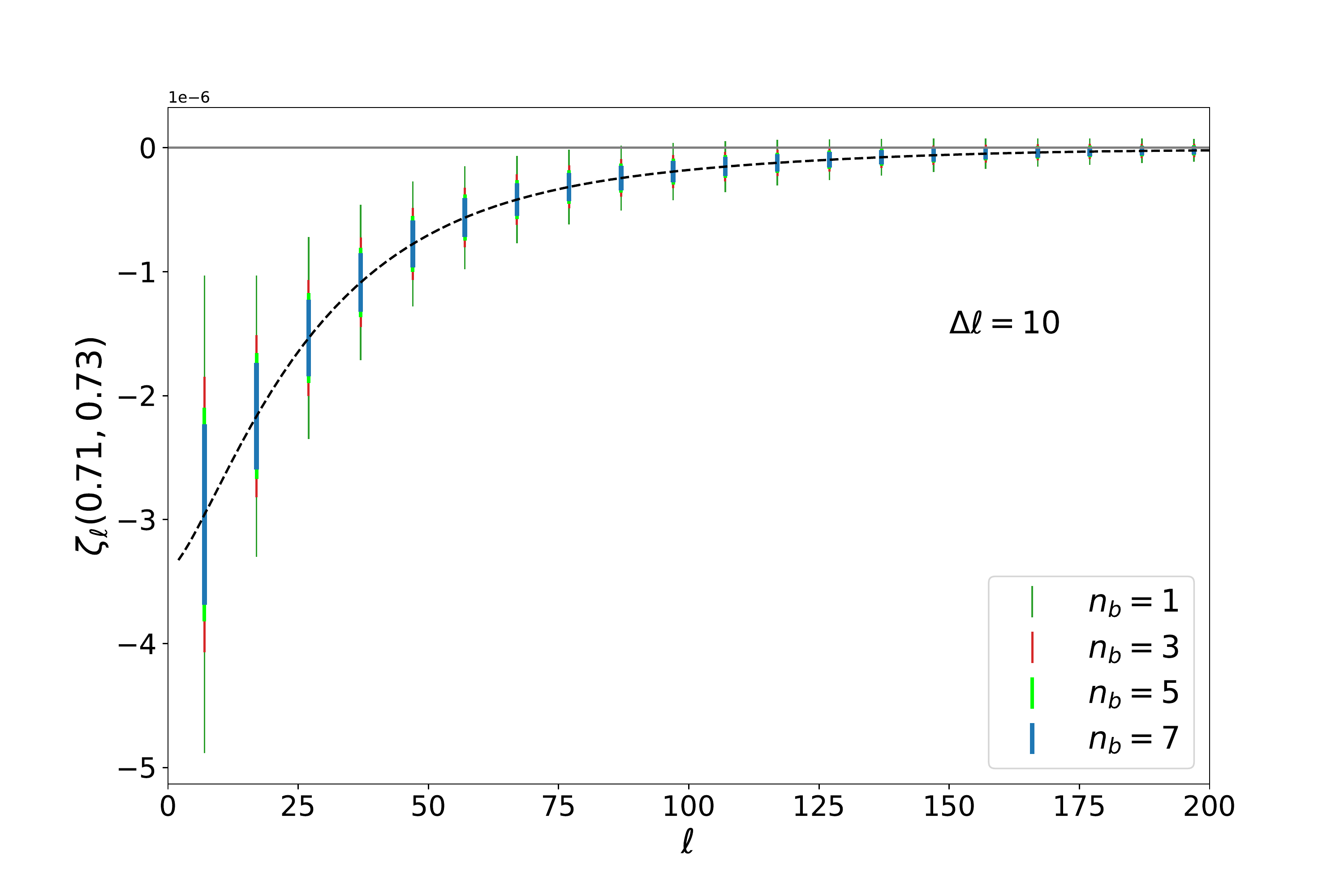}
\caption{{Anti-symmetric estimator around the pair $z_i=0.71$ and $z_j=0.73$ (black dashed line) with error bars for different data manipulations. We show two cases of $\ell$-binning, $\Delta\ell=5$ on the \emph{left} and $\Delta\ell=10$ on the \emph{right}. The size of the error bars depends on the number of contiguous bin pairs used to determine $\zeta(0.71,0.73)$.}}
\label{fig:detectability_zeta_sk2}
\end{figure}

\begin{figure}[]
\centering
\vspace*{-0.5cm}
\includegraphics[width=7.5cm]{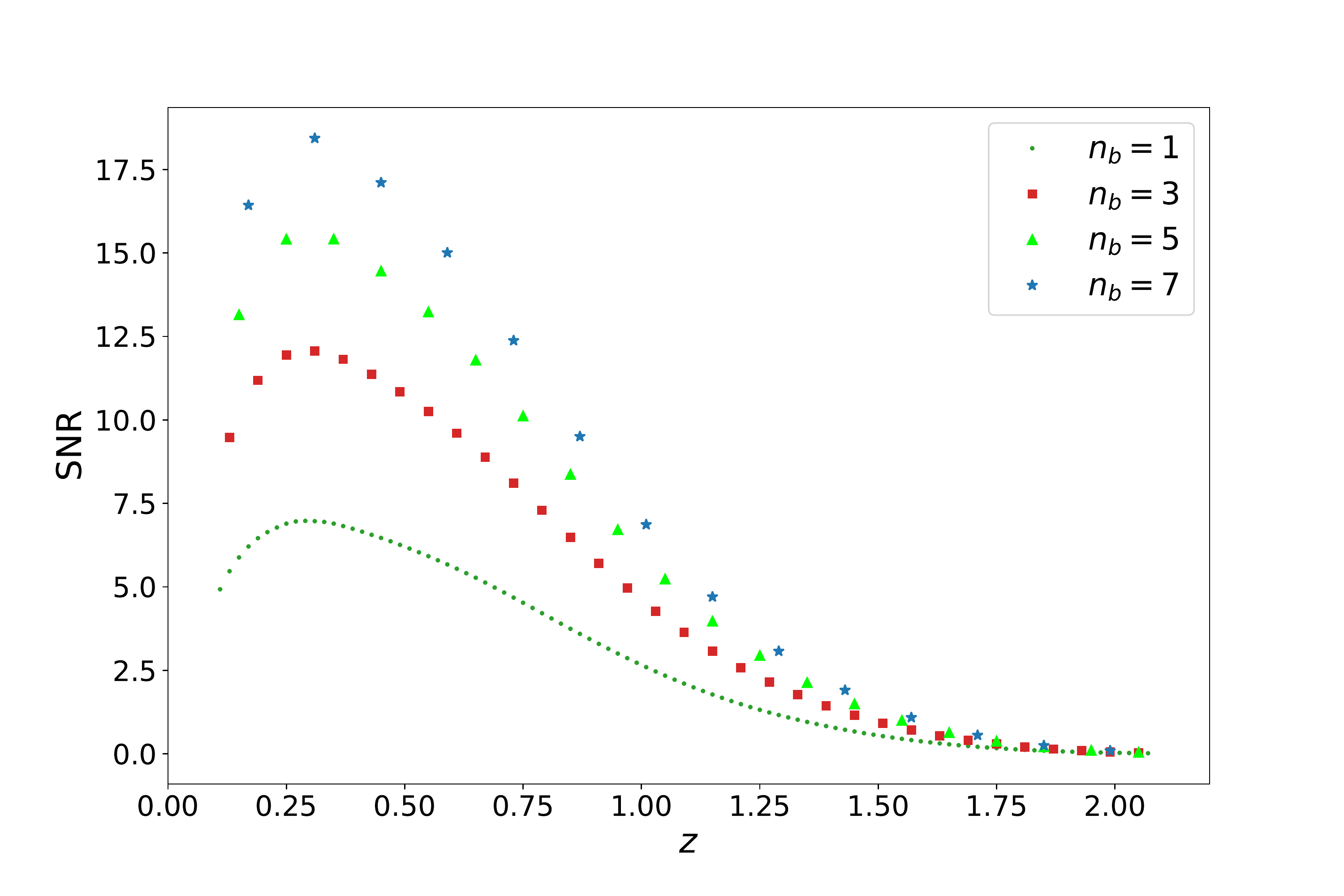}
\caption{Signal-to-noise ratio for $n_b$ averaged $\zeta$'s.}
\label{fig:snr_average}
\end{figure}

The anti-symmetric estimator evolves in redshift as expected but, as we see on the right panel of \autoref{fig:zeta_ska2_072}, this redshift evolution is {only} visible on large scales (low $\ell$), {although such differences are small in comparison with the amplitude of the signal}. We can take this to our advantage to improve the detectability of a single anti-symmetric estimator. We can average a set of {consecutive} $\zeta$'s to improve the error bars as seen in \autoref{fig:detectability_zeta_sk2}. Again we take the example of $\zeta(0.71,0.73)$ and we show the effects of choosing different $\ell$-binning and averaging over consecutive $\zeta$'s. The redshift averaging here is equivalent to the weighting scheme used in \cite{Bacon:2014uja} to enhance the Doppler Dipole effect. As expected having a larger $\Delta\ell$ and average over a few bins decreases the errors bars. {Comparing the left panel of \autoref{fig:zeta_ska2_072} and the right panel of \autoref{fig:detectability_zeta_sk2} one can see by eye that the amplitude on the anti-symmetric estimator can only be explained by the Doppler term, as it excludes ``Den+RSD'' only at a few sigma.} One should note that {these manipulations do} not decrease the overall signal-to-noise, only increases the SNR for a single ``measurement'', as one can see in \autoref{fig:snr_average}. This exercise is important so that we can compare with the SNR presented in previous works \cite{Bonvin:2018ckp}, as well as understand how to use future data to detect the wanted signal. {But fundamentally, such manipulation are crucial to disentangle clustering contributions from density and RSD terms from the ones from the Doppler term.}

\section{Application of $\zeta$: tests of the Euler Eq. and Equivalence principle} \label{sec:testEuler}

The Euler Equation follows from the {spatial} part of the {energy}-momentum tensor conservation
\be
\partial_\mu T^{\mu\nu}=0\,,
\ee
which in a perturbed FRW, with metric {as defined in \autoref{eq:frwpertmetric},} becomes
\be
\label{eq:Euler}
\vec{v}' +{\cal H}\vec{v}+\nabla \Psi =0\,.
\ee
The Doppler term takes advantage of the Euler equation to arrive to the form presented in {\autoref{eq:delta_real}} from its original 
\be 
\delta_{{\rm Doppler}}(z,\hat n)=\frac1{\cal H} ({\hat n}{\cdot}\nabla) \Psi - \left[{A^A_{\mathrm{D}}(z)}-1\right] \vec{v}{\cdot}{\hat n} +\frac1{\cal H} \vec{v}'{\cdot}{\hat n}\,.
\ee
Therefore if the Euler equation is different from \autoref{eq:Euler} then the measured Doppler term will have a different amplitude. Such modifications to the Euler equation have been studied by \cite{Bonvin:2018ckp} and \cite{Umeh:2020cag} providing a test of the equivalence principle. The authors of \cite{Bonvin:2018ckp} go through great detail to explain potential origins of modifications of the Euler equation. Here we will not restate them and refer the reader to their clear and enthusing summary. In practice, one is testing the weak equivalence principle and check if baryons in dark matter haloes feel the potential gravitational effects in the same manner (i.e., if no velocity bias between the baryons and the Dark matter exists). In this paper, we, therefore, follow a theory agnostic approach as in \cite{Bonvin:2018ckp} and use a phenomenological approach to the potential modifications to the Euler equation, i.e, 
\be
\label{eq:Eulermod}
\vec{v}' +{\cal H}\left [1+\Theta({z})\right] \vec{v}+\left [1+\Gamma({z})\right]\nabla \Psi =0\,.
\ee
The generic functions $\Theta$ and $\Gamma$ depend on the particular modified theory of gravity one considers. In this phenomenological approach we will take
\be
\Theta({z})=\frac{1-\Omega_{{\mathrm{m}}}({z})}{1-\Omega_{{\mathrm{m}},0}}\Theta_0 \,, \quad \Gamma({z})=\frac{1-\Omega_{{\mathrm{m}}}({z})}{1-\Omega_{{\mathrm{m}},0}}\Gamma_0
\ee
as in \cite{Bonvin:2018ckp}. Using the modified Euler (ME) equation the real space Doppler term becomes 
\be \label{eq:modDop_eul}
\delta^{{\mathrm{ME}}}_{{\mathrm{doppler}}}(z,\hat n)=- \left[{A^A_{\mathrm{D}}(z)}+\Theta({z})\right] \vec{v}{(z)\cdot}{\hat n}+\frac{\Gamma({z})}{\cal H}{\hat n}{\cdot}\nabla \Psi{(z)}\,.
\ee
Alternatively one can write the generic Doppler transfer function in $\ell$-space as
\be \label{eq:modEul_trans}
\Delta^{{\mathrm{ME,D}}}_\ell(k)=\left[ \l{A^A_{\mathrm{D}}}+\Theta\r  v_k  +\frac{\Gamma}{\cal H}  k\Psi_{k} \right] \frac{\partial j_\ell(k\chi)}{\partial k\chi} \,.\\
\ee
In {appendix} \ref{app:eulermods} we review the derivation of the $\ell$-space transfer function and how we implemented these changes in CAMB\_sources. 

\begin{table}
\caption{\label{tab:results_euler} Marginal {errors (central column) and conditional errors (right column with superscript C)} on Euler equations' modification parameters for two different bin widths ($\Delta z$) assuming a faint and a bright HI galaxy sample from SKAO2.\\}
\centering
\begin{tabular}{l||cc|cc}
 $\Delta z$ & $\sigma_{\Theta_0}$  & $\sigma_{\Gamma_0}$ & $\sigma^{{\mathrm{C}}}_{\Theta_0}$  & $\sigma^{{\mathrm{C}}}_{\Gamma_0}$ \\
 \hline \hline
$0.01   $ & 6.5 & 5.9 &0.50 & 0.45 \\
$0.02   $ & 8.0 & 7.0 &0.61 & 0.54  \\
\hline
\end{tabular}
\end{table}

\begin{figure}[]
\centering
\vspace*{-0.5cm}
\includegraphics[width=10.cm]{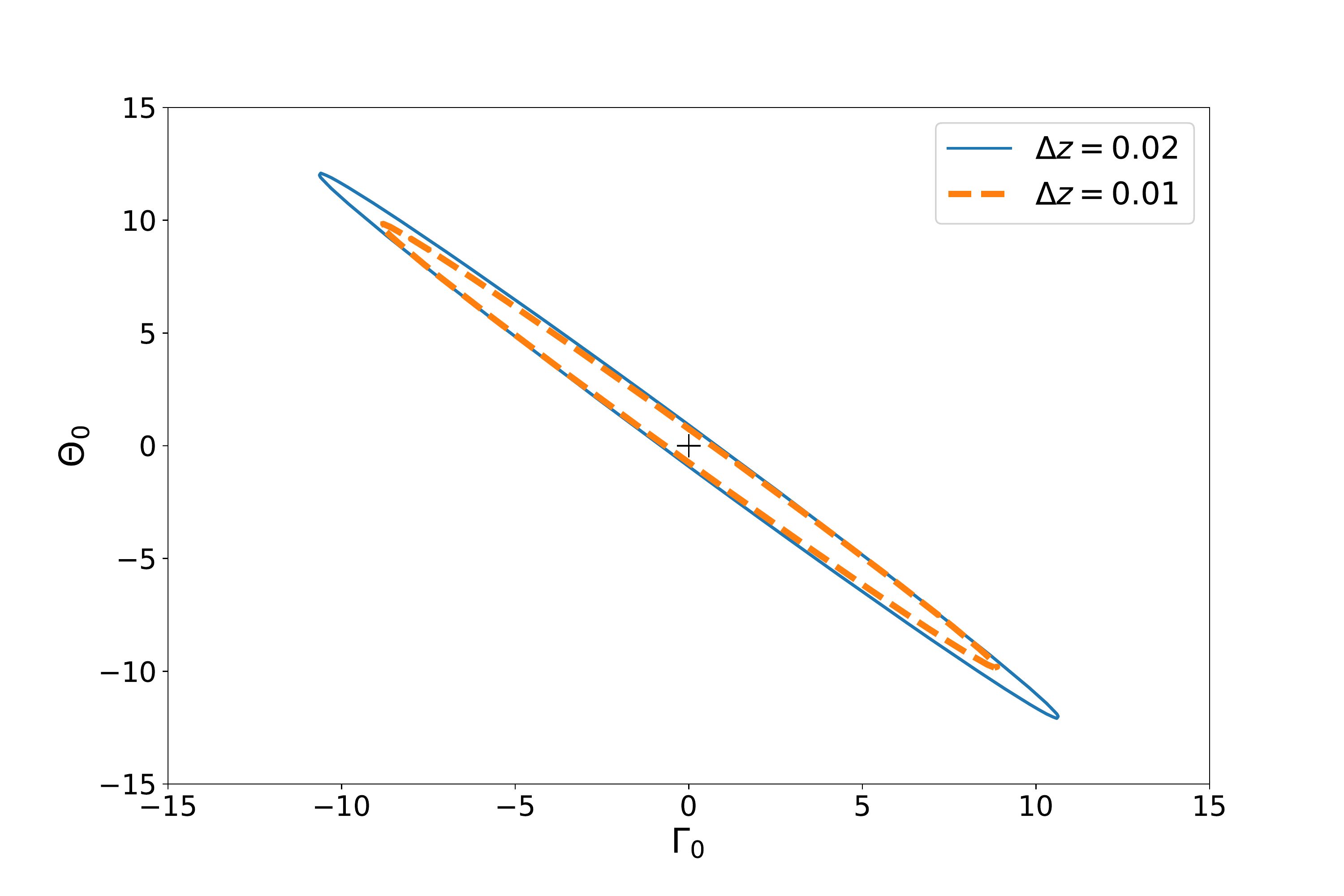}
\caption{One sigma contours for Euler equation modifications' parametrisation.}
\label{fig:contours_modEul}
\end{figure}

With these tools in hand we ran a simple fisher forecast (see \autoref{eq:fisher}) for the set of parameters $\vartheta=\{\Theta_0,\Gamma_0\}$. As before, we will fix the cosmology and assume that tight constraints on them have already been put by other surveys or estimators. From \autoref{tab:snr_comb_surv} one has already seen that the combination of the faint and bright samples of an HI galaxy survey with SKAO2 has the highest SNR. Therefore we will only focus on this combination to estimate future constraints on modifications of the Euler equation. We present our results in {the middle column of} \autoref{tab:results_euler}. The constraints are not as good as expected. Fundamentally, in this parametrisation, there is a substantial degeneracy between $\Theta_0$ and $\Gamma_0$, as it is clearly visible in \autoref{fig:contours_modEul}, despite the high constraining power. Still, they are insightful, as it shows that a thinner resolution breaks ({somewhat}) the degeneracy. Fundamentally this degeneracy indicates that changing the friction term in the Euler equation is equivalent to changing how velocities couple with the potential. As noted in \cite{Bonvin:2018ckp}, different modified theories of gravity only modify how velocities couple with potentials or how peculiar velocities feel the cosmological friction term. Hence, in such cases, one can effectively set one of the parameters to zero, which, in our simple forecast, is equivalent to look at the conditional errors only. We present the conditional errors in {the right column of \autoref{tab:results_euler}} where one sees that such a test becomes a stringent one. 

\section{Conclusion}\label{sec:conclusion}

Our set goal for this paper was to build an estimator based on the observed angular power spectrum which was the angular equivalent to the Doppler dipole proposed by \cite{Bonvin:2013ogt} with the {two-point} function. With this in mind, we adapted the anti-symmetric clustering estimator proposed in \cite{Jalilvand:2019bhk} to spectroscopic galaxy surveys. With these tools, we showed that for high redshift resolution one can reconstruct the Doppler term, even if blended with other terms. Fundamentally, within the angular power spectrum, the Doppler term is subdominant, but it is a leading term in anti-symmetric clustering. We then studied the SNR for combinations of LSS surveys (or galaxy samples) and concluded that although there is some signal in upcoming LSS spectroscopic surveys, one needs to wait for more futuristic galaxy surveys for larger SNRs. We then studied the case of the SKAO2 HI galaxy survey with a split between faint and bright galaxies. We first used this as an example to show that indeed the Doppler term is the main contributor to the anti-symmetric estimator $\zeta$. We also showed how to rebin the data to increase the detectability of the signal. We then made use of our anti-symmetric estimator to {assess how well one can} test General Relativity. In particular, the Euler equation using the same approach as \cite{Bonvin:2018ckp}. Although the parametrisation is {suboptimal}, we could see the constraining power in \autoref{fig:contours_modEul}. Some modified theories of gravity will not have the two modifications $\{\Theta_0,\Gamma_0\}$, therefore one can still apply this to those.

In this paper, we focused on the Doppler term. Despite this, it was clear from \autoref{fig:sbe0} to \autoref{fig:sbeim} that if one can rebins the spectroscopic data into thick bins, the clustering anti-symmetry is reproducing the correlation $\sim\langle\delta \kappa\rangle$, as already studied in \cite{Jalilvand:2019bhk} a photometric survey and HI IM. We will explore these possibilities in the future as a way to extract $\kappa$ from spectroscopic surveys. 

There is one natural follow-up question arising from the present work, as well as from the {Doppler} Dipole estimator \cite{Bonvin:2013ogt}).  \emph{What is the equivalent estimator using the 3D power spectrum (or its multipoles) and how does it perform?}. It already exists in the literature as the imaginary part of the 3D cross-power spectrum \citep{McDonald_2009}, or its non-zero odd multipoles. We will pursue in a future paper the 3D power spectrum equivalent of the angular and 2-point function anti-symmetric correlations using similar combinations of current, forthcoming and futuristic surveys.  

\acknowledgments
We thank Jan-Albert Viljoen, Stefano Camera, and Roy Maartens for useful discussions and comments. We thank Phil Bull for discussions on detecting HI galaxies. JF and CC are supported by the UK Science \& Technology Facilities Council (STFC) Consolidated Grant ST/P000592/1. This work made use of the South African Centre for High-Performance Computing, under the project Cosmology with Radio Telescopes, ASTRO-0945.

\newpage
\appendix

\section{Astrophysical details of the surveys} \label{app:survdets}

For each survey or sample, we need four quantities, 3 biases, and the noise which for galaxy surveys is just the scale-independent shot-noise 
\be
{\cal N}_{\ell,ij} =\frac{\delta_{ij}}{N_i}\,,
\ee 
where $N_i$ is the angular number density of galaxies in bin $i$. We also need to know the overlapping footprint of each combination. For simplicity, we will assume the maximum overlap possible, i.e., the overlapping redshift range and the area of the smallest survey. In the case of SKAO's MID HI IM with DESI-like BGS samples and Euclid-like spectroscopic H$\alpha$ galaxy survey, we will assume a 10 000 $\deg^2$. For more details on the calculation of magnification and evolution bias please refer to \cite{Maartens:2021dqy}.



\subsection{Bright Galaxy Sample of a DESI-like survey}

\begin{figure}[]
\centering
\vspace*{-0.5cm}
\includegraphics[width=7.5cm]{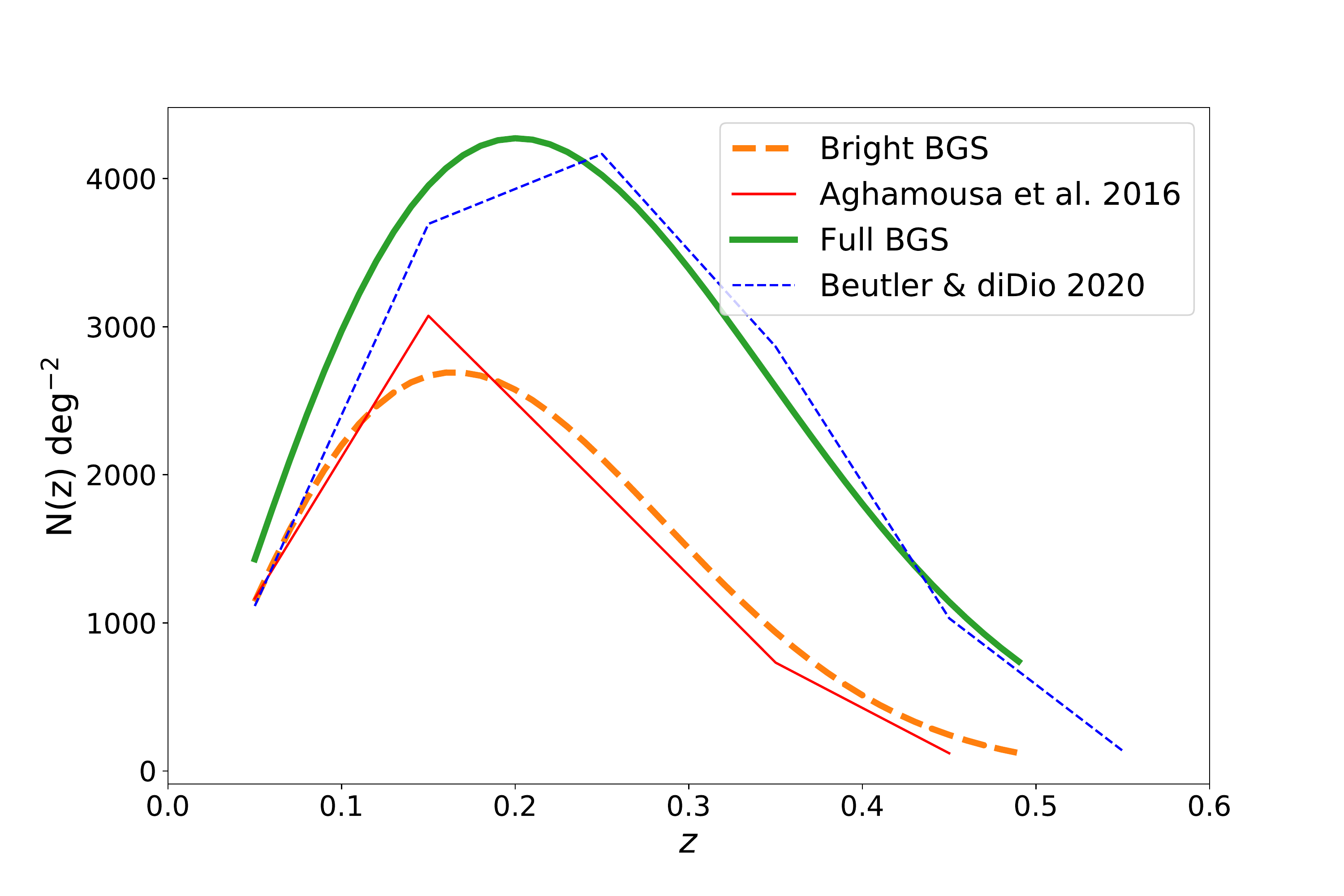}
\caption{Angular number density from luminosity function compared to the ones from presented in papers.}
\label{fig:desi_bgs}
\end{figure}

DESI will observe around 15000 $\deg^2$ of the sky \cite{Aghamousa:2016zmz}. Its lowest redshift sample, the Bright Galaxy Sample, can be approximately described by the luminosity function obtained from the GAMA survey \cite{Loveday_2011}. The luminosity function in terms of the absolute magnitude is written as 
\be
\Phi(M)\ \di M=0.4\ln 10\ \phi(z)\ 10^{0.4(\alpha_G+1)(M_p(z)-M)}\ \exp\{-10^{0.4(M_p-M)}\}\,.
\ee
with 
\bea
\phi(z)&=& \phi_0\ 10^{0.4Pz}\,, \\
M_p(z)&=& M_{p,0}-Q(z-z_0)\,,
\eea
with \cite{Loveday_2011}
\bea
&\alpha_{G}=-1.23\,,\ \phi_0= 9.4\times 10^{-3} ~ ({\mathrm{h/Mpc}})^3\,,\ P=1.8\,,&\nonumber\\
&\ M_{p,0}=-20.70 + 5\log_{10}h\,,\ Q=0.7\,,\ z_0=0.1\,.&
\eea
The absolute magnitude is computed in term of the apparent magnitude using the distance modulus
\be
M=m-25 -5 \log_{10}D_{{\mathrm{L}}}(z)/{\rm Mpc}-K(z)\,,
\ee
where $D_{{\mathrm{L}}}$ is the luminosity distance and $K$ is the k-correction computed. We will take $K=0.87z$ found in \cite{Jelic-Cizmek:2020pkh}. The number density up to the magnitude threshold at a given redshift is 
\be \label{eq:n_lf_mag}
\bar n (z,m<m_*)=\int_{-\infty}^{M(m_*,z)} \Phi(M,z)\ \di M\,.
\ee
We will consider two samples, a ``full'' and a ``bright'' sample, both at $z<0.5$. For the bright sample, we will consider a magnitude threshold of 19.5 with the normalisation of 700 {gal/$\deg^2$} \cite{Aghamousa:2016zmz}. The full sample also includes a further expected 600 gal/$\deg^2$ with $19.5<m<20$ \cite{Beutler:2020evf}. In this sample, we will not consider the redshift range down to redshift zero but truncate it instead at $z=0.1$. In \autoref{fig:desi_bgs} we plot the number density of the full and bright samples computed from the GAMA luminosity function (\autoref{eq:n_lf_mag}) compared to the tabled numbers found in \cite{Aghamousa:2016zmz,Beutler:2020evf}. Therefore we take our approximation to be good enough for this paper. We will use \cite{Aghamousa:2016zmz} for the bias of the Bright sample and \cite{Beutler:2020evf} for the bias of the Full sample. In \autoref{fig:surv_dets} we plot the number density of sources and the biases of the Full BGS-like sample in thick solid green and the Bright BGS-like sample in thick dashed orange. 

\subsection{SKAO MID HI IM}

In the case of HI intensity mapping the angular power spectrum is computed for fluctuations in $\Delta(\hat n,z)=T_{{\mathrm{HI}}}(1+\delta)$ where the temperature is \cite{Bull:2014rha}
\be \label{eq:TvsnHI}
T_{{\mathrm{HI}}}=\frac{3 h c^3 A_{10} }{32\pi \nu_{21}^2k_{\rm B}}\frac{(1+z)^2}{ H(z)} n_{{\mathrm{HI}}}\,.
\ee
Here we will use the has the fit \cite{Bacon:2018dui}
\be
T_{\rm HI}=0.056 + 0.232 z - 0.024 z^2\,.
\ee 
Similarly we will use the fits of \cite{Bacon:2018dui} for the bias, i.e., 
\be
b_{\rm HI}=0.666 + 0.178 z - 0.050 z^2\,.
\ee
For intensity mapping we recover the correct HI transfer function in \autoref{eq:deltatot} by setting $s_{\rm HI}=0.4$. The evolution bias is computed in \cite{Maartens:2021dqy}. 
In \autoref{fig:surv_dets} we plot the HI biases in thick dotted purple. 

In IM the observed angular power spectrum results from the ``true'' power spectrum that has gone through the instrument's optics. This is modelled by the instruments' beam that, to first order, we approximate by a Gaussian. In harmonic space it becomes 
\be
B_\ell(z)=\exp\left\{\frac{-\ell(\ell+1)\theta_{{\rm FWHM}}^2(z)}{16\ln 2}\right\}\,,
\ee
where the angular resolution is
\be
\theta_{{\rm FWHM}}(z)=1.22 \frac{\lambda_{\rm HI}(1+z)}{D_{{\rm dish}}}\,,
\ee
and $D_{{\rm dish}}$ is the diameter of the dishes and $\lambda_{{\mathrm{HI}}}=21$cm the wavelength of the HI line. Then the observed angular power spectrum is given by
\be
C^{{\mathrm{HI}}, {\rm obs}}_\ell \!\l z_i,z_j\r=B_\ell(z_i) B_\ell(z_j) C^{{\rm HI}}_\ell \!\l z_i,z_j\r\,.
\ee
The instrumental noise we assumed uncorrelated Gaussian noise, i.e., 
\be \label{eq:inst_noise_hiim}
{\cal N}_{\ell,ij} =\frac{4\pi\,f_{\rm sky}\, T_{{\rm sys},i}^2}{2\, N_{\rm d}\, \Delta \nu\, t_{\rm tot}}
\, \delta_{ij}\,.
\ee 
We will take the same system temperature specifications as in \citep{Bacon:2018dui} for both band 1 and band 2 of SKAO MID. The SKAO's MID IM survey is planned to cover 20 000 $\deg^2$ over 10 000 hours with 197 dishes detecting neutral hydrogen emission up to $z\lesssim 3$. Here $\Delta \nu$ corresponds to the bin width. When we combine IM with galaxy surveys we keep the scanning ratio constant and adapt the integration time accordingly.  

\subsection{H$\alpha$ spectroscopic sample with a Euclid-like survey}

\begin{figure}[]
\centering
\vspace*{-0.5cm}
\includegraphics[width=7.5cm]{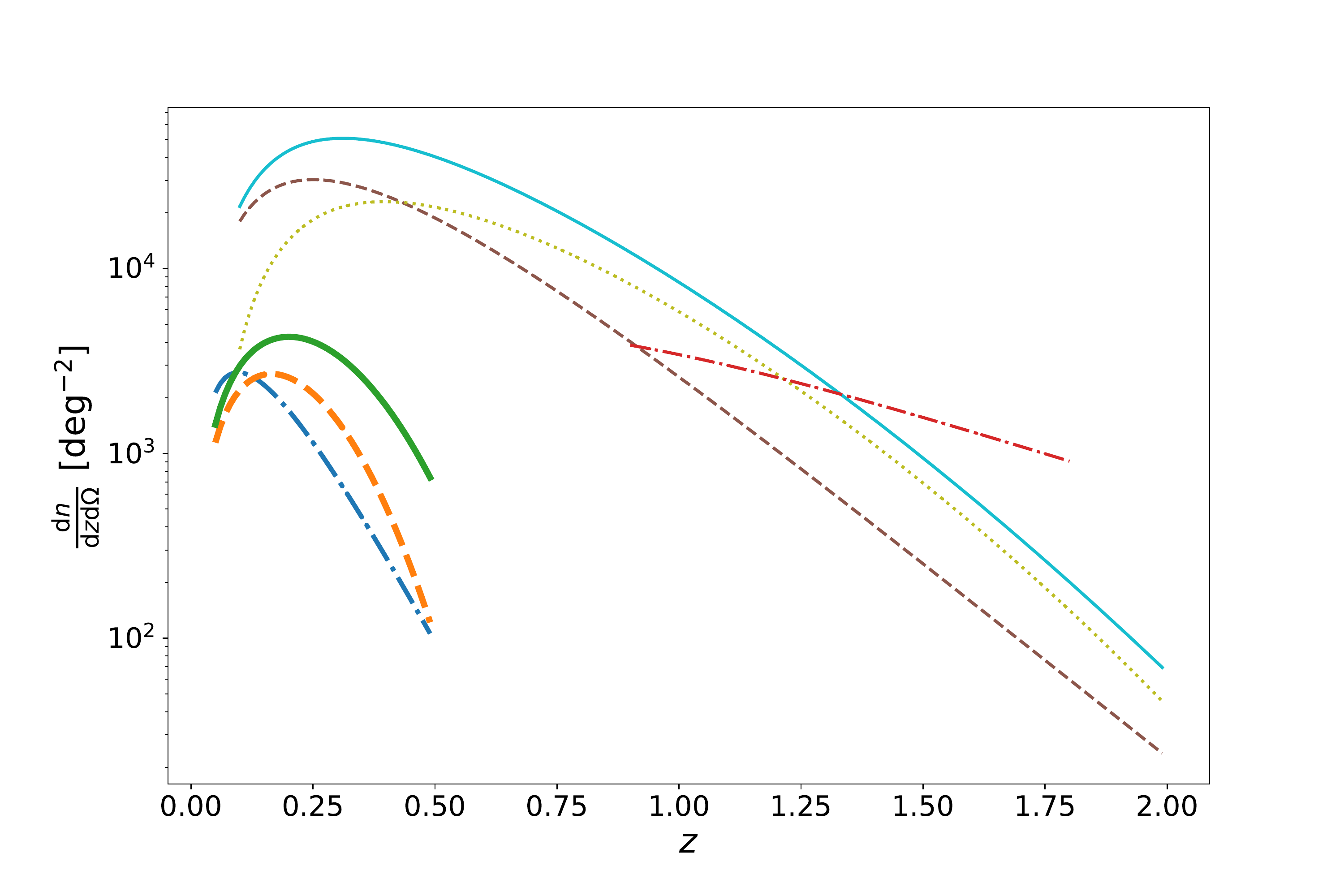}
\includegraphics[width=7.5cm]{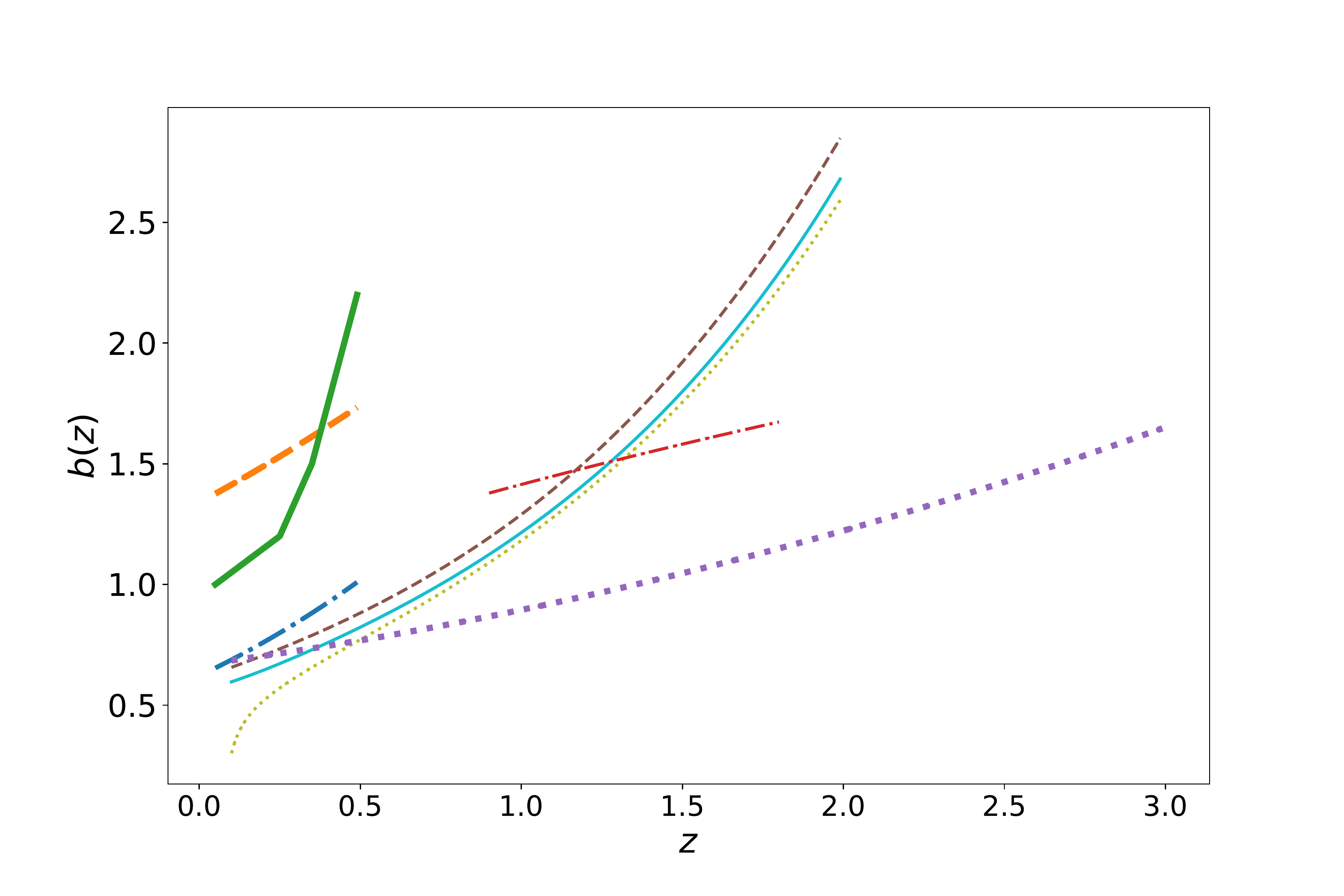}
\includegraphics[width=7.5cm]{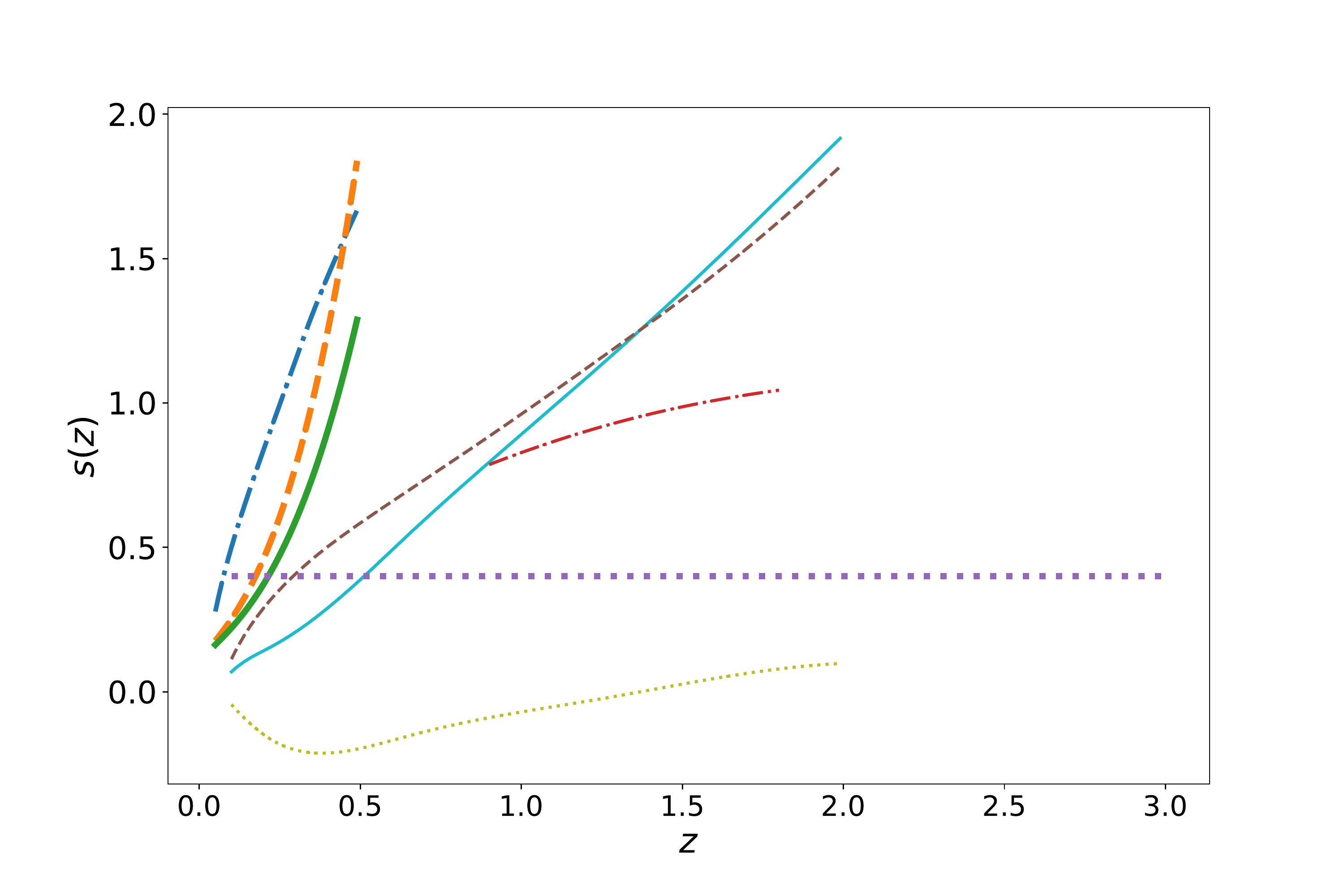}
\includegraphics[width=7.5cm]{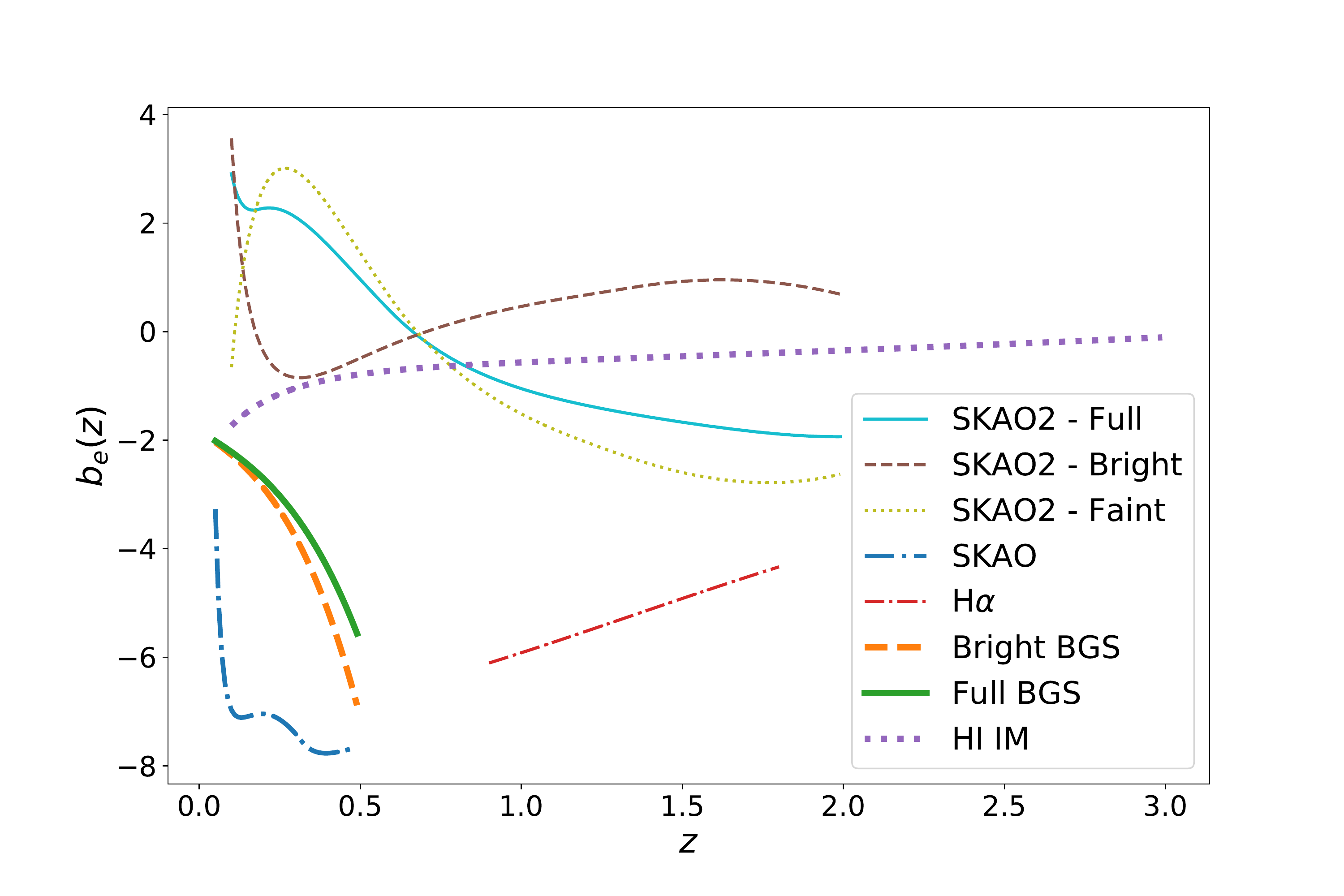}
\caption{\emph{Top Left}: Angular number density for the different surveys considered.\emph{Top Right}: Clustering bias of the considered samples. \emph{Bottom Left}: Magnification bias of the considered samples. \emph {Bottom Right}: Evolution bias of the considered samples}
\label{fig:surv_dets}
\end{figure}

To model the H$\alpha$ spectroscopic sample we will adopt Model 3 of \cite{Pozzetti:2016cch} for the H$\alpha$ Luminosity function, i.e., 
\be
\Phi(L,z)\ \di L=\frac{\phi_\star}{L_\star(z)}\l\frac{L}{L_\star(z)}\r^{\alpha_{{\rm H\alpha}}}\left[1+(e-1)\l \frac{L}{L_\star(z)}\r^\Delta\right]^{-1} \di L\,,
\ee
with
\be
\log_{10} L_\star(z)=\log_{10} L_{\star,\infty}+\l\frac{1.5}{1+z}\r^\beta\log_{10}\frac{L_{\star,0.5}}{L_{\star,\infty}}\,.
\ee
We took their best fit values which are \cite{Pozzetti:2016cch}  
\bea
\alpha_{{\rm H\alpha}}=-1.587\,,\ \log_{10}(\phi_\star {\rm Mpc}^3)= -2.920 \,,\ \Delta=2.288\,,\ \beta=1.615\, \nn\\
\log_{10}(L_{\star,\infty}/{{[}\rm erg\ s^{-1}{]})}=42.956\,,\ \log_{10} (L_{\star,0.5}/{\rm {[}erg\ s^{-1}{]}})=41.733.\nn
\eea
Then the number density is simply given by
\be
\bar n (z,L<L_*)=\int^{\infty}_{L_*(z)} \Phi(L,z)\ \di L\,,
\ee
where the luminosity cut is given by the flux threshold 
\be
L_*(z)=4\pi D_{{\mathrm{L}}}^2(z)\ F_*\,.
\ee
For a Euclid-like survey, we will take the flux threshold of $F_*=2\times 10^{-16}$ erg/s/cm$^2$. We will also assume a footprint of $15000\deg^2$ and a redshift range $z\in[0.9,1.8]$. For the H$\alpha$ galaxies clustering bias we will follow \cite{Amendola:2016saw}. In \autoref{fig:surv_dets} we plot the redshift distribution and biases for the H$\alpha$ spectroscopic survey considered here in HI biases in thin dot-dashed red. 

\subsection{HI galaxies with SKAO}

As for optical galaxies, the number density of HI galaxies is dependent on the sensitivity threshold of the instrument. In \cite{Maartens:2021dqy} we find an explanation how such flux cut $S_{\rm cut}$ is computed from instrumental specifications. We follow their results and descriptions for SKAO and SKAO2.
Once the $S_{\rm cut}$ is computed as a function of redshift for a given survey we use the calibrations of \cite{Yahya:2014yva} to find the correct coefficients $\{c_1,c_2,c_3,c_4,c_5\}$ for the distribution of sources and the bias
\bea
\frac{\di n_{\rm HI\ gal}}{\di z\di \Omega}&=&10^{c_1}z^{c_2}\exp\l-c_3 z\r\,,\\
b_{\rm HI\ gal}&=&c_4z^{c_2}\exp \l c_5 z\r\,.
\eea
Once we obtain $\{c_1,c_2,c_3\}$ as a function of redshift and flux cut we follow \cite{Maartens:2021dqy} to compute the magnification and evolution biases.


For SKAO2 we will further divide into a ``faint'' and ``bright'' sample with the redshift dependent cut 
\be
S_{\rm cut}^{\rm bright}=\frac12 \left( 5 + \frac{21}{1+z}\right)~{\rm \mu Jy}\,,
\ee 
which, although arbitrary, makes the number density of faint and bright HI galaxies the same order of magnitude. One should note that the faint bias can be computed from the full sample and the bright sample, i.e., 
\be
b^{\rm faint}=\frac{b^{\rm all}n^{\rm all}-b^{\rm bright}n^{\rm bright}}{n^{\rm faint}}\,,
\ee
with $n^{\rm faint}=n^{\rm all}-n^{\rm bright}$. The magnification bias of the faint sample should be a balance between the flux cut at the faint and bright ends, which we take it to be
\be
s^{\rm faint}=s^{\rm all}-s^{\rm bright}\,.
\ee
Note that the flux cut between the two samples should also contribute to the faint magnification bias. Such contribution should have the same magnitude, but opposite sign, to the magnification bias of the bright sample. 

In \autoref{fig:surv_dets} we plot the redshift distribution and biases for the SKAO MID HI galaxy sample in thick dot-dashed blue, the full SKAO2 HI galaxy sample in thin solid cyan, the bright SKAO2 HI galaxy sample in thin dashed brown, and the faint SKAO2 HI galaxy sample in thin dotted yellow.


\section{Implementation of Euler Eq. modification in CAMB} \label{app:eulermods}

The modified Doppler term in \autoref{eq:modDop_eul} has two changes with respect to the common implementation. The first change, given by $\Theta(\eta)$ only alters the amplitude of the standard term. We only need to compute the $\alm$ and $\Delta_\ell$ of the gradient of the potential term. Let us call $W_{Nz}(z)$ to the selection function which includes the redshift distribution of sources and the window function. The gradient part we need to still compute the $a_{\ell m}$, which is given by
\bea
\alm^{{\nabla \Psi},A}(z_i)&=& \int_0^{\infty} {\rm d}z\ W_{Nz}(z{,z_i}) \int \don\ \frac{\Gamma}{\cal H} {\hat n}{\cdot}\nabla \Psi\ \ylmc{n}\nn\\
&=&\int {\rm d}z\ W_{Nz}(z{,z_i}) \int \don\ \frac{\Gamma}{\cal H}\frac{\partial}{\partial \chi} \left[\frac1{\l2\pi\r^3}\int \dt k\ \Psi_{\vk}\ e^{i \chi\vec{k}{\cdot}\hat{n}}\right] \ylmc{n}\nn\\
&=&\frac1{\l2\pi\r^3}\int {\rm d}z\ W_{Nz}(z{,z_i}) \frac{\Gamma}{\cal H}\frac{\partial}{\partial \chi} \int \dt k\ \Psi_{\vk}\ \int \don\ e^{i \chi\vec{k}{\cdot}\hat{n}}\ylmc{n}\nn\\
&=&\frac{i^\ell}{2\pi^2}\int {\rm d}z\ W_{Nz}(z{,z_i}) \frac{\Gamma}{\cal H} \int \dt k\ k\Psi_{\vk}\ \frac{\partial}{\partial k\chi}j_\ell(k\chi)\,,
\eea
where we used the fact that 
\be
\int \don\ e^{i \chi\vec{k}{\cdot}\hat{n}}\ylmc{n}= 4\pi i^\ell j_\ell(k\chi)\,.
\ee
The correction to the Doppler {kernel} function is given by
\bea
\Delta^{\nabla \Psi}_\ell (k,z_i)&=& \int_0^{\infty} {\rm d}z\ W_{Nz}(z{,z_i})\frac{\Gamma}{\cal H}\ k\Psi_{\vk}\ \frac{\partial}{\partial k\chi}j_\ell(k\chi)\,,\nn\\
&=&\int_0^{\eta_0} {\rm d}\eta\ H W_{Nz}(z{,z_i})\frac{\Gamma}{\cal H}\ k\Psi_{\vk}\ \frac{\partial}{\partial k\chi}j_\ell(k(\eta_0-\eta))\,,\nn\\
&=&-\int_0^{\eta_0} {\rm d}\eta\ \frac{W_{Nz}{(z_i)}}{a}\Gamma\ \Psi_{\vk}\ \frac{\partial}{\partial \eta}j_\ell(k(\eta_0-\eta))\,,\nn\\
&=&\int_0^{\eta_0} {\rm d}\eta\ \left[\l\frac{W_{Nz}{(z_i)}\Gamma}{a}\r'\ \Psi_{\vk}+\frac{W_{Nz}{(z_i)}\Gamma}{a}\ \Psi_{\vk}'\right]j_\ell(k(\eta_0-\eta))\,,
\eea
where we use the fact that $\di \chi=-\di \eta$ and in the last step we used integration by parts given that $W_{Nz}$ vanishes in the boundaries. Note that we write the transfer function this way as the numerical integrals are all made with respect to $j_\ell$.

\newpage
\bibliographystyle{JHEP}
\bibliography{Bibliography}

\end{document}